\documentclass[aps,prl,twocolumn,a4paper,superscriptaddress
,10pt] {revtex4-1}
\usepackage[colorlinks,citecolor=blue]{hyperref}
\usepackage{hypernat}
\usepackage{graphicx}
\usepackage{amsmath}
\usepackage{amsfonts}
\usepackage{amssymb}
\usepackage{amsthm}
\usepackage{times}
\usepackage{color}

\begin{document}
\title{ Anomalous Heat Conduction and Anomalous Diffusion in
Low Dimensional Nanoscale Systems}

\author{Sha Liu}
\affiliation{
    NUS Graduate School for Integrative Sciences and Engineering,
    Singapore 117456, Republic of Singapore
}

\author{Xiangfan Xu}
\author{Rongguo Xie}
\affiliation{
Department of Physics and Centre for Computational Science and Engineering,
National University of Singapore, Singapore 117546, Republic of Singapore
}
\author{Gang Zhang}
 \email{zhanggang@pku.edu.cn}
\affiliation{
Key Laboratory for the Physics and Chemistry of Nanodevices and Department of
Electronics, Peking University, Beijing 100871, P. R. China
}
\author{Baowen Li}
 \email{phylibw@nus.edu.sg}
\affiliation{
Department of Physics and Centre for Computational Science and Engineering,
National University of Singapore, Singapore 117546, Republic of Singapore \\
NUS-Tongji Center for Phononics and Thermal Energy Science,
Department of
Physics, Tongji University, 200092 Shanghai, P. R. China
}
\date{May 10, 2012}
\begin{abstract}
Thermal transport is an important energy transfer process in nature. Phonon is
the major energy carrier for heat in semiconductor and dielectric materials. In
analogy to Ohm's law for electrical
conductivity, Fourier's law is a fundamental rule of heat transfer in solids. It
states that the thermal conductivity is independent of sample scale and
geometry. Although
Fourier's law has received great success in describing macroscopic thermal
transport in the past two hundreds years, its validity in low dimensional
systems is still an open question. Here we give a brief review of the recent
developments in
experimental, theoretical and numerical studies of heat transport in low
dimensional systems, include lattice models, nanowires, nanotubes and graphenes.
We will demonstrate that the phonon transports in low dimensional systems
super-diffusively, which leads to a size dependent thermal conductivity. In
other words, Fourier's law is breakdown in low dimensional structures.
\end{abstract}

\maketitle

%%THIS PART DEFINES SOME NEW COMMAND
\newcommand{\textcitee}[1]{Ref.~\cite{#1}}
\newcommand{\pt}[2]{\partial_{#2}#1}
\newcommand{\mean}[1]{\left\langle #1 \right\rangle}
\newcommand{\grad}{\nabla}
\newcommand{\kp}{\kappa}
\newcommand{\ep}{\epsilon}
\newcommand{\Dt}{\mathrm{\Delta}}
\newcommand{\dt}{\delta}
\newcommand{\expk}{\beta}
\newcommand{\expx}{\alpha}
\newcommand{\dev}{\mathrm{\Delta}}
\newcommand{\dd}{\mathrm{d}}
\newcommand{\ee}{\mathrm{e}}
\newcommand{\JJ}{\mathcal{J}}
\newcommand{\ii}{\mathrm{i}}
\newcommand{\mbf}[1]{{\bf #1}}
\newcommand{\dif}[2]{\frac{\dd #1}{\dd #2}}
\newcommand{\diff}[3]{\frac{\dd^{#3} #1}{\dd #2^{#3}}}
\newcommand{\msd}[1]{\overline{x^2}(#1)}
\newcommand\emsd{\beta}
\newcommand{\etal}{{\it et~al.}}
\newcommand{\ie}{{\it i.e.}}
\newcommand{\noseh}{Nos\'e-Hoover}
\newcommand{\FPUb}{FPU-$\beta$}
\newcommand{\phif}{$\phi^4$}
\newcommand{\mwith}{\mathrm{\quad with \quad}}
\newcommand{\levy}{L\'evy}
\newcommand{\reffig}[1]{Fig.~\ref{#1}}
\newcommand{\sectionN}[2]{\section{#1}\label{sec:#2}}

\newcommand{\includegraphicsS}[2]{
\begin{center}
\resizebox{#1}{!}{\includegraphics{#2}}
\end{center}
}

%\pacs{44.10.+i, 63.22.-m, 66.70.-f, 05.70.Ln}

\section{Introduction}

The conduction of heat is one fundamental energy transport mechanisms in nature.
In addition to electron and photon, phonon- the heat pulse through lattice, also
carries and processes energy, which has broad applications for heat
control/management in the real world. Actually, for non-metallic materials,
phonon is the most dominated heat energy carrier. Its contribution to heat
conduction is much larger than those from electrons and photons. Traditionally,
the phenomenon of thermal transport is believed to follow the Fourier's law of
heat conduction
\begin{equation}
 \mathbf{J}=-\kappa \grad{T},
 \label{eq:fourier}
\end{equation}
where $\mathbf{J}$ is the heat flux in the system, $\grad{T}$ is the gradient of
temperature. The conductivity $\kappa$ is a geometry-independent
coefficient which mainly depends on the composition and structure of the
material and the temperature.
This empirical law has received great success in describing macroscopic thermal
transport and is widely accepted as a general truth.

In the past decades, low dimensional nano scale systems have been extensively
studied due to their promising potential applications for future electronic,
optoelectronic, and phononic/thermal devices. In addition to the electrical and
optical properties, the thermal properties of nanoscale materials are also
important but less studied compared with electronic and optical properties. The
thermal properties in nanostructures differ significantly from their bulk
counterpart because the phonon characteristic lengths are comparable to the
characteristic length of the nanostructures. However, it is still an open and
much debated question whether the Fourier's law is valid in low dimensional
systems. A rigorous proof for this empirical law from microscopic Hamiltonian
dynamics is still absent even though it is already two-hundred-year old. Here
``invalid'' means the thermal conductivity $\kappa$ will depend on the system
size.

Such size dependent thermal conductivity has
been observed in theoretical models, such as the harmonic chains
\cite{JMP.67.Rieder,PTP.68.Nakazawa},
the
\FPUb{} model \cite{PRL.97.Lepri,EL.98.Lepri} and the hard point gas model
\cite{PRE.99.Hatano,PRL.01.Dhar,PRL.02.Grassberger}.
The discovery of this anomalous behavior in general low dimensional models has
then inspired enormous research studies. It is
found that in these models, the thermal conductivity typically diverges with
system size:
\begin{equation}
 \kappa\sim L^{\expk}
\end{equation}
for 1D models and
\begin{equation}
 \kappa\sim \log L
\end{equation}
for 2D models in thermodynamic limit, which predicts extraordinarily high
thermal conductivity for low dimensional systems with
finite but large sizes. Therefore, not only it is a fundamental demand for the
development of statistical physics to understand
normal and anomalous heat conduction in low dimensional systems, but it is also
of great interest from the
application point of view, since the achievement of modern nano fabrication
technology allows one to access and utilize one dimensional (1D) and two
dimensional (2D) structures with sizes in the range of few nanometers up to few
hundred nanometers.

In recent years, much effort has been devoted to study the anomalous heat
conduction in low dimensional nanostructures by numerical
simulations such as \cite{zhang2005JCP123B} and \cite{Yang2010NT5}. These
studies provide evidence that low dimensional nanostructures are very promising
platforms to verify fundamental thermal transport theories. Inspired by these
theoretical and numerical studies, it is exciting that length dependent thermal
conductivity has been observed in carbon nanotubes and boron-nitride
nanotubes \cite{Chang2008PRL}.
Unfortunately, an
experimental investigation of the quantitative size dependent thermal
conductivity in nanoscale is exceptionally challenging. The difficulty lies
primarily with the technique associated with measuring temperature distribution
and heat flux of low dimensional systems, with the added complexity of
accurately varying the size of the investigated object. Due to these challenges
in experimental measurement, the experimental data need to be complemented by
theoretical study on a microscopic footing.

Generally, heat transport can be studied in the aspects of two different yet
deeply related phenomena. One is to study the heat flux in a non-equilibrium
steady state to get the heat conductivity as we normally do in simulation or in
experiments, in which temperature biases are applied to the system. The other
approach is to study the heat diffusion in a non-equilibrium transient process
in which the system is firstly excited away from equilibrium and then relax to
equilibrium. Both approach should reveal some properties of heat conduction of
the system. Moreover, the connection between thermal conductivity and heat
diffusion, especially for the anomalous cases, is still an open question and
attract wide research interests.

As heat conduction in low dimensional systems is a fast growing area,
combination of experimental, theoretical and numerical investigations are
indispensable to speed up its development. In this articles, we would like to
give a review on the recent development of heat transport in low dimensional
systems, from both experimental and theoretical point of view. The rest of the
article is organized as follows: section \ref{sec:experi}  introduces the
experiments on quasi-1D nanostructures and 2D graphene that display the
anomalous size dependent thermal conductivity. Section \ref{sec:zhang} discusses
the numerical studies on thermal conduction of nanotubes, nanowires and
graphenes, which would shed light on understanding the phenomenon of
size dependent thermal conductivity from a theoretical point of view. Section
\ref{sec:liu} is devoted to the general theories on anomalous heat conduction
and anomalous energy diffusion in low dimensional systems. Finally, in section
\ref{sec:conclusion} and \ref{sec:outlook}, we present a few conclusions and a
short outlook.

There exists a large amount of literatures and studies on different aspects of
thermal property in low dimensional systems. For a comprehensive review on
thermal property of nano materials, please refer to these articles
Refs.~\cite{zhang2010NS2,Pop2010NR3,RMP.12.Li}. There are also a
few reviews
talk about anomalous heat transport in low dimensional systems from the
viewpoint of fundamental statistical physics, such as
Ref.~\cite{PR.03.Lepri} and \cite{AP.08.Dhar}. Due to the limit of space, we
only address the most fundamental physical issue, i.e., the violation of
Fourier's law in this article.

\sectionN{Experimental Observation of Anomalous Heat
Conduction in Real Nano Materials}{experi}

\subsection{Anomalous Heat Conduction in Quasi 1D Nanostructures}

When the dimensions of materials shrink to the nanoscale,
thermal transport properties can be very  different from
those of their counterpart bulk materials. Recent
experimental investigations in (quasi) 1D
structures have revealed two opposite intriguing properties:
unusual high thermal conductivities and significantly
suppressed thermal conductivities. The unusual high thermal
conductivities are attributed to the unique crystalline
structures of the 1D materials, such as carbon
nanotubes  \cite{Nanotube}, and stretched polymer
nanofibres  \cite{Shen2010Nnano}, whereas the significantly
suppressed thermal conductivities are due to an increased
phonon-boundary scattering, which has been observed in
various nanowires, such as
Si
 \cite{Li2003APL,Chen2008PRL,Hochbaum2008Nat451},
Si/SiGe superlattice  \cite{li2003APL_SiGe},
ZnO  \cite{Bui2012small}, Bi  \cite{Moore2009JAP}, etc. More
recently, it has been observed that the thermal conductivity
can be further suppressed by the coherent phonon resonance
%phonon confinement
effect
beyond the phonon-boundary scattering limit in Ge-Si
core-shell ultrathin nanowires  \cite{Wingert2011NL11}.

Although significant experimental efforts have been devoted
to study the diameter-dependence of the thermal
conductivity to understand the phonon-boundary scattering
mechanisms in the nanostructures, the study of length
dependence of the thermal conductivity to address the
validity of Fourier's Law in 1D nanostructure remains rare.
The primary technical challenges lie in the problem of
suspicious thermal contact resistance and the difficulty in
guaranteeing the diameter exactly the same when varying the
length of the samples in different measurements. In 2008,
Chang \etal{} developed a sequential multiprobe method that
can establish the deviation from Fourier's Law behavior in
one-dimensional nanostructures, without suffering from the
problem of suspicious contact resistance
 \cite{Chang2008PRL}. A suspended microelectromechanical
system (MEMS) device was used to measure thermal
conductivity of carbon nanotubes (CNTs) and boron-nitride
nanotubes (BNNTs) as a function of length. Unlike the
traditional method using a movable probe as a local probe,
Chang \etal{} deposited a series of thermal contacts to vary
the length along the same sample, thus ensuring the diameter
of the sample unchanged.

\begin{figure}[tb]
\includegraphicsS{0.7\columnwidth}{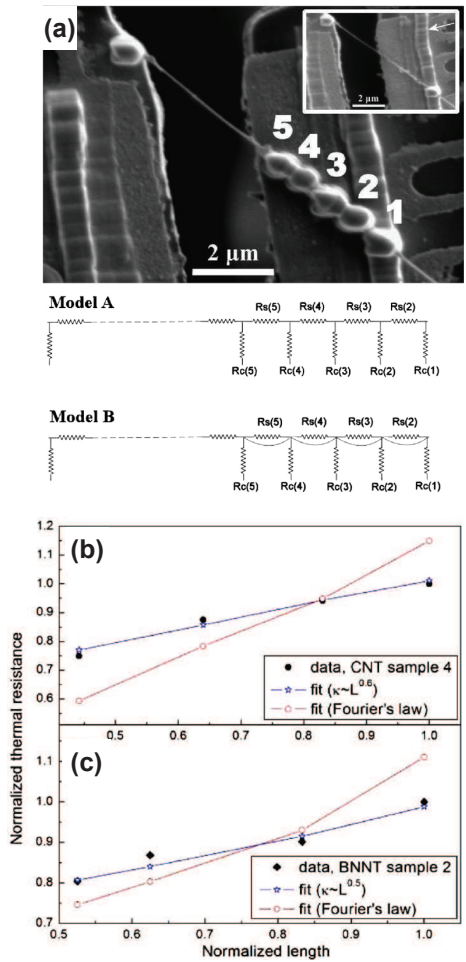}
\caption{\label{fig:xie-1}(a) upper: SEM
image of a thermal conductivity test fixture  with a BNNT
after five sequences of Pt-C deposition. The numbers denote
the nth deposition. The inset shows the SEM image after the
first Pt-C deposition. The arrow denotes the preformed 'rib'
for suspending the BNNT. Lower: Two circuit models for
analyzing the data taking into account of the contact
resistance. Rs(n) and Rc(n)denote the sample resistance and
the contact resistance at each deposition, respectively.(b)
Normalized thermal resistance vs normalized sample length
for CNT sample (solid black circles), best fit assuming
$\beta$=0.6 (open blue stars), and best fit assuming
Fourier's law (open red circles). (c) Normalized  thermal
resistance vs normalized sample length for BNNT sample
(solid black diamonds), best fit assuming
$\beta$=0.4 (open blue stars), and best fit assuming
Fourier's law (open red circles). For further details  see
in  \cite{Chang2008PRL}.}
\end{figure}

Fig.\ref{fig:xie-1}(a) shows experimental procedure, and the
inset in Fig. \ref{fig:xie-1}(a) shows the  nanotube bonded
to the test fixture by Pt-C composite using
electron-beam-induced deposition. The right contact attaches
the nanotube to the top of a preformed vertical "rib" on the
right thermal pad; this elevation ensures that the nanotube
is fully suspended between the two contacts. In order to
vary the length of the suspended segment, a series of
additional thermal contacts, labeled as 1 to 5 in \reffig{fig:xie-1}(a), was
deposited inward of the
original right-hand contact using Pt-C composite.

The thermal conductance of the nanotube as a function of
length was measured by the method previously  described by
L. Shi \etal{}  \cite{Shi2003JHT}. In brief, the Pt loop on
one
of the suspended thermal pads acts as a heater , while the
other loop acts as a sensor (\reffig{fig:xie-4}a). Heat loss from the heater is
through the low-thermal-conductance leads suspending the
thermal pad, and through the sample bridging to the sensor.
By measuring the temperature rises at the heater and the
sensor, taking into account the heat loss through its
suspending leads, the thermal conductance of the sample can
be calculated. From the analysis of nanotube thermal
conductance versus sample length, taking account into the
finite contact resistance between the nanotube and the
thermal pads (Fig.\ref{fig:xie-1}(a)), it is clear that
thermal conductivity of nanotube does not follow the
Fourier's law,
regardless of whether $L \gg\lambda$ ($\lambda$ is phonon
mean free path), as shown in Fig.\ref{fig:xie-1}(b)  and
Fig.\ref{fig:xie-1}(c).
Chang \etal{}. found that the thermal conductivity diverges
with tube length as $\kappa\propto L^\beta ($L is  the
length of the nanotube;
$\beta$=0 corresponds to Fourier's Law). The value of
$\beta$ ranges from $0.6$ to $0.8$ for CNTs, whereas
$\beta$ ranges from $0.4$ to $0.6$ for BNNTs. The different
length behavior might be due to the difference  in
isotropic disorder between BNNTs and CNTs. It is worth
pointing out that the observed
$\beta$ on multi-walled CNTs differs from the theoretical
predictions on single-walled CNTs   \cite{zhang2005JCP123B}.
The possible physical origin for this discrepancy might be
the inter-shell phonon scattering in multi-walled CNTs.

\begin{figure}[tb]
\includegraphicsS{0.7\columnwidth}{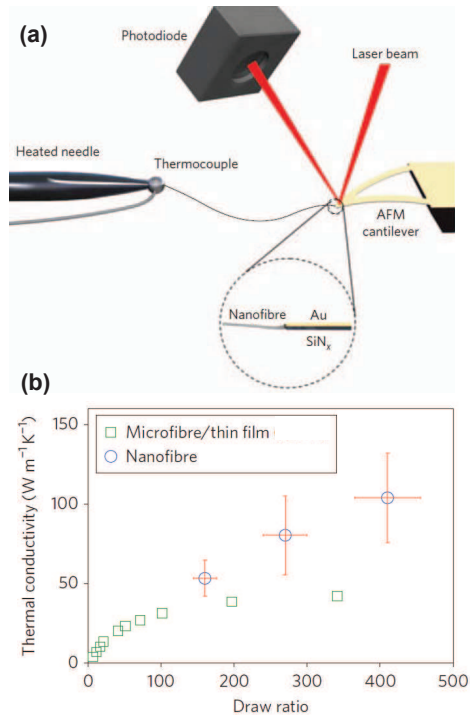}
\caption{\label{fig:xie-2}(a) Schematic of
experimental set-up used to measure the thermal  properties
of a single ultra-drawn nanofibre. The thermal sensor is a
silicon nitride AFM cantilever coated with a 70-nm gold
film. A laser beam (wavelength, 650 nm; output power, 3 mW)
is focused on the tip of the cantilever and reflected onto a
bi-cell photodiode. The nanofibre drawn from the AFM
cantilever is loosely suspended between a micro thermocouple
and the AFM cantilever. (b) Thermal conductivities of three
samples vs their corresponding draw ratios. For further
details see in \textcitee{Shen2010Nnano}.}
\end{figure}

In addition to the experimental observation of the
violation of the Fourier's Law in both BNNTs and CNTs,  the
anomalous heat conduction was also experimentally observed
in polymer nanofibres by Shen \etal{} \cite{Shen2010Nnano}.
Bulk polymers are generally regarded as thermal insulators
because they have very low thermal conductivity on the
order of 0.1 W/mK. Inspired by the theoretical work which
suggests that individual polyethylene chains can have
extremely high thermal conductivity \cite{Henry2008PRL}, Shen
\etal{} fabricated high-quality ultra-drawn polyethylene
nanofibres with diameters of 50-500 nm and lengths up to
tens of millimeters. Shen \etal{} used a sensitive
bi-material AFM cantilever in combination with a micro
thermocouple to measure the thermal conductivities of the
individual nanofibres, as shown in Fig. \ref{fig:xie-2}a. It
was found that the nanofibres have an extraordinary high
thermal conductivity which increases with increasing draw
ratios (Fig. \ref{fig:xie-2}b). The highest measured thermal
conductivity (~104 W/mK) is about three times higher than
that of micrometer-sized fibres and ~300 times that of bulk
polyethylene (~0.35 W/mK). A value of ~104 W/mK is higher
than many metals, including Pt, Fe and Ni. The high thermal
conductivity was attributed to the restructuring of the
polymer chains by stretching, which improves the fibre
quality toward an ideal single crystalline fibre. Very recently, the similar
anomalous high thermal conductivity is also observed in spider silk
\cite{Huang2012AM24}.

\subsection{Anomalous Heat Conduction in Graphene}

\begin{figure}[tb]
\includegraphicsS{0.8\columnwidth}{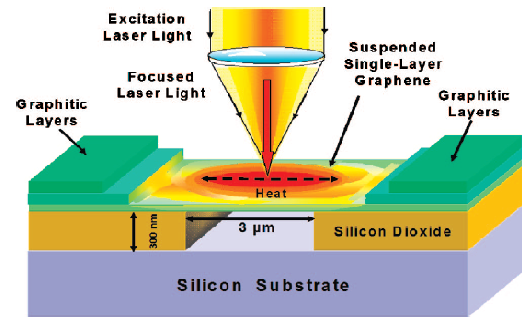}
\caption{Schematic of the Raman based thermal conductivity
measurements. The focused laser light creates a local hot spot and
the temperature rise is measured by Raman spectroscope. For further details see
\textcitee{balandin1}.
}
\label{fig:xu1}
\end{figure}

Although significant progress has been made for 1D systems,
the study of heat conduction in 2D
systems is still in its infancy due to
the lack of true 2D
materials, lack of 2D phonon transport theory and short of computational power
. The discovery of graphene has changed this scenario
 \cite{graphene}. The first experiment, based on Raman spectroscope
measurement (\reffig{fig:xu1}), was carried out by UC
riverside group in
2008  \cite{balandin1,Balandin2008NL8,balandin3}. This experiment was
performed on suspended single layer graphene (SLG) exfoliated from
HOPG. The room temperature thermal conductivity $\kappa$
was measured to reach a high value of $\sim$4800-5300 W/mK, exceeds
that of bulk graphite and carbon nanotube \cite{Nanotube}.
Such a high thermal conductivity is related to the long phonon mean free
path of graphene, which is calculated to be around 750 nm at
room temperature \cite{balandin1}.

The thermal conductivity in fully supported graphene is however
much smaller. The measurement of SiO$_2$-supported graphene
revealed a value of $\sim$600 W/mK at room temperature, which is due to backside
scatterings and the flexural phonons leakage into the substrate
  \cite{LSScience}. By introducing an exact numerical solution for
  Boltzman's transport equation, Seol \etal{} claimed that the
thermal conductivity of suspended graphene should be around 5
times larger, which is $\sim$ 3000 W/mK at room temperature and is consistent
with the results from UC riverside group.

Several following Raman based measurements, performed by other
independent groups, confirmed the high thermal conductivity both
in exfoliated and CVD graphene with corbino geometry
 \cite{Cai,Chen,Lee,Faugeras}. Cai $et$ $al$. reported that the
thermal conductivity of suspended CVD graphene exceeds
$\sim$2500W/mK near 350K and decreases to around $\sim$1400 W/mK
at about 500K  \cite{Cai}. Lee $et$ $al$. found that the value in
suspended exfoliated graphene is $\sim$1800W/mK near 325K, and
decreases to $\sim$700W/mK at around 500K  \cite{Lee}. Another
group repeated the similar experiment on exfoliated graphene and
found a value of $\sim$600W/mK \cite{Faugeras}. The differences can
be explained by the difference in the actual temperature in
graphene under laser heating, optical absorption, laser spot size,
and the well-know uncertainty in the Raman based measurements
 \cite{BalandinReview}. Despite the variation in the measured
values, the thermal conductivity in both suspended and supported
graphene is believed to be much larger than that in silicon and
most of the metals.

\begin{figure}[tb]
\includegraphicsS{0.8\columnwidth}{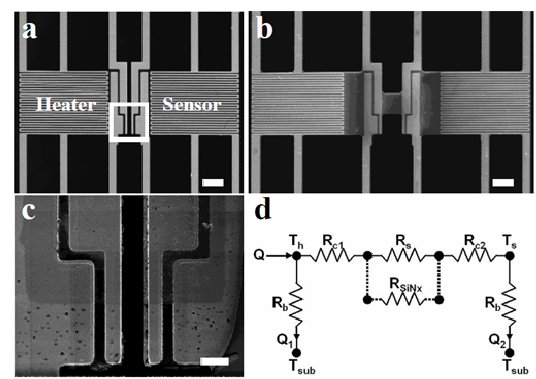}
\caption{(a)-(b) SEM images of the thermal bridge for suspended
and supported graphene samples; scale bar: 5 $\mu$m. (c)Zoom in
image of the thermal bridge, corresponding to (a); scale bar: 1
$\mu$m. (d) Equivalent thermal circuit of the device, R$_{SiNx}$
is the thermal resistance of the nitride platform for supported
samples. For further details see \textcitee{Wang}.
}
\label{fig:xu2}
\end{figure}

Apart from Raman based measurement, typical thermal-bridge method,
using the MEMS membranes mentioned above,
was also used to measure the thermal conductance of graphene by
the group from National University of Singapore \cite{Wang,xu}.
\reffig{fig:xu2}a and \ref{fig:xu2}b show scanning electron
microscope (SEM) images of
the suspended thermal-bridge used for the measurements of
suspended and supported graphene, respectively. For details of
the fabrication process and measurement accuracy, please refer to
L. Shi \etal{} \cite{Shi2003JHT}.

\begin{figure}[tb]
\includegraphicsS{0.8\columnwidth}{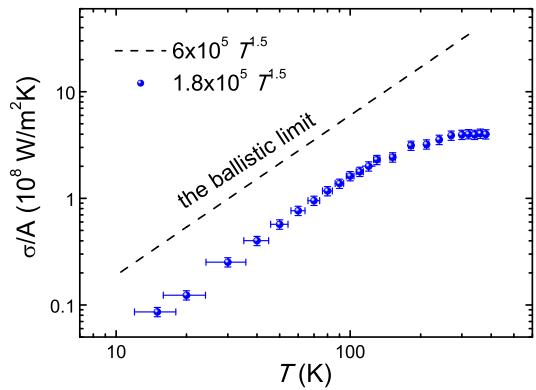}
\caption{Thermal conductance per unit cross section area
$\sigma$/A as a function of temperature. The measured data are
approaching the expected ballistic limit (black dashed
curve). For further details see \textcitee{xu}.}
\label{fig:xu3}
\end{figure}

Suspended CVD SLG was measured with length of 300 nm by Xu \etal{} using the
thermal-bridge method \cite{xu}. $\sigma$/A
(thermal conductance per unit cross section area) was found to
reach a high value of around 1.8$\times$10$^{5}$$T$$^{1.53}$
W/m$^{2}$K below 100K (\reffig{fig:xu3}). Mingo and Broido
 \cite{ballisticMingo} have predicted that the thermal conductance
will approach the ballistic limit in clean devices, with
calculated $\sigma$/A to reach 6$\times$10$^{5}$$T$$^{1.5}$
W/m$^{2}$K. The experimentally observed values are within 30$\%$
of the predicted ballistic thermal conductance in graphene,
demonstrate that the thermal transport in suspended micron
graphene is indeed in the ballistic regime. Furthermore, the
temperature dependence of $\kappa$ can be well fitted with
$\kappa=$b$T^{\alpha}$ ($\alpha$ $\sim$ 1.5). This observed
temperature dependence of $\kappa$($T$) and the high value in
$\sigma$/A clearly identify the dominant phonons contribution to
the thermal conduction in clean graphene. Mingo $et$ $al$. have
argued that the ZA phonons carry most of the heat in SLG; at low
temperatures, the out-of-plane ZA modes are predicted to lead to a
~$T^{1.5}$ behavior of the thermal conductivity  \cite{ZAphonon}.

In another experiment, the room temperature thermal conductivity was measured to
be around ~1000W/mK in suspended SLG with 5 $\mu$m in length and
1.6 $\mu$m in width  \cite{xu2}, which is 40$\%$ to 50$\%$ larger
than that obtained using the same measurement technique in
suspended exfoliated bilayer graphene of similar sample size
 \cite{Pettes}. This is expected due to the lack of interlayer
coupling in SLG  \cite{balandin3}. While the Raman based
measurements in CVD/exfoliated SLG show an average value which is
much larger  \cite{balandin1,Cai,Chen}, a direct comparison is
challenging raised from the uncertainty of Raman measurements
 \cite{BalandinReview}, and, more likely, the differences in sample
geometry.

Efforts have been made to study the size effect of thermal
conductivity in SLG. Chen $et$ $al$. studied the thermal
conductivity in suspended CVD SLG of corbino geometry with
different diameters ranging from 2.9 $\mu$m to 9.7
$\mu$m using Raman method \cite{Chen}. The measured thermal conductivity
randomly
varies from $\sim$ 2600 W/mK to $\sim$ 3100 W/mK with diameter
changes, showing no sample sized dependence. The authors
attributed this to a relatively large measurement uncertainty as
well as grain boundaries, defect or possible polymer residues.

\begin{figure}[tb]
\includegraphicsS{0.8\columnwidth}{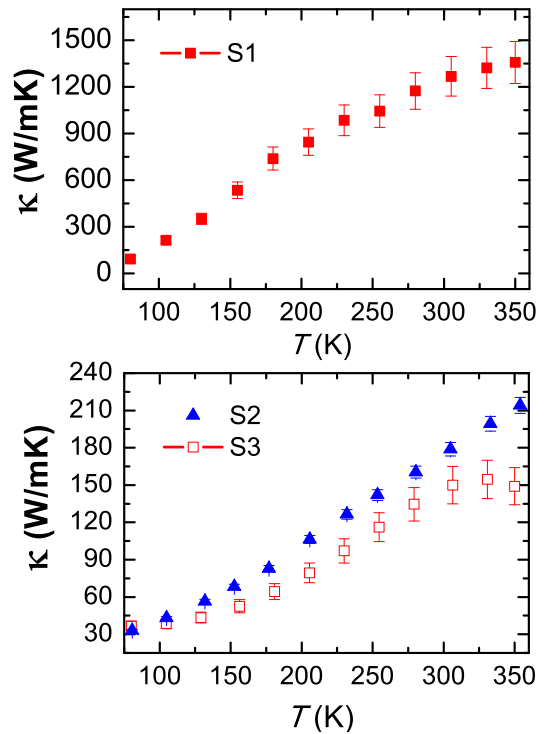}
\caption{Thermal conductivity of multi-layer graphene vs.
temperature. The lengths of samples S1, S2 and S3 are 5 $\mu$m
(three layers), 2 $\mu$m (five layers) and 1 $\mu$m (three
layers), respectively. All the samples have the same width of 5
$\mu$m. Sample S1 and S3 are supported by a thin SiN$_x$ membrane
and S2 is suspended. For further details see \textcitee{Wang}.}
\label{fig:xu4}
\end{figure}

Alternatively, it was found experimentally that the thermal
conductivity of multi-layer graphene changes with sample size
 \cite{Wang}. In this experiment, both the suspended and supported
multi-layer graphene were measured by using the suspended
thermal-bridge configuration mentioned above. A room temperature
thermal conductivity of 1250 W/mK was obtained for a 5 $\mu$m-long
supported flake (\reffig{fig:xu4}). Experimental results
show that thermal
conductivity depends strongly on the graphene size and is lowered
by 85$\%$ when the length of the flake reduces from 5 $\mu$m to 1
$\mu$m (S1 and S3 in the same Figure). This length dependence is
consistent with the calculation in SLG that thermal conductivity
is predicted to be length dependent even when graphene flake is as
long as 100 $\mu$m, due to the emergence of the long wave phonons
when sample size increases  \cite{balandin4}. More interestingly,
theories had envisioned the length dependent behavior in 2D
systems before the thermal conductivity in graphene was carried
out \cite{PRL.02.Narayan,AP.08.Dhar}. A detailed discussion on the size
dependence of the thermal conductivity in a 2D Fermi-Pasta-Ulam
lattice was carried out and a $\sim$ log$L$ behavior is proposed
in a wide range of $L$ and $W/L$ ratio \cite{Yang}.

\sectionN{Atomistic Simulations of Heat Conduction in Nano Materials}{zhang}

It is obvious that the systematic applications of
nanomaterials based devices will be greatly accelerated by
a detailed understanding of their material property. For
nanoscale materials, the fundamental question one may ask is
whether Fourier's law is still valid. The question is not
trivial, since nanoscale structures is of finite number of atoms and far
from the thermodynamic limit. Due to the challenge  in
nanoscale experimental measurement of temperature and heat
flux, atomistic simulations have an important contribution
to the development of this area. In this section, we focus
on the simulation investigations about the thermal
conduction and diffusion in low dimensional nano materials.
Because a large variety of studies on thermal transport of
nanoscale materials have been done in the past decade, here
we only addressed the most fundamental aspects of validity
of Fourier's law in nano materials, in particular, in
nanotubes, nanowires and graphene.

\subsection{Anomalous Heat Conductivity and Energy Diffusion In Nanotubes
and Nanowires}

Carbon nanotube (CNT) is one of the promising nanoscale
materials come to the spotlight of research after it  was
discovered in the 1990s \cite{Iijima1991Nature}. In
addition to electronic and optical properties, thermal
property of CNTs has attracted more and more interests. It
was found experimentally that at room temperature the
thermal conductivity of a single CNT is about 3000 W/mK
 \cite{Nanotube}. Recently, thermal contact resistance of
carbon nanotube junctions have been explored experimentally
 \cite{Yang2010APL96}. Both electron and phonon can be heat
carrier. Yamamoto \etal{} has demonstrated that even
for the metallic nanotubes, the electrons give limited
contribution to thermal conductivity of single-walled CNTs
(SWNTs) at low temperature, and with increase of temperature
this part decreases quickly  \cite{Yamamoto2004PRL92}. It is
found that the phonon mean free path of CNT can exceed the
characteristic length of the structural ripples. This makes
CNT an ideal phonon waveguide which can have superior phonon
transport property to its counterpart of photon in optical
fibers \cite{Chang2007PRL99}. Thus SWNT is ideally suited
for molecular dynamics (MD) investigations of the phonon
thermal conduction law in low dimensional system. By using
MD simulations, it was found that the thermal conductivity
of SWNT diverges with the length of system as $\kappa\sim
L^\expk$ , where the exponent $\expk$ depends on the
temperature and SWNT diameters, and the value of $\expk$ is
between 0.12 and 0.4  \cite{zhang2005JCP123a} and between
0.11 and 0.32  \cite{Maruyama2002PB323}. As discussed in
above section, this length dependent thermal conductivity
has been verified experimentally, although the absolute
value of $\expk$ varies.

\begin{figure}[tb]
\includegraphicsS{1\columnwidth}{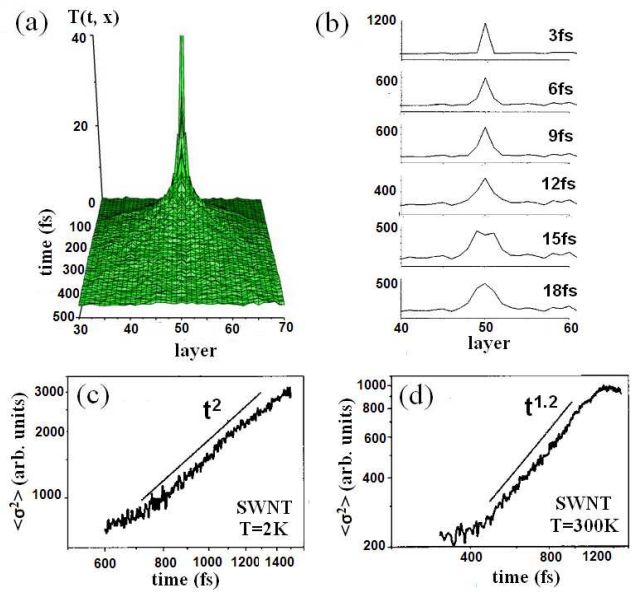}
\vspace{-0.2cm} \caption{\label{fig:Simu-1}(color online).
Energy diffusion in carbon nanotubes. (a)  Representatives
of pulse propagation along the SWNT at 2K. (b) The
snapshots of heat pulse diffusion in SWNT at 300K. (c)
and (d) Energy diffusion in SWNT at 2K and 300K,
respectively.  For further details see in
\textcitee{zhang2005JCP123B}.}
\end{figure}

To understand the physical mechanism for the length
dependent thermal conductivity in SWNT observed both
experimentally and theoretically, Zhang and Li investigated
the vibration energy diffusion along SWNT
 \cite{zhang2005JCP123B}. In the calculation the Tersoff
empirical potential is used to derive the force form, which
has been widely used in the study of heat conduction in
nanostructured carbon systems. They first thermalize the CNT
to a temperature T, then a packet of energy (heat pulse) is
excited in the middle of the tube to study how it spreads
along the system. To suppress statistical fluctuations, an
average over 1000 realizations is performed. Fig.
\ref{fig:Simu-1}(a) shows the representatives of pulse
propagation along the lattice at 2K. The pulse diffusion
profiles at room temperature are very similar to those at
low temperature. In the SWNT, one single peak expands with
time, and it is reflected back from the two ends after the
wave front reaches the tube ends, no peak splits are
observed. Indeed, there exist three time periods in the
diffusion process. At the start period, the pulse spreads
between the two nearby layers, which is denoted as "initial
period." Then, the pulse diffuses along the tube, this is
"diffusion period"; finally, the wave front hits the
boundary and is reflected back and transport in opposite
direction, which is "reflection period." The snapshots of
representative pulse propagation along the CNT at room
temperature is shown in Fig. \ref{fig:Simu-1}(b). It is
clear that in CNT, one peak expands with time, this pulse
spread can be described  quantitatively as below:
\begin{equation}
 \sigma^2(t)=\frac{\sum_i[E_i(t)-E_0](\mathbf{r}
_i(t)-\mathbf{r}_i(0))^2}{ \sum_i[E_i(t)-E_0]}
\end{equation}
here $E_i(t)$ is the energy distribution of atom $i$ at time
$t$, $\mathbf{r}_i(0)$ is the position of  energy pulse at
$t = 0$.

The averaged energy profile spreads as:
\begin{equation}
 \sigma^2(t)= 2D t^\expx, \quad\mathrm{with}\quad
0<\expx\le2,
\end{equation}
where $\mean{.}$ denotes the ensemble average over different
realizations.

In Fig. \ref{fig:Simu-1}(c) and (d) we show
$\mean{\sigma^2(t)}$ versus time in double logarithmic
scale,  so the slope of the curve gives the value of
$\expx$. It is clearly seen that the slope is two at 2K.
This means that phonon energy transports ballistically at
low temperature. This can be understood from the Taylor
expansion of the atomic interaction potential by keeping up
to the second-order term. At low temperature, the vibrations
of atoms are very small, the Tersoff potential can be
approximated by a harmonic one, thus corresponding to a
ballistic transport. However for SWNT at room temperature,
energy transports super-diffusively with $\expx\approx1.2$.
This is slower than ballistic transport ( $\expx = 2$ ) but
faster than normal diffusion ( $\expx = 1$ ). With
temperature increases, the anharmonic terms appear due to
excitation of the transverse vibrational mode. According to
a theory proposed by Wang and Li  \cite{PRL.04.Wang}, it is
known that for a quasi-1D lattice model, the interaction
between the transverse modes and the horizontal modes will
result in a superdiffusive phonon transport. Moreover, Li
and Wang  \cite{PRL.03.Li} demonstrate a connection that
the anomalous energy diffusion induces the divergent thermal
conductivity. Combine the observed super-diffusion with
$\expx\approx1.2$ and their theory, the thermal
conductivity diverges with tube length for the tube length
up to few micrometers, namely, the thermal conduction
of SWNT does not obey Fourier's law. This provides the physical mechanism
for the experimental and theoretical observed length
dependent thermal conductivity of CNTs.

In addition to CNTs, silicon nanowire (SiNW) is another
promising one-dimensional nanomaterial that pushes  the
miniaturization of microelectronics towards a new level and
have drawn significant attentions because of the ideal
interface compatibility with the conventional Si-based
technology. With the progressive research in its
applications  \cite{cui2001Sci293,zhang2008NL8,xiang2006Nat441},
more and more
theoretical efforts
have been made to understand the thermal properties of SiNWs
due to the potential thermoelectric applications in both
power generation and on-chip cooling
 \cite{Hochbaum2008Nat451,Boukai2008Nat451,zhang2009APL94,zhang2009APL95}. The
impacts of temperature
 \cite{Donadio2010NL10}, surfaces structure
 \cite{Markussen2009PRL103,Sansoz2011NL11,Donadio2009PRL102},
diameter  \cite{Shi2009APL95,Yao2009APL94}, tubular
and core-shell structure
 \cite{Wingert2011NL11,Yang2005NL5,Chen2010NL10,Hu2011NL11,Chen2011JCP135} and
doping
 \cite{Yang2008NL8,Chen2009APL95,Shi2010APL96} have been
reported.

\begin{figure}[tb]
\includegraphicsS{0.8\columnwidth}{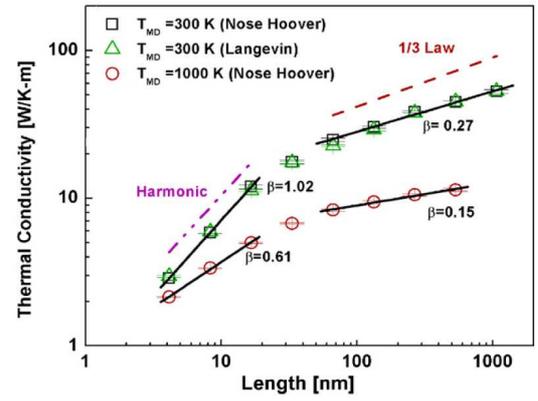}
\vspace{-0.2cm} \caption{\label{fig:Simu-2}(color online).
The dependence of thermal conductivity of SiNWs  on the
longitude length. The black solid curves are the power law
fitting curves (linear in log-log scale). For further
details see in \textcitee{Yang2010NT5}.}
\end{figure}

Like in CNTs, the length dependent thermal
conductivity is also an important physical  question to
SiNWs. By using NEMD simulation, Yang \etal{} have
studied the length dependence of thermal conductivity of
SiNWs  \cite{Yang2010NT5}. In order to establish a
temperature gradient along the SiNW, the atoms close to the
two ends are put into heat bathes with high and low
temperature. \noseh{} and Langevin heat bathes
 \cite{Chen2010JPSJ79} are applied to ensure the conclusions
are independent of heat bath. The dependence of thermal
conductivity on the length of SiNW is
shown in Fig. \ref{fig:Simu-2}. Both types of heat baths
give rise to the same results. It is obvious that  the
thermal conductivity increases with the length as,
$\kappa\propto L^\expk$, even when the wire length is as
long as $1.1\mu m$. Traditionally, the phonon mean free path
($\lambda$) is a characteristic length scale beyond which
phonons scatter and lose their phase coherence. In
three-dimensional systems, it gives Fourier's law when
system scale $L\gg\lambda$ . By using the phonon relaxation
time ($\sim 10 ps$) in SiNWs, and the group velocity of
phonon as $6400 m/s$, the mean free path $\lambda$ is about
$60nm$  \cite{Yang2010NT5}. The maximum SiNW length ($1.1\mu
m$) in this study is obviously much longer than the phonon
mean free path of SiNW. This demonstrates that in SiNWs,
Fourier's law is broken even when the length is obviously
longer than the traditionally mean free path.

More interestingly, it is found that the length dependence
of thermal conductivity is different in  different length
regimes. At room temperature,
when SiNW length is less than about $60 nm$, the thermal
conductivity increases with the length linearly (
$\beta\approx1$ ). For
the NW with length larger than $60 nm$, the divergent
exponent $\expk$ reduces to about $0.27$. This  critical
length ($60nm$) is just the value of mean free path, then
this length dependent divergent exponent can be fully
understood as below. There is weak interaction among phonons
when the length of SiNW is
shorter than mean free path, thus the phonons transport
ballistically, like in the harmonic lattice.  However, when
the length of SiNW is longer than the mean free path,
phonon-phonon scattering dominates the phonon transport and
the phonon cannot flow ballistically.

In addition, the diverged exponent $\expk$ also depends on
temperature. At $1000$K, $\expk$ is only about 0.15 when NW
length is longer than $60nm$. The decrease of $\expk$ can be
understood from the temperature dependent phonon
interaction.  In SiNWs at high temperature, the displacement
of atoms increases, which induces more phonon-phonon
interaction, and reduction in $\expk$.

\begin{figure}[tb]
\includegraphicsS{0.9\columnwidth}{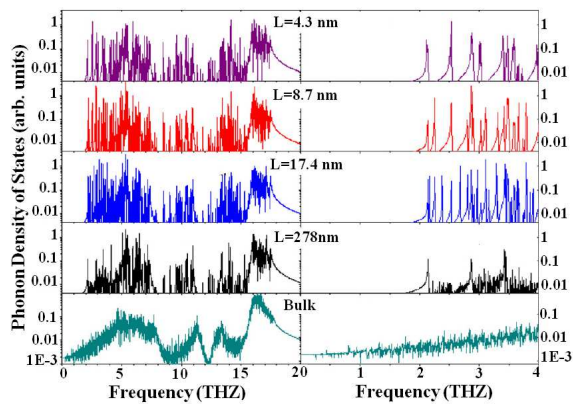}
\vspace{-0.2cm} \caption{\label{fig:Simu-3}(color online).
The phonon density of states along the longitude  direction
of SiNWs with different lengths, and the phonon density of
states of bulk Si. For further details see in
\textcitee{Yang2010NT5}.}
\end{figure}

The divergence of thermal conductivity can be understood
from the phonon density of states spectra (PDOS) as shown
in Fig. \ref{fig:Simu-3}. In the left column of Fig.
\ref{fig:Simu-3}, the higher energy part (short wavelength)
in PDOS spectra is not sensitive to the wire length. The
peak at about $16$ THz in PDOS of bulk silicon appeares
even when the NW length is only $4$ nm. However, as shown in
right column of Fig. \ref{fig:Simu-3}, the lower frequency
part in PDOS spectra is much different from that of bulk Si.
In contrast to the continuous spectra of bulk silicon, there
are many discrete peaks in PDOS spectra of NW. For a short
SiNW, the low energy phonon density is very low, thus the
thermal conductivity is low. With the increase of length,
more and more (long wavelength) phonons are excited, which
leads to the increase of thermal conductivity. However, due
to the size confinement effect, the energy density of
acoustic phonon in PDOS spectra of SiNW is much smaller than
that of bulk Si, and leads to the low thermal conductivity
of SiNW compared to that of bulk Si.

\begin{figure}[tb]
\includegraphicsS{0.8\columnwidth}{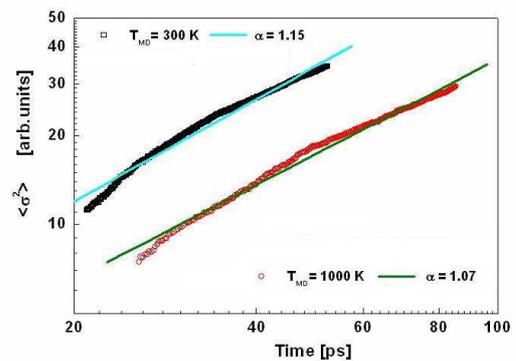}
\vspace{-0.2cm} \caption{\label{fig:Simu-4}(color online).
Energy diffusion in SiNW at room temperature and  at 1000K.
The length of SiNW is 140 nm. For further details see in
\textcitee{Yang2010NT5}.}
\end{figure}

So far we show the length dependent thermal conductivity of
SiNWs. Now let us turn to the energy diffusion  process in
SiNWs. The heat energy diffusion in SiNW was firstly
simulated by Yang \etal{} in 2010  \cite{Yang2010NT5}.
Followed the approach applied in diffusion in CNT by Zhang
and Li  \cite{zhang2005JCP123B}, they first thermalize the
system to an equilibrium state with temperature $T_0$, then
atoms in the middle layer are given a much higher
temperature $T_1$. The evolution of the energy profile along
the chain is then recorded afterwards. Fig. \ref{fig:Simu-4}
shows $\mean{\sigma^2(t)}$ versus time in double logarithmic scale,
so that the slope of the curve gives the value of $\expx$.
For the NWs of length of $140nm$, we obtain $\expx = 1.15$
and $1.07$ at 300K and 1000K, respectively. Different
theory describes the physical connection between energy
diffusion and thermal conductivity  \cite{PRL.03.Li,Densis}. For instance, in
normal diffusion, the phonon
transports diffusively, it corresponds to a size-independent
thermal conductivity. This is what we have in bulk material
and the Fourier's law is valid. However, in the ballistic
transport, the thermal conductivity of the system is
infinite when the system goes to thermodynamic limit. This
is the case for one-dimensional harmonic lattice. In another
case, if $\alpha < 1$, which we call sub-diffusion case,
corresponds to $\beta < 0$, namely, the thermal
conductivity of the system goes to zero. Thus the system is
an insulator. In the super diffusion regime, it predicts
that the thermal conductivity increases as the length of the
system increases. This is what we observed in SiNWs.

Thus the anomalous heat diffusion is responsible for the
length dependent thermal conductivity in SiNWs.  Combine
with the results in CNTs, all these reports provide strong
evidence that Fourier's law of heat conduction is not valid
in one dimensional and quasi one dimentional
nanostructures. Moreover, these
theoretical studies demonstrate that nanowire and nanotube
are promising platforms to verify phonon transport
mechanisms.

\subsection{Computational Study on Thermal Conductivity of Graphenes}

Besides to the one-dimensional nano materials, Graphenes
 \cite{graphene,Zhang2005Nat438}  have attracted immense interests recently,
mostly because of their unusual one-atom-thick structure. Superior thermal
conductivity (as high as 5000 W
m${}^-1$ K${}^-1$) has been observed in graphene
 \cite{Balandin2008NL8}, which has raised the exciting
prospect of using them for thermal devices.

Based on graphene, narrow (sub-10 nm) graphene nanoribbon
(GNR) can be obtained either by cutting  mechanically
exfoliated graphenes, by patterning epitaxially growth
graphenes or by unzipping carbon nanotubes (CNTs)
 \cite{graphene,Berger2006Sci312,Wang2010NC2}. GNRs are materials with
distinctive
electronic and thermal transport properties and are
candidates for field effect transistors of future
carbon-based nanoelectronics. There are rich physical
phenomena about thermal property of GNRs. For instance,
intrinsic anisotropy originate from different boundary
condition at ribbon edges is observed in narrow GNRs,
results to the room temperature thermal conductance of
zigzag GNRs is about thirty percent larger than that of
armchair GNRs  \cite{Xu2009APL95,Xu2010PRB81}. The effects of edge,
roughness, and hydrogen termination have been investigated
by different simulation methods
\cite{Evans2010APL96,Ni2011JPCM23,Chien2011APL98,Pei2011carbon49,Hu2010APL97,Aksamija2011APL98}.
And isotopic doping can reduce
remarkably the thermal conductivity of graphene
 \cite{Ouyang2009EPL88}. In addition to the isotope randomly
doping, the impact of disorder defect on thermal conductance
 \cite{Jiang2011APL98,Xie2011JPCM23} and
thermoelectric figure of merit of GNRs  \cite{Ni2009APL95}
has been investigated by using density functional theory
calculations combined with the nonequilibrium Green's
function method. It is found that the figure of merit can be
remarkable enhanced five times by randomly introducing
vacancies to the graphene. And remarkable thermal
conductivity enhancement has been found in graphene
nanoribbons under homogeneous uniaxial stretching due to a
lot of dispersive phonon modes are converged to the low
frequency region  \cite{Zhai2011EPL96}. And thermal
conductivity of GNRs is very sensitive to tensile strain
 \cite{Wei2011Nano22}. More interesting, the presence of a
substrate  \cite{Guo2011PRB84,Ong2011PRB84} and
layer-layer interaction
\cite{zhang2011NS3,Yang2012APL100,Zhong2011APL98,Cao2012PLA376} lower the
thermal conductivity. The excellent thermal property of GNRs
makes them ideal candidate for nanoscale phononic devices
 \cite{Yang2008APL93,Yang2009APL95,Hu2009NL9,zhang2011NS3}.

Similar to the cases in CNT and SiNW, thermal conductivity
of GNRs also depends on their sizes. By using  equilibrium
molecular dynamics simulations, the effect of GNR width on
thermal conductivity has been studied  \cite{Evans2010APL96}.
They found that the thermal conductivity increases with
width, but for width larger than $5nm$ the thermal
conductivity becomes relatively size independent. The larger
conductivity values at very small sizes is explained from
the limited number of phonons in the system. In their
simulation, the width range is from $2nm$ to $12nm$. The
similar width dependence of thermal conductivity was also
observed by non-equilibrium molecular dynamics simulations
 \cite{Guo2009APL95}. Moreover, thermal conductivity also
depends on GNR length. In their work, for GNR with width of
$2nm$, its thermal conductivity increases from about 200
W/mK to 400W/mK when length increases from $10nm$ to
$60 nm$. Based on the extrapolation procedure, they
predicted the thermal conductivity of GNRs with a length of
$2$ micrometer will be about 3000W/mK, which is close to
the experimental reports. However, as the maximum length in
their simulation  \cite{Guo2009APL95} is only about $60nm$,
it is not long enough to provide detailed information about
the divergent exponent. Further simulation studies on much
larger GNRs will be greatly helpful to explore whether
Fourier's law is still validity in two-dimensional
one-atom-thick systems (See also \cite{poster}).

\sectionN{Heat Transport in Low Dimensional Models}{liu}

The anomalous heat transport phenomena observed in the above
mentioned low dimensional nanostructures can be
partially understood
by general theories on low dimensional
models, which generally demonstrate that the
phenomenological Fourier's law may not be valid.
Actually, ever since Fourier proposed
this law on heat conduction about 200 year ago, enormous
number of studies has been conducted to fully understand it
from fundamental physics. But a rigorous derivation from
microscopic Hamiltonian dynamics is still absent. However,
from these studies, we are already able to draw the
surprising conclusion that heat conduction in low
dimensional systems may not follow the Fourier's law.
In this section, we will briefly review the fundamental
studies on heat transport in low
dimensional systems: from theories to numerical simulations on the toy models.
Especially, we introduce the works that study anomalous heat conduction from
the aspect of anomalous energy diffusion.

In order to understand the realistic models from a rigorous
perspective, it is better to begin with the simplest models
which involve only the basic but important ingredients, such as
linearity and disorder, that will possibly lead to the
expected normal or anomalous transport behaviors.

\subsection{Low Dimensional Harmonic Lattices}

The ordered harmonic lattice is the simplest lattice model
that we can start with. From the early Boltzmann transport
equation, it is not surprising that this linear model
does not have well defined transport coefficients due to
the lack of interaction between phonon modes. Therefore, heat transport
in harmonic models should not obey Fourier's law. Actually,
a clearer clue was found after Debye extended Boltzmann
kinetic theory of ideal gas
and established the expression of heat conductivity $\kappa=cvl$, where
$c$ is the heat capacity, $v$ the phonon velocity,
and $l$ the phonon mean free
path, because the non-interacting phonons should have infinite mean free paths.

The first explicit results on heat transport in
classic harmonic models were given by Rieder, Lebowitz and
Lieb \cite{JMP.67.Rieder}. They studied a harmonic chain
connected to stochastic Langevin heat baths. The lattice has
Hamiltonian $
	H=\sum_{l=1}^N \frac{{p_l^2}}{2m}+ \sum_{l=1}^{N-1}
\frac{1}{2}
k(x_{l+1}-x_l)^2
$. The particles at the ends are subjected to additional
forces $F_{1,N}= \eta_{1,N} - \gamma v_{1,N}$, where the
noise terms $\eta_{1,N}$ are independent Gaussian random processes with mean
zero and variances
$\mean{\eta_{1,N}(t)\eta_{1,N}(0)}=2\gamma k_B  T_{1,N} \delta(t)$.
Rieder \etal{} proved
that, at large $N$, the heat flux will saturate to
\cite{JMP.67.Rieder}
\begin{equation}
	J= \frac{k k_B}{2\gamma} \left[
1+\frac{\nu}{2}-\frac{\nu}{2}\sqrt{1+\frac{4}{\nu}}\right]
(T_1-T_N),
\end{equation}
where $\nu=\frac{mk}{\gamma^2}$. This heat flux is
independent of the length $N$ for large $N$, which
demonstrates the expected ballistic transport. Another
important result on harmonic model is that the non-equilibrium temperature
profile is flat at the center part with temperature
$T=(T_1+T_N)/2$. Temperature jumps occur at the
boundaries. In this sense, temperature gradient cannot
be established in harmonic lattices. Another
work by Nakazawa  extended Rieder \etal's model by
introducing on-site harmonic potential to all sites and a
similar length-independent heat flux was found
\cite{PTP.68.Nakazawa}. In the same work \cite{PTP.68.Nakazawa}, Nakazawa
also proved that high dimensional harmonic
lattices can be reduced to a 1D problem. Therefore the
conclusions are similar.

Due to the absence of scattering mechanism in
harmonic models, heat transport in these lattices is
ballistic. Therefore, it is necessary to introduce
scattering of phonons to get the desired diffusive
transport. Basically, two approaches are
commonly adopted: by introducing
disorder or nonlinearity to the system. Both of these two
methods can cause phonon-phonon interaction. However, as we
will see later, neither method works effectively for
diffusive transport in low dimensional lattices.

For the first approach, we consider the harmonic lattices
with disorder. Disorder can be generally introduced by
randomly assigning varies particle masses or
spring constants, or both. These two types of disorder do not change
the underlying physics very much. The former with
random masses is just like isotopic doping, which is more
interesting and relevant in applications. Therefore, it
is studied intensively. The presence of disorder will
generally cause localization of normal modes,
which could be partially understood as an analogue to the Anderson
localization of electrons. In the Anderson
tight-binding model, all the eigenstates of electrons are
localized in one and two dimensional cases, therefore the
systems are electric insulators. However, in the phonon
case, the picture is much more complicated.

For one dimensional disordered harmonic chains, Allen and Ford first
observed that the thermal conductivity of
an infinite disordered harmonic chain is finite
\cite{PR.68.Allen}, unlike the Anderson model of
electrons in which the conductivity will exponentially decay
to zero. Furthermore, the eigenstates of electrons
in the strongly disordered cases are all exponentially
localized. However, in disordered media, Matsuda and Ishii
\cite{PTPS.70.Matsuda}
showed that only the high frequency phonons has this
property, while phonons with low frequencies $\omega<
\omega_d$ can be considered as delocalized phonons, since the localization
lengths are greater than the system length. The characteristic
frequency $\omega_d$ depends on the variance of mass
distribution $\mean{\Dt m^2}$ as $\omega_d\sim
(\frac{km}{N\mean{\Dt m^2}})^{1/2}$, where $k$, $m$ and
$N$ are the spring constant, the average mass and the lattice
length, respectively. Therefore, low frequency
phonons will contribute to heat transport, which prevents the
disordered harmonic lattices from being a thermal insulator.
Matsuda and Ishii also studied the thermal conductivity of
 disordered chains connected to white noise heat
baths (Langevin baths) and baths modeled by ordered semi-infinite
harmonic chains (Rubin baths), respectively
\cite{PTPS.70.Matsuda}. They found that for both models (a):
disordered harmonic chain connected to Langevin baths with
fixed boundary conditions; and model (b): disordered
harmonic chain connected to Rubin baths with free boundary
conditions, their thermal conductivity will diverge with the
system length as $\kappa\sim N^\expk$ with $\expk=1/2$.
However, Casher and Lebowitz later rigorously proved that
the correct exponent for model (a) should be $\expk =-1/2$
\cite{JMP.71.Casher}, while for model (b) the result
$\expk=1/2$ is supported by numerical simulations
\cite{JMP.71.Rubin} and a rigorous proof was provided
by Verheggen later \cite{CMP.79.Verheggen}.

\begin{figure}[tb]
\includegraphicsS{0.8\columnwidth}{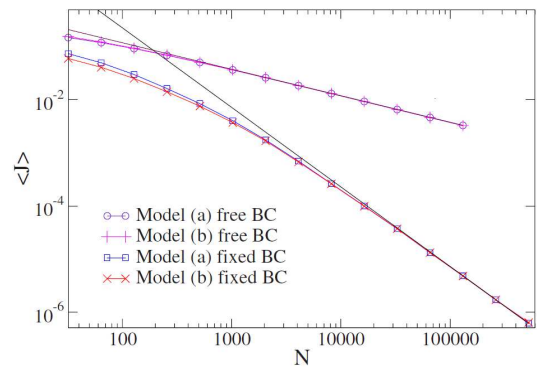}
\caption{$\mean{J}$ vs. $N$ for different boundary
conditions with temperature bias fixed. Results are given
for the Langevin heat baths (model a) and Rubin baths (model b). The two
straight lines correspond to the asymptotic power law
behaviors $J\sim N^{-1/2}$ and $J\sim N^{-3/2}$,
respectively.
For further details see \textcitee{PRE.08.Roy}.}
\label{fig:disorder}
\end{figure}

Altogether, the thermal transport properties of disordered
harmonic chains would probably depend on the both the
boundary conditions and the heat baths. To clarify this point,
Dhar restudied the thermal transport in
disordered harmonic chains
connected to various baths modeled
by generalized Langevin equations \cite{PRL.01.Dhara}.
Using a Langevin equations and Green's function (LEGF)
formalism, Dhar found that in disordered
harmonic chains, the exponent $\expk$ depends on
the spectral properties of
the baths. The previous model (a) and (b) are only
two special cases of the generalized Langevin heat baths,
which have $\expk=-1/2$ and $1/2$, respectively. A special
choice of heat bath spectrum can even lead to normal heat
transport with $\expk=1$, obeying Fourier's law
\cite{PRL.01.Dhara}. A more detailed study on the
effects of the boundary conditions on the transport
properties were carried out by Roy and Dhar later using the
same LEGF formalism \cite{PRE.08.Roy}. The results are a
little contradictory to those of \cite{PRL.01.Dhara}, which
states that the different exponents $\expk$ are indeed
dependent on the different boundary conditions, instead of
the spectral properties of the baths. Clear numerical
evidence also supports this conclusion as shown in Fig.
(\ref{fig:disorder}), \ie{}, for
disordered harmonic chains with free boundary
conditions $\expk=1/2$ and with fixed boundary conditions
$\expk=-1/2$, no matter whether Langevin baths or Rubin
baths are used. Ref.~\cite{PRE.08.Roy} also showed that, in the low
frequency regime, the transmission coefficient of the
disordered chains $\mathcal{T}(\omega)$ can be
approximated by the transmission of the ordered chains
$\mathcal{T}_{\mathrm{ordered}}(\omega)$. Therefore,
heat flux in a disordered harmonic chain can be estimated as
\begin{equation}
	J\sim (T_L-T_R)\int_0^{\omega_d}
\mathcal{T}_{\mathrm{ordered}}(\omega) \dd{\omega},
\end{equation}
where $\omega_d$ is the low frequency boundary
$\omega_d\sim(\frac{km}{N\mean{\Delta m^2}})^{1/2}$.
Considering the fact that the transmission coefficients for low
frequency modes scale as $T_{\mathrm{ordered}}(\omega)\sim
1$ for free boundary conditions and
$T_{\mathrm{ordered}}(\omega)\sim
\omega^2$ for fixed boundary conditions, respectively,
we can obtain $J_{\mathrm{free}}\propto
(T_L-T_R)\,\omega_d\propto
(T_L-T_R)/N^{1/2}$ for free boundary conditions and
$J_{\mathrm{fixed}}\propto (T_L-T_R)\,\omega_d^3\propto
(T_L-T_R)/N^{3/2}$ for fixed
boundary conditions, respectively.

Till now, we only discussed the disordered harmonic chain
without onsite pinning potential (except the case of fixed boundary
conditions, the particles at the ends are pinned). The
absence of pinning in this model makes the system
translational invariant in a coarse grained manner. As a
consequence, the phonon modes are not fully localized and
the low frequency ones can transport heat. When onsite
potentials are involved, the picture is a little different. In
their work
\cite{PRL.08.Dhar}, Dhar and Lebowitz proved that with
onsite potentials, all the phonon modes are localized.
Consequently, the heat flux $J$ decays exponentially with
the chain length and the system is a thermal insulator in
the thermal dynamic limit. This could be understood because the
onsite potentials will typically break the
translational invariance of the chain and eliminate the
low frequency phonon modes in this structure. The remaining phonon modes have
high frequencies and are localized.

Compared with the one dimensional lattices, the studies on two
dimensional disordered ones are much less due to the
mathematical difficulties. In a renormalization group
study by John \etal{} \cite{PRB.83.John}, it was found
that the low frequency modes in 2D disordered harmonic lattices are
not localized either as in 1D cases. The localization length diverges as
$\ee^{1/\omega^2}$ in the low frequency regime. Therefore,
modes with frequency $\omega <\omega_d= \log^{-1/2}(N)$ can
be considered as extended, which will contribute to heat
transport.

In another study by Lee and Dhar
\cite{PRL.05.Lee}, 2D disordered harmonic
lattices connected to two types of stochastic heat baths are
investigated. The first type is modeled by uncorrelated
Gaussian processes and the other by exponentially correlated
Gaussian
processes. In their work, disorder is introduced as isotopic
doping. A mass $M=2$ is randomly assigned to half of the
particles, while the rest have mass $m=1$.
The authors simulated square lattices with widths up to
$L=256$ and found a power law divergent heat conductivity in both two
types of models. For uncorrelated baths the
exponent $\expk\approx 0.41$ while for correlated baths
$\expk\approx0.49$. In the same work
\cite{PRL.05.Lee}, Lee and Dhar also studied a special case
of correlated disorder in which the lattice is only
disordered in the conducting direction. It was found that
this model can be
transformed into an effective one dimensional problem
thus being mathematically tractable. Analytical results showed
that the exponent $\expk=-1/2$ which was also verified by
numerical simulations using square samples with widths
up to $L=512$.
In another work
\cite{PRL.02.Yang}, Yang considered 2D
lattices with bond-missing defects connected to
\noseh{} heat baths. The results showed that when the defect
density is large enough, temperature gradient can build up
and the heat conductivity is finite, while for the case
of small defect density, the conductivity will diverge
logarithmically.

\subsection{Low Dimensional Anharmonic Systems}
It is clear now that disorder alone cannot effectively make the phonon transport
in the harmonic chains diffusively.
As an alternative approach to introduce
phonon-phonon scattering, the effect of
nonlinear interaction on the properties of
heat transport in low dimensional systems has also been
intensively studied, especially by numerical simulations.
Because of the intrinsic non-integrability of the system
Hamiltonian, analytical results are rare.

Basically, there are three different approaches
to study heat transport analytically: the mode coupling theory (MCT)
\cite{EL.98.Lepri,PRE.98.Lepri}, the renormalization group
theory \cite{PRL.02.Narayan} and the Peierls-Boltzmann
kinetic theory \cite{PRE.03.Pereverzev}. All
are based on Green-Kubo formula
\cite{JCP.54.Green,JPSJ.57.Kubo}
\begin{equation}
 \label{eq:GK}
\kp=\frac{1}{k_BT^2}\lim_{t\to\infty}\lim_{L\to\infty}
 \int_0^tC(t) \dd{\tau},
\end{equation}
where $C(t)=\mean{\JJ(t)\JJ(0)}/L$ is the time
correlation of the total heat flux $\JJ$.

Strictly speaking, Green-Kubo formula can only be
applied to infinite systems, as indicated by the limit taken
$\lim_{L\to\infty}$, which should be
performed before $\lim_{t\to\infty}$ strictly. However, as
we already know, in many low dimensional systems, the heat
conductivity in thermal dynamic limit is divergent. The
focus of such anomalous situations is how the
conductivity diverges with the system size. To utilize the Green-Kubo
formula to study the length dependence of heat conductivity
in these cases, a usual procedure is to truncate the upper
limit of the integration in Eq. (\ref{eq:GK}) to $t_c\sim
L/v_s$. This is commonly believed as
that the sound waves propagate to the boundaries at a
finite speed $v_s$, which will lead to a fast decay of
correlation $C(t)$ at time $\sim L/v_s$
\cite{PR.03.Lepri,AP.08.Dhar}. Using this treatment, the
previously mentioned three theories, \ie{} mode coupling
theory, renormalization group theory and Peierls-Boltzmann
equation kinetic theory, are all aimed to calculate the long
time tail of the correlation function $C(t)$. A basic result
shared by all these theories is that the systems with
nonlinear interaction remain anomalous except those subjected
to onsite potentials. For 1D cases without onsite potential,
these theories predict that the conductivity of the system
would diverge as a power law $\kappa\sim N^\expk$. However,
the values of $\expk$ differ from one to another. For 2D cases, the mode
coupling theory and the renormalization group theory predict logarithmic
divergence.

The mode coupling theory approach was first used
by Lepri \etal{} \cite{EL.98.Lepri,PRE.98.Lepri}. The basic
idea is that the divergence of heat conductivity is due to
the slow relaxation of phonon modes with long wavelengths.
A later development
of this theory was done
by Delfini \etal{} \cite{JSM.07.Delfini}. By constructing a
correlation function $G(k,t)= \mean{Q^*(k,t)Q(k,0)}$ of the
normal mode coordinates $Q(k)=\frac{1}{\sqrt{N}}
\sum_{n=1}^{N} x_n\exp(-\ii k n)$, the authors were able to
get the exact evolution equations for $G(k,t)$ using the
Mori-Zwanzig projection approach \cite{PRE.98.Lepri}. Using
some approximations, these equations were solved
self-consistently for the Fermi-Pasta-Ulam (FPU) chains with
interaction potential $U(x)= k_2 x^2/2+ k_3 x^3/3+ k_4
x^4/4$, which gives $G(k,t)=A(k,t)\ee^{\ii\omega(k)t}+\mathrm{c.c.}$ for
a small wavenumber $k$. In this formula, $\omega(k)$ is the temperature
dependent dispersion relation which is obtained using harmonic approximation
$\omega(k)=2|\sin(k/2)|$, and $A(k,t)$ has the form
\begin{equation}
	A(k,t)=\begin{cases}
	        g(\sqrt{3k_3^2k_BT/2\pi}t k^{3/2}) & k_3\ne0; \\
	        g(\sqrt{15(k_4k_BT/2\pi)^2}t k^{2}) & k_3=0 \ \mathrm{and}\ k_4\ne
0.
	       \end{cases}
\end{equation}
Finally, the heat flux correlation function $C(t)$
was obtained from $C(t)\propto \sum_k (\frac{\dd \omega(k)}{\dd k})^2 G^2(k,t)$.
It was shown that
$C(t)\sim t^{-2/3}$ for the case $k_3\ne 0$ and $C(t)\sim
t^{-1/2}$ for $k_3=0$ but $k_4\ne0$. Therefore, by
inserting $C(t)$ into the Green-Kubo formula and adopting the
cut-off time $t\propto N$, we obtain $\expk=1/3$ and
$\expk=1/2$ for each case, respectively. The mode
coupling theory can also provide predictions for higher
dimensional systems. It was shown that \cite{PR.03.Lepri} for
two dimensional systems, $C(t)\sim t^{-1}$, so that the heat
conductivity would diverge logarithmically with system
size. While for three dimensional systems, $C(t)\sim
t^{-3/2}$, with which we obtain normal heat conduction.

In a later work, Wang and Li investigated the effect of
transverse degrees of freedom in one dimensional chains
using this mode coupling theory \cite{PRL.04.Wang}. The
system under investigation has Hamiltonian
\begin{equation}
	H=\sum_i \frac{{\bf p}_i^2}{2m}+
\frac{K_r}{2}(|{\bf r}_{l+1}-{\bf r}_l|-a)^2+ K_\phi
\cos(\phi_l),
\end{equation}
where $({\bf r}_i, {\bf r}_i)$ is the canonical
coordinates, and $\phi_l$ is the angle between vectors
${\bf r}_{i+1}-{\bf r}_i$ and ${\bf r}_{i-1}-{\bf r}_i$. This model
 can be regarded as a simplification
of more realistic polymer chains.
The model is still one dimensional but the motions of
particles are two dimensional. Both numerical simulations
and mode coupling analyses suggest that in the presence of interaction with
transverse modes, $\expk=1/3$. When the coupling between
transverse modes and longitudinal modes is weak,
$\expk=2/5$.

The second approach was proposed by Narayan
and Ramaswamy which uses hydrodynamic
equations and renormalization group theory
\cite{PRL.02.Narayan}. It was argued that in an interacting 1D
system with large system length, the thermal fluctuation will wipe out the
long-range order, and consequently the system will behave
like fluid. Therefore it is expected that this approach can be
used generally to describe heat transport in 1D phenomenon.
By assuming that the only conserved
quantities in the system are the total number of particles,
the total
momentum and the total energy, one can obtain three
hydrodynamic equations describing the evolution of the
particle density field and the velocity field, with
addition terms describing the thermal noise. Narayan
and Ramaswamy then solved these equations using linear response
approximation and finally obtained the thermal flux
correlation by considering the symmetries of the
system. It was demonstrated that the flux correlation
function $C(t)\sim t^{-2/3}$ asymptotically for the 1D case. Therefore, one
obtains $\expk=1/3$ by the
cut-off time reasoning. For 2D systems, the authors showed
that the conductivity would diverge logarithmically with
system sizes \cite{PRL.02.Narayan}.

In another work by Mai and Narayan
\cite{PRE.06.Mai}, the authors claimed that the renormalization group theory
analyses can be applied to lattice models as well, even
for stiff one dimensional oscillator chains including the FPU
model. The conductivity of 1D lattices should also diverges in the same way
as fluid, \ie{} diverges in a power-law with exponent $\expk=1/3$. However, in a
consequent work by Hurtado
\cite{PRL.06.Hurtado}, the breakdown of hydrodynamics in a
simple one dimensional fluid was reported.

The estimation of the exponent $\expk$ in anomalous heat
transport using Peierls-Boltzmann equation is first
carried out by Pereverzev \cite{PRE.03.Pereverzev}.
The Boltzmann equation is originally used to
describe the the phase space density evolution of kinetic
gases. From it we have already known that the heat
conductivity can be written as $\kappa=c\,v\,l$, where
$c$, $v$ and $l$ are the specific heat, sound velocity and
mean free path of the gases, respectively. Peierls developed this theory
to describe the phonon transport in solids and found
a similar formula for the conductivity
$\kappa\sim\int c_k v_k^2\tau_k\dd k$, where $k$ specifies
the phonon modes and $\tau_k$ is the phonon relaxation
time. Using this Peierls-Boltzmann approach, Pereverzev
studied the \FPUb{} mode
\cite{PRE.03.Pereverzev}. Considering that anomalous heat
conduction is mainly due to the ballistic-like
behavior of phonons with small wavenumber $k$.
the dependence of $\tau_k$ on the wavenumber $k$ in this
regime is the main interest. After some approximation,
Pereverzev was able to obtain $\tau_k\sim k^{-5/3}$ for
small $k$. It was also found that heat flux correlation
function can be calculated by $C(t)=
\frac{2k_B^2T^2}{\pi}\int_0^\pi
\ee^{-t/\tau_k} v_k^2\dd k$. Because $v_k$ is
almost constant for small $k$, one obtains $C(t)\sim
t^{-3/5}$ and $\expk=2/5$.
This result was confirmed later by Lukkarinen and Spohn
 using a more rigorous Peierls-Boltzmann
approach \cite{CPAM.08.Lukkarinen}. For momentum
non-conserving cases, a finite heat conductivity was
obtained by Aoki \etal{} \cite{JSP.06.Aoki} using this
approach.

Compared to the few analytical results on low dimensional
heat conduction, numerical results are much more abundance.
%Generally, they are  aimed to figure out two basic
%problems: what are the necessary and sufficient conditions
%for Fourier's law and is there any universality for the
%divergent exponent $\expk$ as predicted by the theories.
Here we only briefly highlight the results on some well
studied lattice oscillator models.

Due to the landmark work by Fermi, Pasta and Ulam, the FPU
model has received great attraction in understanding
nonlinear statistical mechanics. In the \FPUb{} model, interaction between
nearest particles has the form $U(x)=k_2
x^2/2+ k_4 x^4/4$. This model is one of the simplest lattice
models which can be used to investigate the effect of nonlinearity
on heat transport.

The first numerical study on heat conduction of \FPUb{}
model was carried out by Kaburaki and Machida \cite{PLA.181.85}.
In their study,
the ends of \FPUb{} lattices were connected to Boltzmann heat baths at different
temperatures. Detailed calculations showed that linear temperature profiles
can be established in the steady state. Moreover, it was found that the
conductivity tends to saturate when the chain length increases. However, it is
clear now that neither of these findings reveals the true heat transport
behaviors of \FPUb{} lattices.

The first indication of anomalous heat conduction violating Fourier's law in
\FPUb{} lattices was presented by Lepri \etal{}
\cite{PRL.97.Lepri}. \FPUb{} lattices connected to
\noseh{} heat baths with lengths up to $N=400$ was studied.
It was found that the heat conductivity
diverges as $\kappa\sim N^{\expk}$ with exponent $\expk=0.55\pm0.55$.
Another finding in this work is that the temperature
profile is non-linear even when the temperature bias is very small, a result
contradictory to the prediction
from the Fourier's law. In a subsequent work \cite{EL.98.Lepri}, the
divergent exponent was improved to
$\expk=0.37$ with length up to $N=2048$.
These pioneering works in one
dimensional lattices have inspired
enormous subsequent studies on anomalous heat
conduction in low dimensional systems.

It should be mentioned that the system lengths in the studies above are too
small to draw any quantitative conclusions.
Actually, in
non-equilibrium simulations, the heat conductance will be greatly affected  by
the contact resistances between the lattice and heat baths. Therefore, it is
necessary to go to a very large system length to make the contact resistances
negligible. In a recent work
\cite{PRL.07.Mai}, Mai \etal{} have carefully taken this
into consideration. They studied the \FPUb{} chain with length up to
$N=65536$ using two different types of heat bath: the
stochastic Langevin heat baths and the deterministic
\noseh{} heat baths. It was found that when the system size
is small, the calculated thermal conductances differ a
lot when different heat baths are used.
Meanwhile, when the system size is large enough, the discrepancy between the
calculated heat conductances is negligible, which ensures us
that the boundary effects could be omitted. In this situation,
it was then found that the
exponent $\expk$ decreases to a final result
$\expk=0.333\pm 0.004$. The authors claimed that this
result is consistent with the renormalization group theory
\cite{PRL.02.Narayan} which supports its universality.

\begin{figure}[tb]
\includegraphicsS{0.8\columnwidth}{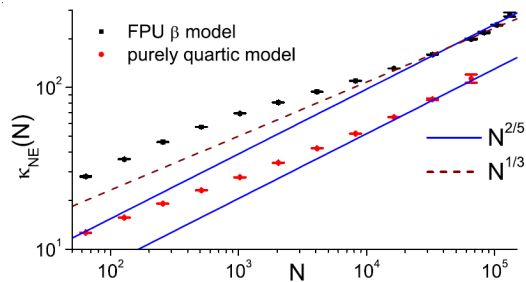}
\caption{
Heat conductivity $\kappa_{\mathrm{NE}}$ vs. lattice length N in log-log scale.
In the rightmost
part of the figure, it can be seen that the running slope increases as the
system length is increased. Solid
and dashed lines with slope 2/5 and 1/3, respectively, are
drawn for reference. For more details please
see \textcitee{EL.11.Wang}.
}
\label{fig:wang}
\end{figure}
However, in a quite recent work,
Wang \etal{} \cite{EL.11.Wang} restudied the \FPUb{} model and the purely
quartic
model with an
even longer length up to $N=131072$. The interaction in purely quartic model
is $U(x)=\frac{1}{4}k_4 x^4$. This model is the high temperature limit of the
\FPUb{} model. They found that the
slope $\dd\ln\kappa/\dd \ln N$ is not monotonically
decreasing as shown by Mai \etal{} \cite{PRL.07.Mai}. The
slope starts to increase when $N$ is
close to $10^5$ and finally reaches $\expk=2/5$ instead of $1/3$ (see
\reffig{fig:wang}). In
order to remove the boundary effects, the authors also performed
equilibrium simulations without heat baths to calculate the
heat flux correlation function $C(t)$ in the Green-Kubo
formula. Again, the slope $\dd\ln C(t)/\dd \ln N$
increases when $t$ is very large. The heat flux
correlation finally saturates at $C(t)\sim t^{-3/5}$, which
supports their non-equilibrium results $\expk=2/5$ and the
Peierls-Boltzmann theory \cite{PRE.03.Pereverzev}.

\begin{figure}[tb]
\includegraphicsS{0.8\columnwidth}{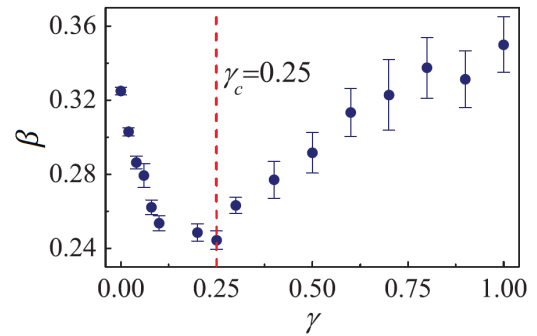}
\caption{
Dependence of $\expk$ on $\gamma$ in the model with
Hamiltonian Eq. (\ref{eq:fpu_nnn}). For more details please
see \textcitee{PRE.12.Xiong}.
}
\label{fig:universal}
\end{figure}
For one-dimensional lattices with
momentum conserving interparticle interactions, all
theories predict a divergent heat conductivity $\kp\propto
L^\expk$ with universal $\expk$. However, in a most recent
study on \FPUb{} lattices with nearest-neighbor (NN) and
next-nearest-neighbor (NNN) coupling \cite{PRE.12.Xiong},
Xiong \etal{} found that the exponent $\expk$ strongly
depends on the ratio of the NNN coupling to the NN coupling
$\gamma$. The Hamiltonian of the system under investigation
is
\begin{equation}
H=\sum_{i} \left[ \frac{p_i}{2m_i}+ V(x_{i+1}-x_i)+
\gamma V(x_{i+2}-x_i) \right],
\label{eq:fpu_nnn}
\end{equation}
where $V(x)$ is of \FPUb{} type $\frac{1}{2} x^2+
\frac{1}{4} x^4$. In their study, the authors
used the reverse nonequilibrium molecular
dynamics method
(RNEMD) \cite{JCP.97.Muller-Plathe} to simulate the system.
Firstly, this method was
tested under condition $\gamma=0$ and length $L=2496$ which
showed that the temperature gradient was well established
and $\expk=0.325\pm 0.002$. Consequent simulations using
different $\gamma$ revealed that the exponent $\expk$ varies
continuously with $\gamma$. Starting from $\expk=0.325$
when $\gamma=0$, $\expk$ keeps decreasing until reaching its
minimum $0.25$ at $\gamma=0.25$. After this point, $\expk$
increases to $0.35$ at $\gamma=1$ as shown in
\reffig{fig:universal}. Therefore, it was suggested that a universal exponent
$\expk$ does not exist. However, it should be mentioned
the result $\expk$ at $\gamma=0$ is quite different from
Wang \etal{}'s which was calculated using both
non-equilibrium method with Langevin heat baths and
equilibrium method with Green-Kubo formula with length up to
$131072$ \cite{EL.11.Wang}.

For momentum non-conserving models, it has been mentioned
that all theories predict a finite heat conductivity.
Numerical simulations also support this conclusion.
Actually, before the three theoretical predictions were announced,
Hu \etal{} had realized that momentum non-conservation
should be a necessary ingredient to obtain the Fourier's
law from their work on the Frenkel-Kontorva (FK) model with a
periodic onsite potential $V(x)\propto\cos(ax)$ in addition
to the harmonic interpartical interaction \cite{PRE.98.Hu},
 Two other studies, one by Aoki \etal{}
\cite{PLA.00.Aoki}, and the other by Hu
\etal{} \cite{PRE.00.Hu}, on
$\phi^4$ model with onsite potential $V(x)= k_4 x^4/4$
further confirms this conclusion.

What is the disorder effect to anharmonic lattices?
This problem was first investigated by Payton \etal{}
\cite{PR.67.Payton}.
They performed non-equilibrium simulations on
disordered anharmonic lattice connected to stochastic
baths. It was found that the anharmonicity could greatly
enhance the heat current. But due to the limited computer
facility at that time, they were not able to verify the Fourier's law.
The first systematic study of the disorder effect in
anharmonic lattice was taken by Li \etal{} \cite{PRL.01.Li}.
In their work, Li \etal{} studied the FPU chain with mass
disorder connected to
\noseh{} heat baths. Their results showed that at low
temperatures, the system obeys the Fourier's law. While at
high temperatures, the conductivity diverges with exponent
$\expk=0.43$. However, in a later study by Dhar and Saito
\cite{PRE.08.Dhar}, the authors found that the exponent
$\expk$ is still 1/3 at low temperatures if one goes to much
larger system size. Therefore, there is no crossover from
diffusive to superdiffusive in disordered FPU chain.
Disorder plays less important role in $\expk$ in anharmonic
chains.

For disordered anharmonic chains with pinning, Dhar and Lebowitz studied the
general Hamiltonian systems \cite{PRL.08.Dhar}
\begin{equation}
\begin{split}
H & =\sum_{l=1}^N \left[\frac{p_l^2}{2m_l}+ \frac{k_0}{2}
x_l^2+\frac{\lambda_0}{2} x_l^4\right] \\
& +\sum_{l=1}^{N+1} \left[\frac{k}{2} (x_l-x_{l-1})^2+\frac{\lambda}{4}
(x_l-x_{l-1})^4\right],
\end{split}
\end{equation}
in which $x_0=x_{N+1}=0$. It was found that when anharmonicity is absent, which
corresponds to the disordered harmonic chain with harmonic onsite potential, the
heat flux decays exponentially with length and the system is a thermal
insulator. However, introduction of a small
will lead to the
$J\sim 1/N$ dependence, resulting in a  diffusive heat transport arises.

Regarding the two dimensional oscillator lattices, only few numerical
studies are carried out to investigate the size
dependent heat conductivity. But the good news is that predictions from
theories do not contradict one another. Both the
renormalization group theory and the mode coupling theory
predict a logarithmic divergence for the heat conductivity.

For numerical investigations, Lippi and Livi studied
two dimensional lattices with size $N_x\times N_y$
\cite{JSP.00.Lippi}. Heat was transported along $x$
direction. The Hamiltonian of the system has the form
\begin{equation}
\label{eq:hmt}
	H\!=\!\sum_{i=1}^{N_x}\sum_{j=1}^{N_y}\!\left[
	\frac{|\mbf{p}_{ij}|}{2m}\!+\!
U(|\mbf{x}_{i+1,j}\!-\!\mbf{x}_{ij}|)+
U(|\mbf{x}_{i,j+1}\!-\!\mbf{x}_{ij}|)
	\!\right],
\end{equation}
where $\mbf{p}_{ij}, \mbf{x}_{ij}$ are the canonical
coordinates. The interaction was taken to be the \FPUb{}
type $U(x)=x^2/2+k_4 x^4/4$ and the Lennard-Jones type
$U(x)=A/x^{12}-B/x^6+B^2/4A$, respectively. The parameters
in the Lennard-Jones potential were chosen so that its
Taylor expansion at the minimum coincide with the \FPUb{}
potential. Using non-equilibrium method with \noseh{} heat
baths, Lippi and Livi firstly demonstrated that with
increasing $N_y$ and fixed $N_x$, the current saturates at a
 small ratio $N_y/N_x$. Subsequent
calculations were performed using fixed ratio $N_y/N_x=1/2$. It was
found that the conductivity will diverge logarithmically
with $N_x$ for both models. Additional equilibrium
simulations showed that the heat flux correlation functions
decay as $t^{-1}$. Using Green-Kubo formula, we again
obtain $\kappa\sim \ln(N_x)$. This result is consistent
with predictions from theories.

There also exists numeric works that contradicts the
prediction of logarithmic divergence.
Yang and Grassberger studied a 2D Hamiltonian system involving FPU
type interaction \cite{A.02.Grassberger}. The Hamiltonian of the system
is similar to Eq.~(\ref{eq:hmt}) but with scalar displacements. The results
showed that when the ratio $N_x/N_y$ is small which corresponds to the 2D
behavior, the conductivity diverges exponentially with $\expk=0.22\pm0.03$.
When $N_y=1$ which is actually a 1D system, $\expk=0.37\pm0.01$.
In another recent work, Shiba and Ito also studied the same
system as Eq.~(\ref{eq:hmt}) with the FPU-type interaction
\cite{JPSJ.08.Shiba}.
However, a power law divergence was obtained by using
lattices with sizes up to $384\times 768$. It is found the conductivity
also diverges in a power-law with $\expk\approx 0.268$, even it was assured that
the boundary effect is negligible in that size.

\subsection{Anomalous Energy Diffusion in Low Dimensions}

The nonlinearity of the interaction in the lattices
prohibits us to analytically solve most of the models.
Instead, a lot of quasi-one-dimensional
gas channel systems consists of non-interacting particles
have been proposed to study heat transport in low
dimensions. These works helped a lot in understanding the
basic ingredients required by the Fourier's law. Moreover,
they helped to understand heat transport from another
aspects of energy diffusion.

\begin{figure}[tb]
 \includegraphicsS{0.8\columnwidth}{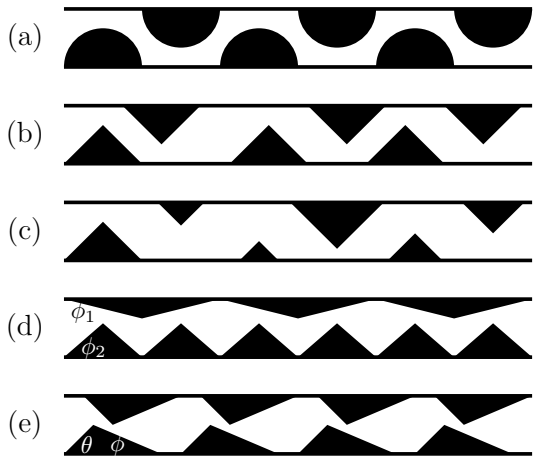}
 \caption{The geometrical configuration of billiard gas
channels. (a) The Lorentz gas channel. (b) The
position-disordered Ehrenfest channel.
(c) The height-disordered Ehrenfest channel. (d) The
polygonal gas channel. (e) The triangle gas channel.}
 \label{fig:billiard}
\end{figure}

Alonso \etal{} first studied the heat
conduction in Lorentz gas channels \cite{PRL.99.Alonso}.
The model consists of a
quasi-1D billiard with periodically distributed
semicircular scatterers as shown in
Fig.~\ref{fig:billiard}(a). Particles carrying heat move
along the billiard and exchange energy with Maxwell thermal
reservoirs at the ends. In this model, no particle can move
between the two reservoirs without being scattered by the
semicircles. Due to the exponential separation of
trajectories, this system has a positive Lyanpunov exponent, \ie{}, it is
chaotic. Alonso \etal{} verified that the heat conduction obeys
Fourier's law which was explained by deterministic
particle diffusion with energy dependent diffusivity
$D(E)\sim E^{1/2}$.

Later, in order to resolve the role of dynamic chaos on
determining the heat transport behavior, Li \etal{} studied
the Ehrenfest gas channels with isosceles right angles
replacing the semicircles in the Lorentz gas channels
\cite{PRL.02.Li}. Their configurations are shown in
Figs.~\ref{fig:billiard}(a) and \ref{fig:billiard}(b). The
linear separation of nearby trajectories in these models
make them different from the Lorentz gas channels in
underlying dynamics. It was found that when the scatterers
are periodic, the heat conductivity diverges with system
length as $\kp\sim N^{0.81}$. Normal heat conduction
$\kp\sim N^0$ can only be reached when either position
disorder (Fig.~\ref{fig:billiard}(b)) or size disorder
(Fig.~\ref{fig:billiard}(c)) is introduced.
These results imply that chaos is not necessary
for normal heat conduction and Fourier's law. Moreover, the
diffusive behavior of the particles are also investigated
using the mean square displacement. Numerical simulations
showed that for the disordered case, the mean square
displacement increase linearly with time, while the periodic
case is superdiffusive $\mean{\Dt
x(t)^2}\sim t^\expx$ with $\expx=1.672$.

Subsequently, two additional works on quasi-1D billiard
model but with different scatterers are studied by Alonso
\etal{} \cite{PRE.02.Alonso} and by Li \etal{}
\cite{PRE.03.Li}, respectively. In the first one, the
scatterers have polygonal shape (Fig.~\ref{fig:billiard}d).
By fixing $\phi_1=(\sqrt{5}-1)\pi/8$ and adjusting
$\phi_2=\pi/q$ with $q=3,4,5,6,7,8,9$, Alonso \etal{} found
that the time dependence of the MSD $\mean{\Dt x(t)^2}\sim
t^\beta$ is superdiffusive for $q=3$ ($\emsd=1.30$),
subdiffusive for $q=4$ ($\emsd$=0.86) and normal for
$q=5,6,7,8,9$ ($\emsd\approx1$). Numerical simulations also
show that for $q=5,6,7,8,9$, the heat conductivity of the
system converges to a non-zero constant in thermal dynamical
limit, while it diverges for $q=3$ and decays to zero for
$q=4$ \cite{PRE.02.Alonso}. In the second work
\cite{PRE.03.Li}, the scatterers are triangles
(Fig.~\ref{fig:billiard}). It was found that the transport
behaviors of this model depend sensitively on whether the
angles, measured in unit of $\pi$, are rational numbers or
irrational numbers. For the rational cases, the conductivity
diverge as power law with exponent $\expk=0.22$ and
the MSD $\expx=1.178$. While for the
irrational case, the conductivity saturates to $0.225$ and
the diffusivity to $0.15$.

All of these works in the billiard models lead us to think
about the intrinsic relation between normal (anomalous)
particle diffusion and normal (anomalous) heat conduction.
Consequently, there were two independent works trying to
relate the diffusion exponent $\expx$ to $\expk$, one by Li
and Wang \cite{PRL.03.Li} and the other by Denisov \etal{}
\cite{Densis}. Both the studies concentrated
on the gas models. Particles transport between the ends of
dynamic channels and exchange energy with the heat baths
connected to the ends. In both models, particles do
not interact just as the billiard models described above.
The first one \cite{PRL.03.Li} used a general continuous
time random walk description. By studying the mean first
passage time of the particles, Li and Wang finally obtained
the relation $\expk=2-2/\expx$.

In the second study
\cite{Densis}, Denisov \etal{} focused on the
special \levy{} type random walk, which generally assumes
that a particle waits at each point for a waiting time
following distribution $\psi_w(t)\sim t^{-\gamma_w-1}\
(\gamma_w>1)$ and then make a jump with distance following
distribution $\psi_f(x,t)$. For \levy{} walk
\cite{PRA.89.Blumen}, $\psi_f(t)\sim t^{{-\gamma_t-1}}$ and $\psi_f(x,t)\sim
\psi_f(t)\dt (|x|-vt)$. This means that
during a single flight, the particles travel at a constant speed $v$. Depending
on $\gamma_t$, the particles diffuse according to
$\mean{x^2(t)}\sim t^\expx$ with $\expx=2$ for
$0<\gamma_f<1$; $\expx=3-\gamma_f$ for $ 1<\gamma_f<2$ and
$\expx=1$ for $\gamma_f>2$. For the superdiffusive cases
$1<\gamma_f<2$, the authors obtained the mean first passage
time $\tau\propto L^{\gamma_f}$. The heat flux
contributed from a single particle then scales as $J\propto
L^{-\gamma_f}=L^{\expx-3}$. Considering that the temperature
gradient $\grad{T} \propto L^{-1}$ and the total number of
particles in the channel $N \propto L$, it was obtained that
the heat conductivity $\kappa\propto
\frac{NJ}{\grad{T}}\propto L^{\expx-1}$ which means
$\expk=\expx-1$. Denisov \etal{} also considered \levy{}
flight with jump length distribution
$\psi_f(x,t)=\psi_f(x)\dt (t-t_f)$ where $\psi_f(x)\sim
x^{-\gamma_f-1}$. However, due to the divergence of mean
square displacement, $\mean{|x|}^2\sim t^{\expx}$ was used
instead of the mean square displacement which gave
$\expk=2/\gamma_f-1=2/(3-\expx)-1$. For the subdiffusive
cases, the conductivity vanishes because the mean passage
time diverges.

Although the billiard gas models helped us
clarify a few puzzles on the conditions for
diffusive heat transport, we should admit that due to the
lacking of interactions and local thermal equilibrium as
pointed out by A. Dhar and D. Dhar
 \cite{PRL.99.Dhar}, conclusions
drawn from them could not be directly transferred to
oscillating lattices. The properties of the
billiard gas models are dominated by their dynamic
properties, or even simpler, geometric properties. while
for lattices, thermodynamic properties are
more important and of interest. Moreover,
heat flux in lattices takes place by phonon transport
without net particle flow, which also make the lattice
systems different from the billiard gases in underlying
physics.

For energy diffusion in lattices, Cipriani \etal{} studied
a one dimensional diatomic hard-point model
\cite{PRL.05.Cipriani}. The
model consists of a chain of hard-point particles with
alternating masses, $m_{2i}=m$ and $m_{2i+1}=rm$, lying on a
line segment with
length $L$. The particles move in one dimension and
elastically collide with other nearest neighbors when they
meet at the same point. Therefore, the order of the
particles are conserved and the system is more like
a lattice model compared to gas models. In
their simulation, the average kinetic
energy is chosen to be $\mean{m_i v_i^2}/2= 1$ and $m=1$.

It is clear that when $r=1$, the collisions only lead to
exchange of velocities and the dynamics is integrable. When
$r$ deviates from 1, the system is non-integrable but still
remains non-chaotic because the evolution
equations is linear: if the velocities of two
successive particles before collision are $v_{i,j}$, then
after collision they will change according to a linear form
\begin{equation}
	v'_i=v_j\pm \frac{1-r}{1+r}(v_i-v_j),
\end{equation}
where plus (minus) sign in $\pm$ corresponds to $i$
even (odd) and $j=i\pm 1$.

\begin{figure}[tb]
\includegraphicsS{0.8\columnwidth}{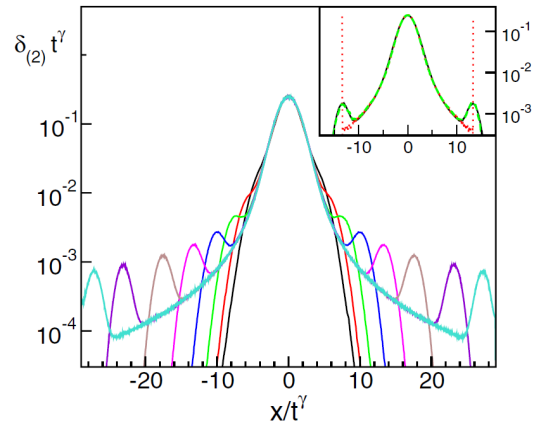}
\caption{\label{fig:hpg}
The rescaled perturbation profiles $\dt_{(2)}(i,t)$ for the
HPG model at $t=$ 40, 80, 160, 320, 640, 1280, 2560, and
3840 for $\gamma=3/5$. Inset: the perturbation profile of
at $t=640$ (solid line) is compared with the
propagators of a Levy walk with exponent $\mu=5/3$ and
velocity $u=1$ (dotted line).
For further details see \textcitee{PRL.05.Cipriani}.}
\end{figure}

After introducing an
infinitesimal perturbation
$\dt v_i(t=0)$ to the $i$'th particle, the spread of
the energy perturbation at later time can be characterized
by
$
\dt_{(2)}(i,t)= m_i (\dt v_i(t))^2.
$
This quantity can be regarded as the energy distribution in
the energy diffusion process. It is observed by
numerical simulation that the profile of $\dt_{(2)}(i,t)$
(See Fig.~\ref{fig:hpg})
satisfies the power-law ansatz
$
	\dt_{(2)}(i,t)\approx t^{-\gamma}
\dt_{(2)}(i/t^{\gamma})
$
very well. A best fit for $i=0$ gives
$\gamma=0.606\pm0.008$. The evolution of the profile
$\dt_{(2)}(i,t)$ was then compared with \levy{} walk with
waiting time distribution $\psi(t)=t^{-\mu-1}$
\cite{PRA.89.Blumen}. The scaling law of \levy{}
walk is \cite{PA.93.Klafter}
\begin{equation}
\label{eq:scaling}
P(x,t)=\frac{1}{t^{1/\mu}}P(\frac{x}{t^{1/\mu}})
\end{equation}
 for the
central part. Therefore,
$\mu=1/\gamma\approx 5/3$. Numerical simulations showed a
good agreement between this two models except that the
ballistic wavefronts are slightly broadened
(Fig.~\ref{fig:hpg}). Since for \levy{} walk, the mean
square displacement $\mean{x^2(t)}\propto t^\expx$ with
$\expx= 3-\mu$, therefore $\expx$ should be $4/3$
approximately. Numerical calculation of mean square
displacement of their diatomic gas model
 showed that $\beta\approx 1.35$,
which supported the connection of their model with \levy{}
walk.

The heat conduction of this model had been
separately studied earlier by Hatano \cite{PRE.99.Hatano}, Dhar
\cite{PRL.01.Dhar} and by Grassberger \etal{}
\cite{PRL.02.Grassberger}. The first work
\cite{PRE.99.Hatano} used $r=1.22$ and found that
$\expk=0.33\sim 0.37$. The second one \cite{PRL.01.Dhar}
observed a small divergence of the heat conductivity with
$\expk\approx0.17$ when $r=1.22$ using length $N$ up to
$1281$. The last one \cite{PRL.02.Grassberger} used a very
efficient event driven algorithm and simulated chains with
length up to $16383$. It was found that for very large $N$,
the divergent exponent tends to
$\expk=0.32_{-0.01}^{+0.03}$ for all mass ratio
between $1$ and $\sim 5$, which was claimed to be
consistent with the prediction of renormalization group
theory, $\expk=1/3$, by Narayan and Ramaswamy
\cite{PRL.02.Narayan}.

\begin{figure}[tb]
\includegraphicsS{1\columnwidth}{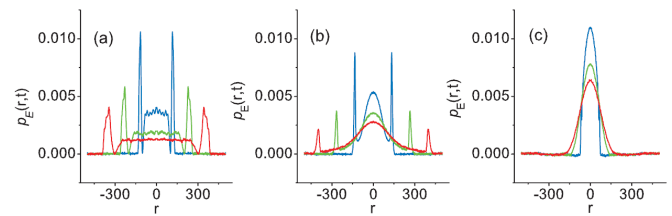}
\caption{
$\rho(i,t)$ versus $t$ for (a) the Toda lattice, (b)
the \FPUb{} lattice and (c) the $\phi^4$ lattice at time
$t=100$ (blue), 200 (green) and 300 (red), respectively.
For further details see \textcitee{PRL.06.Zhao}.}
\label{fig:zhao}
\end{figure}

The hard-point gas model cannot mimic the real lattice
models because the interactions between particles are
present only when they collide. For interacting lattices
connected by springs, another important systematic study
on energy diffusion were carried out by Zhao
\cite{PRL.06.Zhao}. In this
work, Zhao showed a heuristic approach to investigate the
energy distribution in a diffusion process, denoted as
$\rho(i,t)$, using
equilibrium energy spatiotempral correlations, \ie{}
\begin{equation}
\label{eq:rho_zhao}
	\rho(i,t)=\frac{\mean{\Dt E_i(t)\Dt E_0(0)}}{\mean{\Dt
E_0(0)\Dt E_0(0)}}.
\end{equation}
This quantity is in equivalence with $\dt_{(2)}(i,t)$ in
Cipriani \etal{}'s work \cite{PRL.05.Cipriani}.
Using this method, Zhao studied three
well known lattices showing different behaviors of heat
conduction: the Toda lattice which is ballistic, the
\FPUb{} lattice which is superdiffusive and the $\phi^4$
lattice which is diffusive. It was found that for all these
models, the energy correlation functions can well capture
the characteristics of the corresponding energy diffusion
(Fig.~\ref{fig:zhao}).
Specifically, for Toda lattice, the energy distribution
is dominated by two clear ballistic wave front; for
$\phi^4$ lattice, the distribution is Gaussian-like; for
\FPUb{} lattice, it is a combination of the two above, two
seemingly ballistic peak and a Gaussian-like central part.
Quantitative results for the mean square displacement
$\mean{x^2(t)}= \sum_i i^2 \rho(i,t)$ were calculated,
which shows power law dependence on time $\mean{x^2(t)}\sim
 t^\expx$. The exponent $\expx\approx2$ for Toda lattice,
$\expx\approx1.4$ for \FPUb{} lattice and $\expx\approx1$
for $\phi^4$ lattice, which coincide with their ballistic,
superdiffusive and diffusive behaviors, respectively.
Moreover, it was argued that this supports the
$\expk=\expx-1$ relation because the exponent $\expk$ for
\FPUb{} model is $2/5$ as predicted by the Peierls-Boltzmann theory.

In a recent work, Zaburdaev \etal{} restudied the
energy diffusion in one dimensional systems using the
phenomenological \levy{} walk approach
\cite{PRL.11.Zaburdaev}. They first generalized the
\levy{}
walk model to involve random fluctuation in the movement of
particles. Specifically, compared to the standard \levy{}
walk, it was additionally assumed that during a
single flight between two successive collisions, a
particle's
position evolves according to Langevin equation $\dot{x}(t)=
v_0+ \xi(t)$, where $\xi(t)$ is a Gaussian random process
with correlation $\mean{\xi(t)\xi(s)}=D_v \dt (t-s)$. When
the randomness is absent ($D_v=0$), this model reduces to the
standard \levy{} walk model with moving speed $v_0$.
Analytic analysis showed that
for their generalized \levy{} walk model, the central part
of the particle density profile $P(x,t)$ follows the
scaling of standard \levy{} walk, \ie{},
Eq.~\eqref{eq:scaling}, while the two ballistic humps
follow another scaling
\begin{equation}
	\label{eq:scaling2}
	P_{\mathrm{hump}}(\bar{x},t)= \frac{1}{t^{1/2}}
	P_{\mathrm{hump}}(\frac{\bar{x}}{t^{1/2}}),
\end{equation}
where $\bar{x}= x-v_0 t$.

\begin{figure}[tb]
\includegraphicsS{0.8\columnwidth}{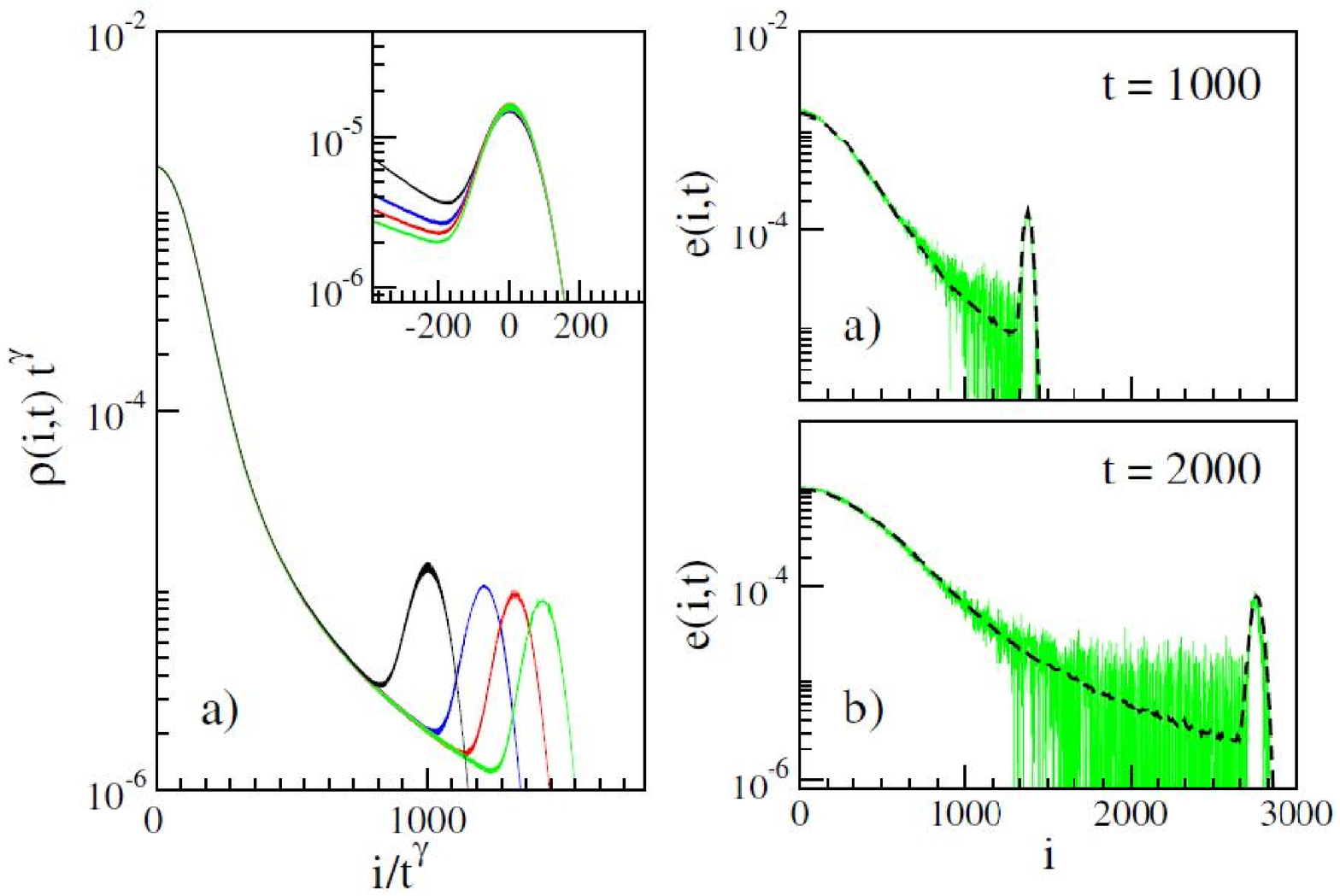}
\caption{
Left: rescaled perturbation profiles of the HPG model at
time $t=1000$,2000,4000,6600 respectively. The profiles
are scaled under Eq.~(\ref{eq:scaling}). The inset shows
the ballistic humps after the scaling transformation
Eq.~(\ref{eq:scaling2}). Right: energy correlation
functions $e(i,t)$ for the \FPUb{} model (thin solid
lines) and the propagators of the generalized \levy{} walk
model (thick dashed lines).
For further details see \textcitee{PRL.11.Zaburdaev}.}
\label{fig:zaburdaev}
\end{figure}

Using this model, Zaburdaev \etal{} restudied the energy
diffusion in hard-point particles (HPG)
model \cite{PRE.99.Hatano,PRL.01.Dhar,PRL.02.Grassberger,PRL.05.Cipriani} and
the \FPUb{} model using
Eq.~\eqref{eq:rho_zhao} by Zhao \cite{PRL.06.Zhao}. For the
hard-point particles model, Zaburdaev \etal{} calculated
the evolution of an infinitesimal perturbation $\rho(i,t)$
following Ref.~\cite{PRL.05.Cipriani}. It was found that the
scaling relations Eq.~\eqref{eq:scaling} with $\gamma=5/3$
and Eq.~\eqref{eq:scaling2} are fulfilled
separately for the central part and the ballistic part,
respectively (Fig.~\ref{fig:zaburdaev}). For
the \FPUb{} model, the authors performed a direct comparison
between the energy spatiotempral correlation functions
$e(i,t)$ as in
Eq.~\eqref{eq:rho_zhao} and the probability density for the
generalized \levy{} walk model. The comparison was made at
time $t=1000$ and $2000$ respectively
(Fig.~\ref{fig:zaburdaev}). A good agreement was
observed by choosing the average energy per particle
$\epsilon= 1$, $v_0=1.384$ and $D_v=0.49$.

Although there are a few numerical
studies showing that energy diffusion in some one
dimensional systems can be described by (generalized) \levy{}
walk, it can be argued
that a fundamental theory connecting these two phenomena
does not exist in general. All existing works are based on
phenomenological approaches. It is not known either for which kind of
systems, the \levy{} walk description is applicable.

However, regarding the connection between
(anomalous) heat conduction and (anomalous)
energy diffusion, in a most recent work, Liu \etal{} have
rigorously proved an equality that can be generally applied
to all homogeneous one dimensional systems near thermal
equilibrium \cite{s.12.Liu}. In their work, Liu \etal{}
considered a general one dimensional Hamiltonian system
which was initially ($t=-\infty$) at thermal
equilibrium with temperature $T$. Then by assuming local thermal equilibrium,
a small perturbation with perturbative Hamiltonian
$H_{ext}=-\frac{1}{T} \int \epsilon(x,t)\Dt T(x)\dd
x$ was imposed on the system from $t=-\infty$, where
$\epsilon(x,t)$ is the local energy density. This form of
perturbation will drive the system
away from equilibrium and finally reach a steady state with
temperature profile $T(x)=T+\Dt T(x)$, provided that local
thermal equilibrium is satisfied.
After removing the perturbation at $t=0$, the system will
start to relax to equilibrium at temperature $T$ again.
This relaxation process is interpreted as energy diffusion.
In their work, Liu \etal{} first analytically proved that
in the diffusion process ($t\ge0$), the normalized distribution of the energy
profile can be written as
\begin{equation}
\rho(x,t)=\frac{\delta\bar\epsilon(x,t)}{\int
\delta \bar\epsilon(x,t)\dd x}= \frac{\int
C_{\epsilon\epsilon}(x-x',t)\Dt T(x')\dd x'}{k_B T^2
c_V\int \Dt T(x)\dd x},
\end{equation}
where $C_{\epsilon\epsilon}=\mean{\Dt\epsilon(x,t)\Dt\epsilon(0,0)
}$ is the autocorrelation function of energy density
fluctuation. Consequently, it was proved
that the mean square displacement of energy diffusion should satisfy a very
general equality independent of the system Hamiltonian
and the initial temperature profile $\Dt T(x)$, which reads
\begin{equation}
\label{eq:msd}
 \diff{\msd{t}}{t}{2}= \frac{C(t)}{k_BT^2c_V},
\end{equation}
where $C(t)$ is again the heat flux autocorrelation
function appearing in the Green-Kubo formula
Eq.~\eqref{eq:GK}. Because in their derivation there is no
any explicit assumption except the local thermal
equilibrium, this equality should be universal, even
for isotropic systems in higher dimensions.

The authors also
showed a few corollaries of this equality for different
diffusive behaviors. When the system is normal diffusive,
\ie{} $\msd{t}\sim 2Dt$ in large $t$ limit, by inserting
Eq.~\eqref{eq:msd} into the Green-Kubo formula, we will get
$\kappa=c_V D$. For the superdiffusive case $\msd{t}\sim
t^\expx$ with $1<\expx<2$, the Green-Kubo formula with a
cut-off time $t_c\propto L$ will lead to
$\kappa\sim L^{\expx-1}$, which means $\expk=\expx-1$. While
for subdiffusive cases ($0<\expx<1$), the heat conductivity
decays to 0 as $\kappa\sim 1/L^{1-\expx}$
\cite{s.12.Liu}.

\sectionN{Conclusions}{conclusion}
In this Colloquium we took the reader on a tour presenting
the state of the art of the topic about anomalous thermal
conduction and heat diffusion in low dimensional nanoscale structures,
an emerging research direction which is expected to shade
light for renewable energy and thermal management for future
nano devices. Particularly, we surveyed and explained
physical mechanisms that respond to the anomalous thermal
conductivity.

We firstly reviewed recent experiments on
thermal conductivity of nanotube, nanowire and graphene.
Then, we provided a review on the computational
investigations of thermal conduction in nanascale,
concentrating on atomistic simulations. At last, various
theoretical models to explain the anomalous thermal
conductivity and the connection between thermal conduction
and heat diffusion were overviewed. Both numerical and
experimental studies in nanostructures ranging from
nanotubes \cite{Chang2008PRL,zhang2005JCP123a,zhang2005JCP123B}
nanowires \cite{Yang2010NT5}, to polyethylene nanofibers
\cite{Henry2008PRL,PRB.09.Henry,Shen2010Nnano}
have shown
that heat conduction in low-dimensional systems does not
obey Fourier's law even though the system length is much
longer than the phonon mean free path. In these structures
phonons transports superdiffusively---a process faster than
random walk but slower than ballistic motion, which leads to
a length dependent thermal conductivity. Thus nano
structured materials are promising platforms to testify
fundamental phonon transport theories.

Moreover, given the
fact that thermal management devices generally can be
considered for applications that require power ranging from
milli-watts up to several thousand watts, the general range
of applications where they are indispensable is stupendous.
Compared with decades ago, we have a better and clearer
understanding of the thermal conduction and heat diffusion
in nanoscale materials.

\sectionN{Outlook and Challenges}{outlook}

However, there are still many open
questions. On the experimental side, experimental
demonstration of low-dimensional phonon transport theory
depends on how to measure its thermal conductivity
accurately. To this end, one needs to make the sample
smaller, for example towards a dozens of nanometers or even
a few nanometers only. Difficulties in integrating nanomaterials with suspended
structures suitable for thermal measurements have resulted in rare experimental
studies of size-dependent thermal conductivity in 1D and 2D systems. And another
challenge is to get rid of the contact thermal resistance. The unknown thermal
contact resistance remains the primary technical challenge in testing the
Fourier's law. More efforts are needed to explore a brand new method of
measuring thermal contact resistance directly.

Because of the
complexities inherent in thermal contact resistance
measurement, this task needs the help from theoretical
predictions. However, so far a truly comprehensive theory
for this effect is lacking. The current approaches for
thermal transport across an interface, such as the acoustic
mismatch (AMM) theory and the diffusive mismatch (DMM)
theory, are based on the assumption that phonon transport
via a combination of either ballistic or diffusive
transport on either side of the interface. Both schemes
offer limited accuracy for nanoscale interfacial resistance
predictions, due to as we have shown above, phonons
transport super-diffusively in low-dimensional systems.
Therefore, it is necessary to establish an improved theory
describing thermal transport across the interface by taking
into account the anomalous thermal transport characteristics
of nanostructures. To provide quantitative prediction
from theory and simulation, how to set up a transport theory
by incorporating nonlinearity in the quantum regime is still
a challenge. The nonequilibrium Green's (NEGF) function
method \cite{PRB.08.Xu} certainly serves as an elegant
mathematical framework. However, when including the
phonon-phonon interactions, the NEGF presents a cumbersome
task. Thus further systematical investigations combine
experimental and theoretical efforts on fundamental
mechanisms will be helpful to advance the field.

\sectionN{Acknowledgement}{acknowledgement}
This work is supported by grants from Ministry of Education, Singapore,
Science and Engineering Research Council, Singapore, National University of
Singapore, and Asian Office of Aerospace R\&D (AOARD) of the US Air Force by
grants, R-144-000-285-646, R-144-000-280-305, R-144-000-289-597, respectively;
and the Ministry of Science and Technology of China, Grant No. 2011CB933001
(G.Z.),

\bibliographystyle{apsrev4-1}
\bibliography{refs}

%Merlin.mbs v4.21 2009-07-09.
\begin{thebibliography}{100}%
\makeatletter
\providecommand \@ifxundefined [1]{%
 \ifx #1\undefined \expandafter \@firstoftwo
 \else \expandafter \@secondoftwo
\fi
}%
\providecommand \@ifnum [1]{%
 \ifnum #1\expandafter \@firstoftwo
 \else \expandafter \@secondoftwo
\fi
}%
\providecommand \enquote [1]{``#1''}%
\providecommand \bibnamefont  [1]{#1}%
\providecommand \bibfnamefont [1]{#1}%
\providecommand \citenamefont [1]{#1}%
\providecommand\href[0]{\@sanitize\@href}%
\providecommand\@href[1]{\endgroup\@@startlink{#1}\endgroup\@@href}%
\providecommand\@@href[1]{#1\@@endlink}%
\providecommand \@sanitize [0]{\begingroup\catcode`\&12\catcode`\#12\relax}%
\@ifxundefined \pdfoutput {\@firstoftwo}{%
 \@ifnum{\z@=\pdfoutput}{\@firstoftwo}{\@secondoftwo}%
}{%
 \providecommand\@@startlink[1]{\leavevmode\special{html:<a href="#1">}}%
 \providecommand\@@endlink[0]{\special{html:</a>}}%
}{%
 \providecommand\@@startlink[1]{%
  \leavevmode
  \pdfstartlink
   attr{/Border[0 0 1 ]/H/I/C[0 1 1]}%
   user{/Subtype/Link/A<</Type/Action/S/URI/URI(#1)>>}%
  \relax
 }%
 \providecommand\@@endlink[0]{\pdfendlink}%
}%
\providecommand \url  [0]{\begingroup\@sanitize \@url }%
\providecommand \@url [1]{\endgroup\@href {#1}{\urlprefix}}%
\providecommand \urlprefix [0]{URL }%
\providecommand \Eprint[0]{\href }%
\@ifxundefined \urlstyle {%
  \providecommand \doi [1]{doi:\discretionary{}{}{}#1}%
}{%
  \providecommand \doi [0]{doi:\discretionary{}{}{}\begingroup
  \urlstyle{rm}\Url }%
}%
\providecommand \doibase [0]{http://dx.doi.org/}%
\providecommand \Doi[1]{\href{\doibase#1}}%
\providecommand \bibAnnote [3]{%
  \BibitemShut{#1}%
  \begin{quotation}\noindent
    \textsc{Key:}\ #2\\\textsc{Annotation:}\ #3%
  \end{quotation}%
}%
\providecommand \bibAnnoteFile [2]{%
  \IfFileExists{#2}{\bibAnnote {#1} {#2} {\input{#2}}}{}%
}%
\providecommand \typeout [0]{\immediate \write \m@ne }%
\providecommand \selectlanguage [0]{\@gobble}%
\providecommand \bibinfo [0]{\@secondoftwo}%
\providecommand \bibfield [0]{\@secondoftwo}%
\providecommand \translation [1]{[#1]}%
\providecommand \BibitemOpen[0]{}%
\providecommand \bibitemStop [0]{}%
\providecommand \bibitemNoStop [0]{.\EOS\space}%
\providecommand \EOS [0]{\spacefactor3000\relax}%
\providecommand \BibitemShut [1]{\csname bibitem#1\endcsname}%
%</preamble>
\bibitem{JMP.67.Rieder}%
  \BibitemOpen
  \bibfield{author}{%
  \bibinfo {author} {\bibfnamefont{Z.}~\bibnamefont{Rieder}}, \bibinfo {author}
  {\bibfnamefont{J.~L.}\ \bibnamefont{Lebovitz}},\ and\ \bibinfo {author}
  {\bibfnamefont{E.}~\bibnamefont{Lieb}},\ }%
  \bibfield{journal}{%
  \Doi{10.1063/1.1705319}{\bibinfo {journal} {J. Math. Phys.}}\ }%
  \textbf{\bibinfo {volume} {8}},\ \bibinfo {pages} {1073} (\bibinfo {year}
  {1967})%
  \bibAnnoteFile{NoStop}{JMP.67.Rieder}%
\bibitem{PTP.68.Nakazawa}%
  \BibitemOpen
  \bibfield{author}{%
  \bibinfo {author} {\bibfnamefont{H.}~\bibnamefont{Nakazawa}},\ }%
  \bibfield{journal}{%
  \Doi{10.1143/PTP.39.236}{\bibinfo {journal} {Prog. Theor. Phys.}}\ }%
  \textbf{\bibinfo {volume} {39}},\ \bibinfo {pages} {236} (\bibinfo {year}
  {1968})%
  \bibAnnoteFile{NoStop}{PTP.68.Nakazawa}%
\bibitem{PRL.97.Lepri}%
  \BibitemOpen
  \bibfield{author}{%
  \bibinfo {author} {\bibfnamefont{S.}~\bibnamefont{Lepri}}, \bibinfo {author}
  {\bibfnamefont{R.}~\bibnamefont{Livi}},\ and\ \bibinfo {author}
  {\bibfnamefont{A.}~\bibnamefont{Politi}},\ }%
  \bibfield{journal}{%
  \Doi{10.1103/PhysRevLett.78.1896}{\bibinfo {journal} {Phys. Rev. Lett.}}\ }%
  \textbf{\bibinfo {volume} {78}},\ \bibinfo {pages} {1896} (\bibinfo {year}
  {1997})%
  \bibAnnoteFile{NoStop}{PRL.97.Lepri}%
\bibitem{EL.98.Lepri}%
  \BibitemOpen
  \bibfield{author}{%
  \bibinfo {author} {\bibfnamefont{S.}~\bibnamefont{Lepri}}, \bibinfo {author}
  {\bibfnamefont{R.}~\bibnamefont{Livi}},\ and\ \bibinfo {author}
  {\bibfnamefont{A.}~\bibnamefont{Politi}},\ }%
  \bibfield{journal}{%
  \bibinfo {journal} {Europhys. Lett.}\ }%
  \textbf{\bibinfo {volume} {43}},\ \bibinfo {pages} {271} (\bibinfo {year}
  {1998})%
  \bibAnnoteFile{NoStop}{EL.98.Lepri}%
\bibitem{PRE.99.Hatano}%
  \BibitemOpen
  \bibfield{author}{%
  \bibinfo {author} {\bibfnamefont{T.}~\bibnamefont{Hatano}},\ }%
  \bibfield{journal}{%
  \Doi{10.1103/PhysRevE.59.R1}{\bibinfo {journal} {Phys. Rev. E}}\ }%
  \textbf{\bibinfo {volume} {59}},\ \bibinfo {pages} {R1} (\bibinfo {year}
  {1999})%
  \bibAnnoteFile{NoStop}{PRE.99.Hatano}%
\bibitem{PRL.01.Dhar}%
  \BibitemOpen
  \bibfield{author}{%
  \bibinfo {author} {\bibfnamefont{A.}~\bibnamefont{Dhar}},\ }%
  \bibfield{journal}{%
  \Doi{10.1103/PhysRevLett.86.3554}{\bibinfo {journal} {Phys. Rev. Lett.}}\ }%
  \textbf{\bibinfo {volume} {86}},\ \bibinfo {pages} {3554} (\bibinfo {year}
  {2001})%
  \bibAnnoteFile{NoStop}{PRL.01.Dhar}%
\bibitem{PRL.02.Grassberger}%
  \BibitemOpen
  \bibfield{author}{%
  \bibinfo {author} {\bibfnamefont{P.}~\bibnamefont{Grassberger}}, \bibinfo
  {author} {\bibfnamefont{W.}~\bibnamefont{Nadler}},\ and\ \bibinfo {author}
  {\bibfnamefont{L.}~\bibnamefont{Yang}},\ }%
  \bibfield{journal}{%
  \Doi{10.1103/PhysRevLett.89.180601}{\bibinfo {journal} {Phys. Rev. Lett.}}\
  }%
  \textbf{\bibinfo {volume} {89}},\ \bibinfo {pages} {180601} (\bibinfo {year}
  {2002})%
  \bibAnnoteFile{NoStop}{PRL.02.Grassberger}%
\bibitem{zhang2005JCP123B}%
  \BibitemOpen
  \bibfield{author}{%
  \bibinfo {author} {\bibfnamefont{G.}~\bibnamefont{Zhang}}\ and\ \bibinfo
  {author} {\bibfnamefont{B.}~\bibnamefont{Li}},\ }%
  \bibfield{journal}{%
  \bibinfo {journal} {J. Chem. Phys.}\ }%
  \textbf{\bibinfo {volume} {123}},\ \bibinfo {pages} {014705} (\bibinfo {year}
  {2005})%
  \bibAnnoteFile{NoStop}{zhang2005JCP123B}%
\bibitem{Yang2010NT5}%
  \BibitemOpen
  \bibfield{author}{%
  \bibinfo {author} {\bibfnamefont{N.}~\bibnamefont{Yang}}, \bibinfo {author}
  {\bibfnamefont{G.}~\bibnamefont{Zhang}},\ and\ \bibinfo {author}
  {\bibfnamefont{B.}~\bibnamefont{Li}},\ }%
  \bibfield{journal}{%
  \bibinfo {journal} {Nano Today}\ }%
  \textbf{\bibinfo {volume} {5}},\ \bibinfo {pages} {85} (\bibinfo {year}
  {2010})%
  \bibAnnoteFile{NoStop}{Yang2010NT5}%
\bibitem{Chang2008PRL}%
  \BibitemOpen
  \bibfield{author}{%
  \bibinfo {author} {\bibfnamefont{C.~W.}\ \bibnamefont{Chang}}, \bibinfo
  {author} {\bibfnamefont{D.}~\bibnamefont{Okawa}}, \bibinfo {author}
  {\bibfnamefont{H.}~\bibnamefont{Garcia}}, \bibinfo {author}
  {\bibfnamefont{A.}~\bibnamefont{Majumdar}},\ and\ \bibinfo {author}
  {\bibfnamefont{A.}~\bibnamefont{Zettl}},\ }%
  \bibfield{journal}{%
  \Doi{10.1103/PhysRevLett.101.075903}{\bibinfo {journal} {Phys. Rev. Lett.}}\
  }%
  \textbf{\bibinfo {volume} {101}},\ \bibinfo {pages} {075903} (\bibinfo {year}
  {2008})%
  \bibAnnoteFile{NoStop}{Chang2008PRL}%
\bibitem{zhang2010NS2}%
  \BibitemOpen
  \bibfield{author}{%
  \bibinfo {author} {\bibfnamefont{G.}~\bibnamefont{Zhang}}\ and\ \bibinfo
  {author} {\bibfnamefont{B.}~\bibnamefont{Li}},\ }%
  \bibfield{journal}{%
  \bibinfo {journal} {NanoScale}\ }%
  \textbf{\bibinfo {volume} {2}},\ \bibinfo {pages} {1058} (\bibinfo {year}
  {2010})%
  \bibAnnoteFile{NoStop}{zhang2010NS2}%
\bibitem{Pop2010NR3}%
  \BibitemOpen
  \bibfield{author}{%
  \bibinfo {author} {\bibfnamefont{E.}~\bibnamefont{Pop}},\ }%
  \bibfield{journal}{%
  \bibinfo {journal} {NanoResearch}\ }%
  \textbf{\bibinfo {volume} {3}},\ \bibinfo {pages} {147} (\bibinfo {year}
  {2010})%
  \bibAnnoteFile{NoStop}{Pop2010NR3}%
\bibitem{RMP.12.Li}%
  \BibitemOpen
  \bibfield{author}{%
  \bibinfo {author} {\bibfnamefont{N.}~\bibnamefont{Li}}, \bibinfo {author}
  {\bibfnamefont{J.}~\bibnamefont{Ren}}, \bibinfo {author}
  {\bibfnamefont{L.}~\bibnamefont{Wang}}, \bibinfo {author}
  {\bibfnamefont{G.}~\bibnamefont{Zhang}}, \bibinfo {author}
  {\bibfnamefont{P.}~\bibnamefont{Hanggi}},\ and\ \bibinfo {author}
  {\bibfnamefont{B.}~\bibnamefont{Li}},\ }%
  \bibfield{journal}{%
  \bibinfo {journal} {arXiv:1108.6120 (Rev. Mod. Phys. in press)}}%
   (\bibinfo {year} {2012})%
  \bibAnnoteFile{NoStop}{RMP.12.Li}%
\bibitem{PR.03.Lepri}%
  \BibitemOpen
  \bibfield{author}{%
  \bibinfo {author} {\bibfnamefont{S.}~\bibnamefont{Lepri}}, \bibinfo {author}
  {\bibfnamefont{R.}~\bibnamefont{Livi}},\ and\ \bibinfo {author}
  {\bibfnamefont{A.}~\bibnamefont{Politi}},\ }%
  \bibfield{journal}{%
  \Doi{DOI: 10.1016/S0370-1573(02)00558-6}{\bibinfo {journal} {Phy. Rep.}}\ }%
  \textbf{\bibinfo {volume} {377}},\ \bibinfo {pages} {1} (\bibinfo {year}
  {2003})%
  \bibAnnoteFile{NoStop}{PR.03.Lepri}%
\bibitem{AP.08.Dhar}%
  \BibitemOpen
  \bibfield{author}{%
  \bibinfo {author} {\bibfnamefont{A.}~\bibnamefont{Dhar}},\ }%
  \bibfield{journal}{%
  \Doi{10.1080/00018730802538522}{\bibinfo {journal} {Adv. Phys.}}\ }%
  \textbf{\bibinfo {volume} {57}},\ \bibinfo {pages} {457} (\bibinfo {year}
  {2008})%
  \bibAnnoteFile{NoStop}{AP.08.Dhar}%
\bibitem{Nanotube}%
  \BibitemOpen
  \bibfield{author}{%
  \bibinfo {author} {\bibfnamefont{P.}~\bibnamefont{Kim}}, \bibinfo {author}
  {\bibfnamefont{L.}~\bibnamefont{Shi}}, \bibinfo {author}
  {\bibfnamefont{A.}~\bibnamefont{Majumdar}},\ and\ \bibinfo {author}
  {\bibfnamefont{P.~L.}\ \bibnamefont{Mceuen}},\ }%
  \bibfield{journal}{%
  \Doi{10.1103/PhysRevLett.87.215502}{\bibinfo {journal} {Phys. Rev. Lett.}}\
  }%
  \textbf{\bibinfo {volume} {87}},\ \bibinfo {pages} {215502} (\bibinfo {year}
  {2001})%
  \bibAnnoteFile{NoStop}{Nanotube}%
\bibitem{Shen2010Nnano}%
  \BibitemOpen
  \bibfield{author}{%
  \bibinfo {author} {\bibfnamefont{S.}~\bibnamefont{Shen}}, \bibinfo {author}
  {\bibfnamefont{A.}~\bibnamefont{Henry}}, \bibinfo {author}
  {\bibfnamefont{J.}~\bibnamefont{Tong}}, \bibinfo {author}
  {\bibfnamefont{R.}~\bibnamefont{Zheng}},\ and\ \bibinfo {author}
  {\bibfnamefont{G.}~\bibnamefont{Chen}},\ }%
  \bibfield{journal}{%
  \Doi{10.1038/nnano.2010.27}{\bibinfo {journal} {Nat. Nanotech.}}\ }%
  \textbf{\bibinfo {volume} {5}},\ \bibinfo {pages} {251} (\bibinfo {year}
  {2010})%
  \bibAnnoteFile{NoStop}{Shen2010Nnano}%
\bibitem{Li2003APL}%
  \BibitemOpen
  \bibfield{author}{%
  \bibinfo {author} {\bibfnamefont{D.}~\bibnamefont{Li}}, \bibinfo {author}
  {\bibfnamefont{Y.}~\bibnamefont{Wu}}, \bibinfo {author}
  {\bibfnamefont{P.}~\bibnamefont{Kim}}, \bibinfo {author}
  {\bibfnamefont{L.}~\bibnamefont{Shi}}, \bibinfo {author}
  {\bibfnamefont{P.}~\bibnamefont{Yang}},\ and\ \bibinfo {author}
  {\bibfnamefont{A.}~\bibnamefont{Majumdar}},\ }%
  \bibfield{journal}{%
  \Doi{10.1063/1.1616981}{\bibinfo {journal} {Appl. Phys. Lett.}}\ }%
  \textbf{\bibinfo {volume} {83}},\ \bibinfo {pages} {2934} (\bibinfo {year}
  {2003})%
  \bibAnnoteFile{NoStop}{Li2003APL}%
\bibitem{Chen2008PRL}%
  \BibitemOpen
  \bibfield{author}{%
  \bibinfo {author} {\bibfnamefont{R.}~\bibnamefont{Chen}}, \bibinfo {author}
  {\bibfnamefont{A.~I.}\ \bibnamefont{Hochbaum}}, \bibinfo {author}
  {\bibfnamefont{P.}~\bibnamefont{Murphy}}, \bibinfo {author}
  {\bibfnamefont{J.}~\bibnamefont{Moore}}, \bibinfo {author}
  {\bibfnamefont{P.}~\bibnamefont{Yang}},\ and\ \bibinfo {author}
  {\bibfnamefont{A.}~\bibnamefont{Majumdar}},\ }%
  \bibfield{journal}{%
  \Doi{10.1103/PhysRevLett.101.105501}{\bibinfo {journal} {Phys. Rev. Lett.}}\
  }%
  \textbf{\bibinfo {volume} {101}},\ \bibinfo {pages} {105501} (\bibinfo {year}
  {2008})%
  \bibAnnoteFile{NoStop}{Chen2008PRL}%
\bibitem{Hochbaum2008Nat451}%
  \BibitemOpen
  \bibfield{author}{%
  \bibinfo {author} {\bibfnamefont{A.~I.}\ \bibnamefont{Hochbaum}}, \bibinfo
  {author} {\bibfnamefont{R.}~\bibnamefont{Chen}}, \bibinfo {author}
  {\bibfnamefont{R.~D.}\ \bibnamefont{Delgado}}, \bibinfo {author}
  {\bibfnamefont{W.}~\bibnamefont{Liang}}, \bibinfo {author}
  {\bibfnamefont{E.~C.}\ \bibnamefont{Garnett}}, \bibinfo {author}
  {\bibfnamefont{M.}~\bibnamefont{Najarian}}, \bibinfo {author}
  {\bibfnamefont{A.}~\bibnamefont{Majumdar}},\ and\ \bibinfo {author}
  {\bibfnamefont{P.}~\bibnamefont{Yang}},\ }%
  \bibfield{journal}{%
  \bibinfo {journal} {Nature}\ }%
  \textbf{\bibinfo {volume} {451}},\ \bibinfo {pages} {163} (\bibinfo {year}
  {2008})%
  \bibAnnoteFile{NoStop}{Hochbaum2008Nat451}%
\bibitem{li2003APL_SiGe}%
  \BibitemOpen
  \bibfield{author}{%
  \bibinfo {author} {\bibfnamefont{D.}~\bibnamefont{Li}}, \bibinfo {author}
  {\bibfnamefont{Y.}~\bibnamefont{Wu}}, \bibinfo {author}
  {\bibfnamefont{R.}~\bibnamefont{Fan}}, \bibinfo {author}
  {\bibfnamefont{P.}~\bibnamefont{Yang}},\ and\ \bibinfo {author}
  {\bibfnamefont{A.}~\bibnamefont{Majumdar}},\ }%
  \bibfield{journal}{%
  \Doi{10.1063/1.1619221}{\bibinfo {journal} {Appl. Phys. Lett.}}\ }%
  \textbf{\bibinfo {volume} {83}},\ \bibinfo {pages} {3186} (\bibinfo {year}
  {2003})%
  \bibAnnoteFile{NoStop}{li2003APL_SiGe}%
\bibitem{Bui2012small}%
  \BibitemOpen
  \bibfield{author}{%
  \bibinfo {author} {\bibfnamefont{C.~T.}\ \bibnamefont{Bui}}, \bibinfo
  {author} {\bibfnamefont{R.}~\bibnamefont{Xie}}, \bibinfo {author}
  {\bibfnamefont{M.}~\bibnamefont{Zheng}}, \bibinfo {author}
  {\bibfnamefont{Q.}~\bibnamefont{Zhang}}, \bibinfo {author}
  {\bibfnamefont{C.~H.}\ \bibnamefont{Sow}}, \bibinfo {author}
  {\bibfnamefont{B.}~\bibnamefont{Li}},\ and\ \bibinfo {author}
  {\bibfnamefont{J.~T.~L.}\ \bibnamefont{Thong}},\ }%
  \bibfield{journal}{%
  \Doi{10.1002/smll.201102046}{\bibinfo {journal} {Small}}\ }%
  \textbf{\bibinfo {volume} {8}},\ \bibinfo {pages} {738} (\bibinfo {year}
  {2012})%
  \bibAnnoteFile{NoStop}{Bui2012small}%
\bibitem{Moore2009JAP}%
  \BibitemOpen
  \bibfield{author}{%
  \bibinfo {author} {\bibfnamefont{A.~L.}\ \bibnamefont{Moore}}, \bibinfo
  {author} {\bibfnamefont{M.~T.}\ \bibnamefont{Pettes}}, \bibinfo {author}
  {\bibfnamefont{F.}~\bibnamefont{Zhou}},\ and\ \bibinfo {author}
  {\bibfnamefont{L.}~\bibnamefont{Shi}},\ }%
  \bibfield{journal}{%
  \Doi{10.1063/1.3191657}{\bibinfo {journal} {J. Appl. Phys.}}\ }%
  \textbf{\bibinfo {volume} {106}},\ \bibinfo {eid} {034310} (\bibinfo {year}
  {2009})%
  \bibAnnoteFile{NoStop}{Moore2009JAP}%
\bibitem{Wingert2011NL11}%
  \BibitemOpen
  \bibfield{author}{%
  \bibinfo {author} {\bibfnamefont{M.~C.}\ \bibnamefont{Wingert}}, \bibinfo
  {author} {\bibfnamefont{Z.~C.~Y.}\ \bibnamefont{Chen}}, \bibinfo {author}
  {\bibfnamefont{E.}~\bibnamefont{Dechaumphai}}, \bibinfo {author}
  {\bibfnamefont{J.}~\bibnamefont{Moon}}, \bibinfo {author}
  {\bibfnamefont{J.-H.}\ \bibnamefont{Kim}}, \bibinfo {author}
  {\bibfnamefont{J.}~\bibnamefont{Xiang}},\ and\ \bibinfo {author}
  {\bibfnamefont{R.}~\bibnamefont{Chen}},\ }%
  \bibfield{journal}{%
  \bibinfo {journal} {Nano Lett.}\ }%
  \textbf{\bibinfo {volume} {11}},\ \bibinfo {pages} {5507} (\bibinfo {year}
  {2011})%
  \bibAnnoteFile{NoStop}{Wingert2011NL11}%
\bibitem{Shi2003JHT}%
  \BibitemOpen
  \bibfield{author}{%
  \bibinfo {author} {\bibfnamefont{L.}~\bibnamefont{Shi}}, \bibinfo {author}
  {\bibfnamefont{D.}~\bibnamefont{Li}}, \bibinfo {author}
  {\bibfnamefont{C.}~\bibnamefont{Yu}}, \bibinfo {author}
  {\bibfnamefont{W.}~\bibnamefont{Jang}}, \bibinfo {author}
  {\bibfnamefont{D.}~\bibnamefont{Kim}}, \bibinfo {author}
  {\bibfnamefont{Z.}~\bibnamefont{Yao}}, \bibinfo {author}
  {\bibfnamefont{P.}~\bibnamefont{Kim}},\ and\ \bibinfo {author}
  {\bibfnamefont{A.}~\bibnamefont{Majumdar}},\ }%
  \bibfield{journal}{%
  \Doi{10.1115/1.1597619}{\bibinfo {journal} {J. Heat. Trans.-T. ASME}}\ }%
  \textbf{\bibinfo {volume} {125}},\ \bibinfo {pages} {881} (\bibinfo {year}
  {2003})%
  \bibAnnoteFile{NoStop}{Shi2003JHT}%
\bibitem{Henry2008PRL}%
  \BibitemOpen
  \bibfield{author}{%
  \bibinfo {author} {\bibfnamefont{A.}~\bibnamefont{Henry}}\ and\ \bibinfo
  {author} {\bibfnamefont{G.}~\bibnamefont{Chen}},\ }%
  \bibfield{journal}{%
  \Doi{10.1103/PhysRevLett.101.235502}{\bibinfo {journal} {Phys. Rev. Lett.}}\
  }%
  \textbf{\bibinfo {volume} {101}},\ \bibinfo {pages} {235502} (\bibinfo {year}
  {2008})%
  \bibAnnoteFile{NoStop}{Henry2008PRL}%
\bibitem{Huang2012AM24}%
  \BibitemOpen
  \bibfield{author}{%
  \bibinfo {author} {\bibfnamefont{X.}~\bibnamefont{Huang}}, \bibinfo {author}
  {\bibfnamefont{G.}~\bibnamefont{Liu}},\ and\ \bibinfo {author}
  {\bibfnamefont{X.}~\bibnamefont{Wang}},\ }%
  \bibfield{journal}{%
  \bibinfo {journal} {Adv. Mater.}\ }%
  \textbf{\bibinfo {volume} {24}},\ \bibinfo {pages} {1482} (\bibinfo {year}
  {2012})%
  \bibAnnoteFile{NoStop}{Huang2012AM24}%
\bibitem{graphene}%
  \BibitemOpen
  \bibfield{author}{%
  \bibinfo {author} {\bibfnamefont{K.~S.}\ \bibnamefont{Novoselov}}, \bibinfo
  {author} {\bibfnamefont{A.~K.}\ \bibnamefont{Geim}}, \bibinfo {author}
  {\bibfnamefont{S.~V.}\ \bibnamefont{Morozov}}, \bibinfo {author}
  {\bibfnamefont{D.}~\bibnamefont{Jiang}}, \bibinfo {author}
  {\bibfnamefont{Y.}~\bibnamefont{Zhang}}, \bibinfo {author}
  {\bibfnamefont{S.~V.}\ \bibnamefont{Dubonos}}, \bibinfo {author}
  {\bibfnamefont{I.~V.}\ \bibnamefont{Grigorieva}},\ and\ \bibinfo {author}
  {\bibfnamefont{A.~A.}\ \bibnamefont{Firsov}},\ }%
  \bibfield{journal}{%
  \Doi{10.1126/science.1102896}{\bibinfo {journal} {Science}}\ }%
  \textbf{\bibinfo {volume} {306}},\ \bibinfo {pages} {666} (\bibinfo {year}
  {2004})%
  \bibAnnoteFile{NoStop}{graphene}%
\bibitem{balandin1}%
  \BibitemOpen
  \bibfield{author}{%
  \bibinfo {author} {\bibfnamefont{S.}~\bibnamefont{Ghosh}}, \bibinfo {author}
  {\bibfnamefont{I.}~\bibnamefont{Calizo}}, \bibinfo {author}
  {\bibfnamefont{D.}~\bibnamefont{Teweldebrhan}}, \bibinfo {author}
  {\bibfnamefont{E.~P.}\ \bibnamefont{Pokatilov}}, \bibinfo {author}
  {\bibfnamefont{D.~L.}\ \bibnamefont{Nika}}, \bibinfo {author}
  {\bibfnamefont{A.~A.}\ \bibnamefont{Balandin}}, \bibinfo {author}
  {\bibfnamefont{W.}~\bibnamefont{Bao}}, \bibinfo {author}
  {\bibfnamefont{F.}~\bibnamefont{Miao}},\ and\ \bibinfo {author}
  {\bibfnamefont{C.~N.}\ \bibnamefont{Lau}},\ }%
  \bibfield{journal}{%
  \Doi{10.1063/1.2907977}{\bibinfo {journal} {Appl. Phys. Lett.}}\ }%
  \textbf{\bibinfo {volume} {92}},\ \bibinfo {pages} {151911} (\bibinfo {year}
  {2008})%
  \bibAnnoteFile{NoStop}{balandin1}%
\bibitem{Balandin2008NL8}%
  \BibitemOpen
  \bibfield{author}{%
  \bibinfo {author} {\bibfnamefont{A.~A.}\ \bibnamefont{Balandin}}, \bibinfo
  {author} {\bibfnamefont{S.}~\bibnamefont{Ghosh}}, \bibinfo {author}
  {\bibfnamefont{W.}~\bibnamefont{Bao}}, \bibinfo {author}
  {\bibfnamefont{I.}~\bibnamefont{Calizo}}, \bibinfo {author}
  {\bibfnamefont{D.}~\bibnamefont{Teweldebrhan}}, \bibinfo {author}
  {\bibfnamefont{F.}~\bibnamefont{Miao}},\ and\ \bibinfo {author}
  {\bibfnamefont{C.~N.}\ \bibnamefont{Lau}},\ }%
  \bibfield{journal}{%
  \bibinfo {journal} {Nano Lett.}\ }%
  \textbf{\bibinfo {volume} {8}},\ \bibinfo {pages} {902} (\bibinfo {year}
  {2008})%
  \bibAnnoteFile{NoStop}{Balandin2008NL8}%
\bibitem{balandin3}%
  \BibitemOpen
  \bibfield{author}{%
  \bibinfo {author} {\bibfnamefont{S.}~\bibnamefont{Ghosh}}, \bibinfo {author}
  {\bibfnamefont{W.}~\bibnamefont{Bao}}, \bibinfo {author}
  {\bibfnamefont{D.~L.}\ \bibnamefont{Nika}}, \bibinfo {author}
  {\bibfnamefont{S.}~\bibnamefont{Subrina}}, \bibinfo {author}
  {\bibfnamefont{E.~P.}\ \bibnamefont{Pokatilov}}, \bibinfo {author}
  {\bibfnamefont{C.~N.}\ \bibnamefont{Lau}},\ and\ \bibinfo {author}
  {\bibfnamefont{A.~A.}\ \bibnamefont{Balandin}},\ }%
  \bibfield{journal}{%
  \Doi{10.1038/nmat2753}{\bibinfo {journal} {Nature Materials}}\ }%
  \textbf{\bibinfo {volume} {9}},\ \bibinfo {pages} {555} (\bibinfo {year}
  {2010})%
  \bibAnnoteFile{NoStop}{balandin3}%
\bibitem{LSScience}%
  \BibitemOpen
  \bibfield{author}{%
  \bibinfo {author} {\bibfnamefont{J.~H.}\ \bibnamefont{Seol}}, \bibinfo
  {author} {\bibfnamefont{I.}~\bibnamefont{Jo}}, \bibinfo {author}
  {\bibfnamefont{A.~L.}\ \bibnamefont{Moore}}, \bibinfo {author}
  {\bibfnamefont{L.}~\bibnamefont{Lindsay}}, \bibinfo {author}
  {\bibfnamefont{Z.~H.}\ \bibnamefont{Aitken}}, \bibinfo {author}
  {\bibfnamefont{M.~T.}\ \bibnamefont{Pettes}}, \bibinfo {author}
  {\bibfnamefont{X.}~\bibnamefont{Li}}, \bibinfo {author}
  {\bibfnamefont{Z.}~\bibnamefont{Yao}}, \bibinfo {author}
  {\bibfnamefont{R.}~\bibnamefont{Huang}}, \bibinfo {author}
  {\bibfnamefont{D.}~\bibnamefont{Broido}}, \bibinfo {author}
  {\bibfnamefont{N.}~\bibnamefont{Mingo}}, \bibinfo {author}
  {\bibfnamefont{R.~S.}\ \bibnamefont{Ruoff}},\ and\ \bibinfo {author}
  {\bibfnamefont{L.}~\bibnamefont{Shi}},\ }%
  \bibfield{journal}{%
  \Doi{10.1126/science.1184014}{\bibinfo {journal} {Science}}\ }%
  \textbf{\bibinfo {volume} {328}},\ \bibinfo {pages} {213} (\bibinfo {year}
  {2010})%
  \bibAnnoteFile{NoStop}{LSScience}%
\bibitem{Cai}%
  \BibitemOpen
  \bibfield{author}{%
  \bibinfo {author} {\bibfnamefont{W.}~\bibnamefont{Cai}}, \bibinfo {author}
  {\bibfnamefont{A.~L.}\ \bibnamefont{Moore}}, \bibinfo {author}
  {\bibfnamefont{Y.}~\bibnamefont{Zhu}}, \bibinfo {author}
  {\bibfnamefont{X.}~\bibnamefont{Li}}, \bibinfo {author}
  {\bibfnamefont{S.}~\bibnamefont{Chen}}, \bibinfo {author}
  {\bibfnamefont{L.}~\bibnamefont{Shi}},\ and\ \bibinfo {author}
  {\bibfnamefont{R.~S.}\ \bibnamefont{Ruoff}},\ }%
  \bibfield{booktitle}{%
  \emph{\bibinfo {booktitle} {Nano Letters}},\ }%
  \bibfield{journal}{%
  \Doi{10.1021/nl9041966}{\bibinfo {journal} {Nano Lett.}}\ }%
  \textbf{\bibinfo {volume} {10}},\ \bibinfo {pages} {1645} (\bibinfo {year}
  {2010})%
  \bibAnnoteFile{NoStop}{Cai}%
\bibitem{Chen}%
  \BibitemOpen
  \bibfield{author}{%
  \bibinfo {author} {\bibfnamefont{S.}~\bibnamefont{Chen}}, \bibinfo {author}
  {\bibfnamefont{A.~L.}\ \bibnamefont{Moore}}, \bibinfo {author}
  {\bibfnamefont{W.}~\bibnamefont{Cai}}, \bibinfo {author}
  {\bibfnamefont{J.~W.}\ \bibnamefont{Suk}}, \bibinfo {author}
  {\bibfnamefont{J.}~\bibnamefont{An}}, \bibinfo {author}
  {\bibfnamefont{C.}~\bibnamefont{Mishra}}, \bibinfo {author}
  {\bibfnamefont{C.}~\bibnamefont{Amos}}, \bibinfo {author}
  {\bibfnamefont{C.~W.}\ \bibnamefont{Magnuson}}, \bibinfo {author}
  {\bibfnamefont{J.}~\bibnamefont{Kang}}, \bibinfo {author}
  {\bibfnamefont{L.}~\bibnamefont{Shi}},\ and\ \bibinfo {author}
  {\bibfnamefont{R.~S.}\ \bibnamefont{Ruoff}},\ }%
  \bibfield{journal}{%
  \Doi{10.1021/nn102915x}{\bibinfo {journal} {ACS Nano}}\ }%
  \textbf{\bibinfo {volume} {5}},\ \bibinfo {pages} {321} (\bibinfo {year}
  {2010})%
  \bibAnnoteFile{NoStop}{Chen}%
\bibitem{Lee}%
  \BibitemOpen
  \bibfield{author}{%
  \bibinfo {author} {\bibfnamefont{J.-U.}\ \bibnamefont{Lee}}, \bibinfo
  {author} {\bibfnamefont{D.}~\bibnamefont{Yoon}}, \bibinfo {author}
  {\bibfnamefont{H.}~\bibnamefont{Kim}}, \bibinfo {author}
  {\bibfnamefont{S.~W.}\ \bibnamefont{Lee}},\ and\ \bibinfo {author}
  {\bibfnamefont{H.}~\bibnamefont{Cheong}},\ }%
  \bibfield{journal}{%
  \Doi{10.1103/PhysRevB.83.081419}{\bibinfo {journal} {Phys. Rev. B}}\ }%
  \textbf{\bibinfo {volume} {83}},\ \bibinfo {pages} {081419} (\bibinfo {year}
  {2011})%
  \bibAnnoteFile{NoStop}{Lee}%
\bibitem{Faugeras}%
  \BibitemOpen
  \bibfield{author}{%
  \bibinfo {author} {\bibfnamefont{C.}~\bibnamefont{Faugeras}}, \bibinfo
  {author} {\bibfnamefont{B.}~\bibnamefont{Faugeras}}, \bibinfo {author}
  {\bibfnamefont{M.}~\bibnamefont{Orlita}}, \bibinfo {author}
  {\bibfnamefont{M.}~\bibnamefont{Potemski}}, \bibinfo {author}
  {\bibfnamefont{R.~R.}\ \bibnamefont{Nair}},\ and\ \bibinfo {author}
  {\bibfnamefont{A.~K.}\ \bibnamefont{Geim}},\ }%
  \bibfield{journal}{%
  \Doi{10.1021/nn9016229}{\bibinfo {journal} {ACS Nano}}\ }%
  \textbf{\bibinfo {volume} {4}},\ \bibinfo {pages} {1889} (\bibinfo {year}
  {2010})%
  \bibAnnoteFile{NoStop}{Faugeras}%
\bibitem{BalandinReview}%
  \BibitemOpen
  \bibfield{author}{%
  \bibinfo {author} {\bibfnamefont{A.~A.}\ \bibnamefont{Balandin}},\ }%
  \bibfield{journal}{%
  \Doi{10.1038/nmat3064}{\bibinfo {journal} {Nature Materials}}\ }%
  \textbf{\bibinfo {volume} {10}},\ \bibinfo {pages} {569} (\bibinfo {year}
  {2011})%
  \bibAnnoteFile{NoStop}{BalandinReview}%
\bibitem{Wang}%
  \BibitemOpen
  \bibfield{author}{%
  \bibinfo {author} {\bibfnamefont{Z.}~\bibnamefont{Wang}}, \bibinfo {author}
  {\bibfnamefont{R.}~\bibnamefont{Xie}}, \bibinfo {author}
  {\bibfnamefont{C.~T.}\ \bibnamefont{Bui}}, \bibinfo {author}
  {\bibfnamefont{D.}~\bibnamefont{Liu}}, \bibinfo {author}
  {\bibfnamefont{X.}~\bibnamefont{Ni}}, \bibinfo {author}
  {\bibfnamefont{B.}~\bibnamefont{Li}},\ and\ \bibinfo {author}
  {\bibfnamefont{J.~T.~L.}\ \bibnamefont{Thong}},\ }%
  \bibfield{journal}{%
  \Doi{10.1021/nl102923q}{\bibinfo {journal} {Nano Lett.}}\ }%
  \textbf{\bibinfo {volume} {11}},\ \bibinfo {pages} {113} (\bibinfo {year}
  {2011})%
  \bibAnnoteFile{NoStop}{Wang}%
\bibitem{xu}%
  \BibitemOpen
  \bibfield{author}{%
  \bibinfo {author} {\bibfnamefont{X.}~\bibnamefont{Xu}}, \bibinfo {author}
  {\bibfnamefont{Y.}~\bibnamefont{Wang}}, \bibinfo {author}
  {\bibfnamefont{K.}~\bibnamefont{Zhang}}, \bibinfo {author}
  {\bibfnamefont{X.}~\bibnamefont{Zhao}}, \bibinfo {author}
  {\bibfnamefont{S.}~\bibnamefont{Bae}}, \bibinfo {author}
  {\bibfnamefont{M.}~\bibnamefont{Heinrich}}, \bibinfo {author}
  {\bibfnamefont{C.~T.}\ \bibnamefont{Bui}}, \bibinfo {author}
  {\bibfnamefont{R.}~\bibnamefont{Xie}}, \bibinfo {author}
  {\bibfnamefont{J.~T.~L.}\ \bibnamefont{Thong}}, \bibinfo {author}
  {\bibfnamefont{B.~H.}\ \bibnamefont{Hong}}, \bibinfo {author}
  {\bibfnamefont{K.~P.}\ \bibnamefont{Loh}}, \bibinfo {author}
  {\bibfnamefont{B.}~\bibnamefont{Li}},\ and\ \bibinfo {author}
  {\bibfnamefont{B.}~\bibnamefont{Oezyilmaz}},\ }%
  \bibfield{journal}{%
  \bibinfo {journal} {arXiv:1012.2937}}%
   (\bibinfo {year} {2010})%
  \bibAnnoteFile{NoStop}{xu}%
\bibitem{ballisticMingo}%
  \BibitemOpen
  \bibfield{author}{%
  \bibinfo {author} {\bibfnamefont{N.}~\bibnamefont{Mingo}}\ and\ \bibinfo
  {author} {\bibfnamefont{D.~A.}\ \bibnamefont{Broido}},\ }%
  \bibfield{journal}{%
  \Doi{10.1103/PhysRevLett.95.096105}{\bibinfo {journal} {Phys. Rev. Lett.}}\
  }%
  \textbf{\bibinfo {volume} {95}},\ \bibinfo {pages} {096105} (\bibinfo {year}
  {2005})%
  \bibAnnoteFile{NoStop}{ballisticMingo}%
\bibitem{ZAphonon}%
  \BibitemOpen
  \bibfield{author}{%
  \bibinfo {author} {\bibfnamefont{L.}~\bibnamefont{Lindsay}}, \bibinfo
  {author} {\bibfnamefont{D.~A.}\ \bibnamefont{Broido}},\ and\ \bibinfo
  {author} {\bibfnamefont{N.}~\bibnamefont{Mingo}},\ }%
  \bibfield{journal}{%
  \Doi{10.1103/PhysRevB.82.115427}{\bibinfo {journal} {Phys. Rev. B}}\ }%
  \textbf{\bibinfo {volume} {82}},\ \bibinfo {pages} {115427} (\bibinfo {year}
  {2010})%
  \bibAnnoteFile{NoStop}{ZAphonon}%
\bibitem{xu2}%
  \BibitemOpen
  \bibfield{author}{%
  \bibinfo {author} {\bibfnamefont{X.}~\bibnamefont{Xu}}, \bibinfo {author}
  {\bibfnamefont{Y.}~\bibnamefont{Wang}}, \bibinfo {author}
  {\bibfnamefont{K.}~\bibnamefont{Zhang}}, \bibinfo {author}
  {\bibfnamefont{X.}~\bibnamefont{Zhao}}, \bibinfo {author}
  {\bibfnamefont{S.}~\bibnamefont{Bae}}, \bibinfo {author}
  {\bibfnamefont{M.}~\bibnamefont{Heinrich}}, \bibinfo {author}
  {\bibfnamefont{C.~T.}\ \bibnamefont{Bui}}, \bibinfo {author}
  {\bibfnamefont{R.}~\bibnamefont{Xie}}, \bibinfo {author}
  {\bibfnamefont{J.~T.~L.}\ \bibnamefont{Thong}}, \bibinfo {author}
  {\bibfnamefont{B.~H.}\ \bibnamefont{Hong}}, \bibinfo {author}
  {\bibfnamefont{K.~P.}\ \bibnamefont{Loh}}, \bibinfo {author}
  {\bibfnamefont{B.}~\bibnamefont{Li}},\ and\ \bibinfo {author}
  {\bibfnamefont{B.}~\bibnamefont{Oezyilmaz}},\ }%
  \bibfield{journal}{%
  \bibinfo {journal} {Unpublished}}%
   (\bibinfo {year} {2012})%
  \bibAnnoteFile{NoStop}{xu2}%
\bibitem{Pettes}%
  \BibitemOpen
  \bibfield{author}{%
  \bibinfo {author} {\bibfnamefont{M.~T.}\ \bibnamefont{Pettes}}, \bibinfo
  {author} {\bibfnamefont{I.}~\bibnamefont{Jo}}, \bibinfo {author}
  {\bibfnamefont{Z.}~\bibnamefont{Yao}},\ and\ \bibinfo {author}
  {\bibfnamefont{L.}~\bibnamefont{Shi}},\ }%
  \bibfield{journal}{%
  \Doi{10.1021/nl104156y}{\bibinfo {journal} {Nano Lett.}}\ }%
  \textbf{\bibinfo {volume} {11}},\ \bibinfo {pages} {1195} (\bibinfo {year}
  {2011})%
  \bibAnnoteFile{NoStop}{Pettes}%
\bibitem{balandin4}%
  \BibitemOpen
  \bibfield{author}{%
  \bibinfo {author} {\bibfnamefont{D.~L.}\ \bibnamefont{Nika}}, \bibinfo
  {author} {\bibfnamefont{S.}~\bibnamefont{Ghosh}}, \bibinfo {author}
  {\bibfnamefont{E.~P.}\ \bibnamefont{Pokatilov}},\ and\ \bibinfo {author}
  {\bibfnamefont{A.~A.}\ \bibnamefont{Balandin}},\ }%
  \bibfield{journal}{%
  \Doi{10.1063/1.3136860}{\bibinfo {journal} {Appl. Phys. Lett.}}\ }%
  \textbf{\bibinfo {volume} {94}},\ \bibinfo {eid} {203103} (\bibinfo {year}
  {2009})%
  \bibAnnoteFile{NoStop}{balandin4}%
\bibitem{PRL.02.Narayan}%
  \BibitemOpen
  \bibfield{author}{%
  \bibinfo {author} {\bibfnamefont{O.}~\bibnamefont{Narayan}}\ and\ \bibinfo
  {author} {\bibfnamefont{S.}~\bibnamefont{Ramaswamy}},\ }%
  \bibfield{journal}{%
  \Doi{10.1103/PhysRevLett.89.200601}{\bibinfo {journal} {Phys. Rev. Lett.}}\
  }%
  \textbf{\bibinfo {volume} {89}},\ \bibinfo {pages} {200601} (\bibinfo {year}
  {2002})%
  \bibAnnoteFile{NoStop}{PRL.02.Narayan}%
\bibitem{Yang}%
  \BibitemOpen
  \bibfield{author}{%
  \bibinfo {author} {\bibfnamefont{L.}~\bibnamefont{Yang}}, \bibinfo {author}
  {\bibfnamefont{P.}~\bibnamefont{Grassberger}},\ and\ \bibinfo {author}
  {\bibfnamefont{B.}~\bibnamefont{Hu}},\ }%
  \bibfield{journal}{%
  \Doi{10.1103/PhysRevE.74.062101}{\bibinfo {journal} {Phys. Rev. E}}\ }%
  \textbf{\bibinfo {volume} {74}},\ \bibinfo {pages} {062101} (\bibinfo {year}
  {2006})%
  \bibAnnoteFile{NoStop}{Yang}%
\bibitem{Iijima1991Nature}%
  \BibitemOpen
  \bibfield{author}{%
  \bibinfo {author} {\bibfnamefont{S.}~\bibnamefont{Iijima}},\ }%
  \bibfield{journal}{%
  \bibinfo {journal} {Nature}\ }%
  \textbf{\bibinfo {volume} {354}},\ \bibinfo {pages} {56} (\bibinfo {year}
  {1991})%
  \bibAnnoteFile{NoStop}{Iijima1991Nature}%
\bibitem{Yang2010APL96}%
  \BibitemOpen
  \bibfield{author}{%
  \bibinfo {author} {\bibfnamefont{J.}~\bibnamefont{Yang}}, \bibinfo {author}
  {\bibfnamefont{S.}~\bibnamefont{Waltermire}}, \bibinfo {author}
  {\bibfnamefont{Y.}~\bibnamefont{Chen}}, \bibinfo {author}
  {\bibfnamefont{A.~A.}\ \bibnamefont{Zinn}}, \bibinfo {author}
  {\bibfnamefont{T.~T.}\ \bibnamefont{Xu}},\ and\ \bibinfo {author}
  {\bibfnamefont{D.}~\bibnamefont{Li}},\ }%
  \bibfield{journal}{%
  \bibinfo {journal} {Appl. Phys. Lett.}\ }%
  \textbf{\bibinfo {volume} {96}},\ \bibinfo {pages} {023109} (\bibinfo {year}
  {2010})%
  \bibAnnoteFile{NoStop}{Yang2010APL96}%
\bibitem{Yamamoto2004PRL92}%
  \BibitemOpen
  \bibfield{author}{%
  \bibinfo {author} {\bibfnamefont{T.}~\bibnamefont{Yamamoto}}, \bibinfo
  {author} {\bibfnamefont{S.}~\bibnamefont{Watanabe}},\ and\ \bibinfo {author}
  {\bibfnamefont{K.}~\bibnamefont{Watanabe}},\ }%
  \bibfield{journal}{%
  \bibinfo {journal} {Phys. Rev. Lett.}\ }%
  \textbf{\bibinfo {volume} {92}},\ \bibinfo {pages} {075502} (\bibinfo {year}
  {2004})%
  \bibAnnoteFile{NoStop}{Yamamoto2004PRL92}%
\bibitem{Chang2007PRL99}%
  \BibitemOpen
  \bibfield{author}{%
  \bibinfo {author} {\bibfnamefont{C.~W.}\ \bibnamefont{Chang}}, \bibinfo
  {author} {\bibfnamefont{D.}~\bibnamefont{Okawa}}, \bibinfo {author}
  {\bibfnamefont{H.}~\bibnamefont{Garcia}}, \bibinfo {author}
  {\bibfnamefont{A.}~\bibnamefont{Majumdar}},\ and\ \bibinfo {author}
  {\bibfnamefont{A.}~\bibnamefont{Zettl}},\ }%
  \bibfield{journal}{%
  \bibinfo {journal} {Phys. Rev. Lett.}\ }%
  \textbf{\bibinfo {volume} {99}},\ \bibinfo {pages} {045901} (\bibinfo {year}
  {2007})%
  \bibAnnoteFile{NoStop}{Chang2007PRL99}%
\bibitem{zhang2005JCP123a}%
  \BibitemOpen
  \bibfield{author}{%
  \bibinfo {author} {\bibfnamefont{G.}~\bibnamefont{Zhang}}\ and\ \bibinfo
  {author} {\bibfnamefont{B.}~\bibnamefont{Li}},\ }%
  \bibfield{journal}{%
  \bibinfo {journal} {J. Chem. Phys.}\ }%
  \textbf{\bibinfo {volume} {123}},\ \bibinfo {pages} {114714} (\bibinfo {year}
  {2005})%
  \bibAnnoteFile{NoStop}{zhang2005JCP123a}%
\bibitem{Maruyama2002PB323}%
  \BibitemOpen
  \bibfield{author}{%
  \bibinfo {author} {\bibfnamefont{S.}~\bibnamefont{Maruyama}},\ }%
  \bibfield{journal}{%
  \bibinfo {journal} {Physica B}\ }%
  \textbf{\bibinfo {volume} {323}},\ \bibinfo {pages} {193} (\bibinfo {year}
  {2002})%
  \bibAnnoteFile{NoStop}{Maruyama2002PB323}%
\bibitem{PRL.04.Wang}%
  \BibitemOpen
  \bibfield{author}{%
  \bibinfo {author} {\bibfnamefont{J.-S.}\ \bibnamefont{Wang}}\ and\ \bibinfo
  {author} {\bibfnamefont{B.}~\bibnamefont{Li}},\ }%
  \bibfield{journal}{%
  \Doi{10.1103/PhysRevLett.92.074302}{\bibinfo {journal} {Phys. Rev. Lett.}}\
  }%
  \textbf{\bibinfo {volume} {92}},\ \bibinfo {pages} {074302} (\bibinfo {year}
  {2004})%
  \bibAnnoteFile{NoStop}{PRL.04.Wang}%
\bibitem{PRL.03.Li}%
  \BibitemOpen
  \bibfield{author}{%
  \bibinfo {author} {\bibfnamefont{B.}~\bibnamefont{Li}}\ and\ \bibinfo
  {author} {\bibfnamefont{J.}~\bibnamefont{Wang}},\ }%
  \bibfield{journal}{%
  \Doi{10.1103/PhysRevLett.91.044301}{\bibinfo {journal} {Phys. Rev. Lett.}}\
  }%
  \textbf{\bibinfo {volume} {91}},\ \bibinfo {pages} {044301} (\bibinfo {year}
  {2003})%
  \bibAnnoteFile{NoStop}{PRL.03.Li}%
\bibitem{cui2001Sci293}%
  \BibitemOpen
  \bibfield{author}{%
  \bibinfo {author} {\bibfnamefont{Y.}~\bibnamefont{Cui}}, \bibinfo {author}
  {\bibfnamefont{Q.}~\bibnamefont{Wei}}, \bibinfo {author}
  {\bibfnamefont{H.}~\bibnamefont{Park}},\ and\ \bibinfo {author}
  {\bibfnamefont{C.~M.}\ \bibnamefont{Lieber}},\ }%
  \bibfield{journal}{%
  \bibinfo {journal} {Science}\ }%
  \textbf{\bibinfo {volume} {293}},\ \bibinfo {pages} {1289} (\bibinfo {year}
  {2001})%
  \bibAnnoteFile{NoStop}{cui2001Sci293}%
\bibitem{zhang2008NL8}%
  \BibitemOpen
  \bibfield{author}{%
  \bibinfo {author} {\bibfnamefont{G.-J.}\ \bibnamefont{Zhang}}, \bibinfo
  {author} {\bibfnamefont{G.}~\bibnamefont{Zhang}}, \bibinfo {author}
  {\bibfnamefont{H.~J.}\ \bibnamefont{Chua}}, \bibinfo {author}
  {\bibfnamefont{R.-E.}\ \bibnamefont{Chee}}, \bibinfo {author}
  {\bibfnamefont{E.~H.}\ \bibnamefont{Wong}}, \bibinfo {author}
  {\bibfnamefont{A.}~\bibnamefont{Agarwal}}, \bibinfo {author}
  {\bibfnamefont{K.~D.}\ \bibnamefont{Buddharaju}}, \bibinfo {author}
  {\bibfnamefont{N.}~\bibnamefont{Singh}}, \bibinfo {author}
  {\bibfnamefont{Z.}~\bibnamefont{Gao}},\ and\ \bibinfo {author}
  {\bibfnamefont{N.}~\bibnamefont{Balasubramanian}},\ }%
  \bibfield{journal}{%
  \bibinfo {journal} {Nano Lett.}\ }%
  \textbf{\bibinfo {volume} {8}},\ \bibinfo {pages} {1066} (\bibinfo {year}
  {2008})%
  \bibAnnoteFile{NoStop}{zhang2008NL8}%
\bibitem{xiang2006Nat441}%
  \BibitemOpen
  \bibfield{author}{%
  \bibinfo {author} {\bibfnamefont{J.}~\bibnamefont{Xiang}}, \bibinfo {author}
  {\bibfnamefont{W.}~\bibnamefont{Lu}}, \bibinfo {author}
  {\bibfnamefont{Y.}~\bibnamefont{Hu}}, \bibinfo {author}
  {\bibfnamefont{Y.}~\bibnamefont{Wu}}, \bibinfo {author}
  {\bibfnamefont{H.}~\bibnamefont{Yan}},\ and\ \bibinfo {author}
  {\bibfnamefont{C.~M.}\ \bibnamefont{Lieber}},\ }%
  \bibfield{journal}{%
  \bibinfo {journal} {Nature}\ }%
  \textbf{\bibinfo {volume} {441}},\ \bibinfo {pages} {489} (\bibinfo {year}
  {2006})%
  \bibAnnoteFile{NoStop}{xiang2006Nat441}%
\bibitem{Boukai2008Nat451}%
  \BibitemOpen
  \bibfield{author}{%
  \bibinfo {author} {\bibfnamefont{A.~I.}\ \bibnamefont{Boukai}}, \bibinfo
  {author} {\bibfnamefont{Y.}~\bibnamefont{Bunimovich}}, \bibinfo {author}
  {\bibfnamefont{J.}~\bibnamefont{Tahir-Kheli}}, \bibinfo {author}
  {\bibfnamefont{J.-K.}\ \bibnamefont{Yu}}, \bibinfo {author}
  {\bibfnamefont{W.~A.}\ \bibnamefont{Goddard~III}},\ and\ \bibinfo {author}
  {\bibfnamefont{J.~R.}\ \bibnamefont{Heath}},\ }%
  \bibfield{journal}{%
  \bibinfo {journal} {Nature}\ }%
  \textbf{\bibinfo {volume} {451}},\ \bibinfo {pages} {168} (\bibinfo {year}
  {2008})%
  \bibAnnoteFile{NoStop}{Boukai2008Nat451}%
\bibitem{zhang2009APL94}%
  \BibitemOpen
  \bibfield{author}{%
  \bibinfo {author} {\bibfnamefont{G.}~\bibnamefont{Zhang}}, \bibinfo {author}
  {\bibfnamefont{Q.}~\bibnamefont{Zhang}}, \bibinfo {author}
  {\bibfnamefont{C.-T.}\ \bibnamefont{Bui}}, \bibinfo {author}
  {\bibfnamefont{G.-Q.}\ \bibnamefont{Lo}},\ and\ \bibinfo {author}
  {\bibfnamefont{B.}~\bibnamefont{Li}},\ }%
  \bibfield{journal}{%
  \bibinfo {journal} {Appl. Phys. Lett.}\ }%
  \textbf{\bibinfo {volume} {94}},\ \bibinfo {pages} {213108} (\bibinfo {year}
  {2009})%
  \bibAnnoteFile{NoStop}{zhang2009APL94}%
\bibitem{zhang2009APL95}%
  \BibitemOpen
  \bibfield{author}{%
  \bibinfo {author} {\bibfnamefont{G.}~\bibnamefont{Zhang}}, \bibinfo {author}
  {\bibfnamefont{Q.}~\bibnamefont{Zhang}}, \bibinfo {author}
  {\bibfnamefont{D.}~\bibnamefont{Kavitha}},\ and\ \bibinfo {author}
  {\bibfnamefont{G.-Q.}\ \bibnamefont{Lo}},\ }%
  \bibfield{journal}{%
  \bibinfo {journal} {Appl. Phys. Lett.}\ }%
  \textbf{\bibinfo {volume} {95}},\ \bibinfo {pages} {243104} (\bibinfo {year}
  {2009})%
  \bibAnnoteFile{NoStop}{zhang2009APL95}%
\bibitem{Donadio2010NL10}%
  \BibitemOpen
  \bibfield{author}{%
  \bibinfo {author} {\bibfnamefont{D.}~\bibnamefont{Donadio}}\ and\ \bibinfo
  {author} {\bibfnamefont{G.}~\bibnamefont{Galli}},\ }%
  \bibfield{journal}{%
  \bibinfo {journal} {Nano Lett.}\ }%
  \textbf{\bibinfo {volume} {10}},\ \bibinfo {pages} {847} (\bibinfo {year}
  {2010})%
  \bibAnnoteFile{NoStop}{Donadio2010NL10}%
\bibitem{Markussen2009PRL103}%
  \BibitemOpen
  \bibfield{author}{%
  \bibinfo {author} {\bibfnamefont{T.}~\bibnamefont{Markussen}}, \bibinfo
  {author} {\bibfnamefont{A.-P.}\ \bibnamefont{Jauho}},\ and\ \bibinfo {author}
  {\bibfnamefont{M.}~\bibnamefont{Brandbyge}},\ }%
  \bibfield{journal}{%
  \bibinfo {journal} {Phys. Rev. Lett.}\ }%
  \textbf{\bibinfo {volume} {103}},\ \bibinfo {pages} {055502} (\bibinfo {year}
  {2009})%
  \bibAnnoteFile{NoStop}{Markussen2009PRL103}%
\bibitem{Sansoz2011NL11}%
  \BibitemOpen
  \bibfield{author}{%
  \bibinfo {author} {\bibfnamefont{F.}~\bibnamefont{Sansoz}},\ }%
  \bibfield{journal}{%
  \bibinfo {journal} {Nano Lett.}\ }%
  \textbf{\bibinfo {volume} {11}},\ \bibinfo {pages} {5378} (\bibinfo {year}
  {2011})%
  \bibAnnoteFile{NoStop}{Sansoz2011NL11}%
\bibitem{Donadio2009PRL102}%
  \BibitemOpen
  \bibfield{author}{%
  \bibinfo {author} {\bibfnamefont{D.}~\bibnamefont{Donadio}}\ and\ \bibinfo
  {author} {\bibfnamefont{G.}~\bibnamefont{Galli}},\ }%
  \bibfield{journal}{%
  \bibinfo {journal} {Phys. Rev. Lett.}\ }%
  \textbf{\bibinfo {volume} {102}},\ \bibinfo {pages} {195901} (\bibinfo {year}
  {2009})%
  \bibAnnoteFile{NoStop}{Donadio2009PRL102}%
\bibitem{Shi2009APL95}%
  \BibitemOpen
  \bibfield{author}{%
  \bibinfo {author} {\bibfnamefont{L.}~\bibnamefont{Shi}}, \bibinfo {author}
  {\bibfnamefont{D.}~\bibnamefont{Yao}}, \bibinfo {author}
  {\bibfnamefont{G.}~\bibnamefont{Zhang}},\ and\ \bibinfo {author}
  {\bibfnamefont{B.}~\bibnamefont{Li}},\ }%
  \bibfield{journal}{%
  \bibinfo {journal} {Appl. Phys. Lett.}\ }%
  \textbf{\bibinfo {volume} {95}},\ \bibinfo {pages} {063102} (\bibinfo {year}
  {2009})%
  \bibAnnoteFile{NoStop}{Shi2009APL95}%
\bibitem{Yao2009APL94}%
  \BibitemOpen
  \bibfield{author}{%
  \bibinfo {author} {\bibfnamefont{D.}~\bibnamefont{Yao}}, \bibinfo {author}
  {\bibfnamefont{G.}~\bibnamefont{Zhang}}, \bibinfo {author}
  {\bibfnamefont{G.-Q.}\ \bibnamefont{Lo}},\ and\ \bibinfo {author}
  {\bibfnamefont{B.}~\bibnamefont{Li}},\ }%
  \bibfield{journal}{%
  \bibinfo {journal} {Appl. Phys. Lett.}\ }%
  \textbf{\bibinfo {volume} {94}},\ \bibinfo {pages} {113113} (\bibinfo {year}
  {2009})%
  \bibAnnoteFile{NoStop}{Yao2009APL94}%
\bibitem{Yang2005NL5}%
  \BibitemOpen
  \bibfield{author}{%
  \bibinfo {author} {\bibfnamefont{R.}~\bibnamefont{Yang}}, \bibinfo {author}
  {\bibfnamefont{G.}~\bibnamefont{Chen}},\ and\ \bibinfo {author}
  {\bibfnamefont{M.~S.}\ \bibnamefont{Dresselhaus}},\ }%
  \bibfield{journal}{%
  \bibinfo {journal} {Nano Lett.}\ }%
  \textbf{\bibinfo {volume} {5}},\ \bibinfo {pages} {1111} (\bibinfo {year}
  {2005})%
  \bibAnnoteFile{NoStop}{Yang2005NL5}%
\bibitem{Chen2010NL10}%
  \BibitemOpen
  \bibfield{author}{%
  \bibinfo {author} {\bibfnamefont{J.}~\bibnamefont{Chen}}, \bibinfo {author}
  {\bibfnamefont{G.}~\bibnamefont{Zhang}},\ and\ \bibinfo {author}
  {\bibfnamefont{B.}~\bibnamefont{Li}},\ }%
  \bibfield{journal}{%
  \bibinfo {journal} {Nano Lett.}\ }%
  \textbf{\bibinfo {volume} {10}},\ \bibinfo {pages} {3978} (\bibinfo {year}
  {2010})%
  \bibAnnoteFile{NoStop}{Chen2010NL10}%
\bibitem{Hu2011NL11}%
  \BibitemOpen
  \bibfield{author}{%
  \bibinfo {author} {\bibfnamefont{M.}~\bibnamefont{Hu}}, \bibinfo {author}
  {\bibfnamefont{K.~P.}\ \bibnamefont{Giapis}}, \bibinfo {author}
  {\bibfnamefont{J.~V.}\ \bibnamefont{Goicochea}}, \bibinfo {author}
  {\bibfnamefont{X.}~\bibnamefont{Zhang}},\ and\ \bibinfo {author}
  {\bibfnamefont{D.}~\bibnamefont{Poulikakos}},\ }%
  \bibfield{journal}{%
  \bibinfo {journal} {Nano Lett.}\ }%
  \textbf{\bibinfo {volume} {11}},\ \bibinfo {pages} {618} (\bibinfo {year}
  {2011})%
  \bibAnnoteFile{NoStop}{Hu2011NL11}%
\bibitem{Chen2011JCP135}%
  \BibitemOpen
  \bibfield{author}{%
  \bibinfo {author} {\bibfnamefont{J.}~\bibnamefont{Chen}}, \bibinfo {author}
  {\bibfnamefont{G.}~\bibnamefont{Zhang}},\ and\ \bibinfo {author}
  {\bibfnamefont{B.}~\bibnamefont{Li}},\ }%
  \bibfield{journal}{%
  \bibinfo {journal} {J. Chem. Phys.}\ }%
  \textbf{\bibinfo {volume} {135}},\ \bibinfo {pages} {204705} (\bibinfo {year}
  {2011})%
  \bibAnnoteFile{NoStop}{Chen2011JCP135}%
\bibitem{Yang2008NL8}%
  \BibitemOpen
  \bibfield{author}{%
  \bibinfo {author} {\bibfnamefont{N.}~\bibnamefont{Yang}}, \bibinfo {author}
  {\bibfnamefont{G.}~\bibnamefont{Zhang}},\ and\ \bibinfo {author}
  {\bibfnamefont{B.}~\bibnamefont{Li}},\ }%
  \bibfield{journal}{%
  \bibinfo {journal} {Nano Lett.}\ }%
  \textbf{\bibinfo {volume} {8}},\ \bibinfo {pages} {276} (\bibinfo {year}
  {2008})%
  \bibAnnoteFile{NoStop}{Yang2008NL8}%
\bibitem{Chen2009APL95}%
  \BibitemOpen
  \bibfield{author}{%
  \bibinfo {author} {\bibfnamefont{J.}~\bibnamefont{Chen}}, \bibinfo {author}
  {\bibfnamefont{G.}~\bibnamefont{Zhang}},\ and\ \bibinfo {author}
  {\bibfnamefont{B.}~\bibnamefont{Li}},\ }%
  \bibfield{journal}{%
  \bibinfo {journal} {Appl. Phys. Lett.}\ }%
  \textbf{\bibinfo {volume} {95}},\ \bibinfo {pages} {073117} (\bibinfo {year}
  {2009})%
  \bibAnnoteFile{NoStop}{Chen2009APL95}%
\bibitem{Shi2010APL96}%
  \BibitemOpen
  \bibfield{author}{%
  \bibinfo {author} {\bibfnamefont{L.}~\bibnamefont{Shi}}, \bibinfo {author}
  {\bibfnamefont{D.}~\bibnamefont{Yao}}, \bibinfo {author}
  {\bibfnamefont{G.}~\bibnamefont{Zhang}},\ and\ \bibinfo {author}
  {\bibfnamefont{B.}~\bibnamefont{Li}},\ }%
  \bibfield{journal}{%
  \bibinfo {journal} {Appl. Phys. Lett.}\ }%
  \textbf{\bibinfo {volume} {96}},\ \bibinfo {pages} {173108} (\bibinfo {year}
  {2010})%
  \bibAnnoteFile{NoStop}{Shi2010APL96}%
\bibitem{Chen2010JPSJ79}%
  \BibitemOpen
  \bibfield{author}{%
  \bibinfo {author} {\bibfnamefont{J.}~\bibnamefont{Chen}}, \bibinfo {author}
  {\bibfnamefont{G.}~\bibnamefont{Zhang}},\ and\ \bibinfo {author}
  {\bibfnamefont{B.}~\bibnamefont{Li}},\ }%
  \bibfield{journal}{%
  \bibinfo {journal} {J. Phys. Soc. Jpn.}\ }%
  \textbf{\bibinfo {volume} {79}},\ \bibinfo {pages} {074604} (\bibinfo {year}
  {2010})%
  \bibAnnoteFile{NoStop}{Chen2010JPSJ79}%
\bibitem{Densis}%
  \BibitemOpen
  \bibfield{author}{%
  \bibinfo {author} {\bibfnamefont{S.}~\bibnamefont{Denisov}}, \bibinfo
  {author} {\bibfnamefont{J.}~\bibnamefont{Klafter}},\ and\ \bibinfo {author}
  {\bibfnamefont{M.}~\bibnamefont{Urbakh}},\ }%
  \bibfield{journal}{%
  \Doi{10.1103/PhysRevLett.91.194301}{\bibinfo {journal} {Phys. Rev. Lett.}}\
  }%
  \textbf{\bibinfo {volume} {91}},\ \bibinfo {pages} {194301} (\bibinfo {year}
  {2003})%
  \bibAnnoteFile{NoStop}{Densis}%
\bibitem{Zhang2005Nat438}%
  \BibitemOpen
  \bibfield{author}{%
  \bibinfo {author} {\bibfnamefont{Y.~B.}\ \bibnamefont{Zhang}}, \bibinfo
  {author} {\bibfnamefont{Y.-W.}\ \bibnamefont{Tan}}, \bibinfo {author}
  {\bibfnamefont{H.~L.}\ \bibnamefont{Stormer}},\ and\ \bibinfo {author}
  {\bibfnamefont{P.}~\bibnamefont{Kim}},\ }%
  \bibfield{journal}{%
  \bibinfo {journal} {Nature}\ }%
  \textbf{\bibinfo {volume} {438}},\ \bibinfo {pages} {201} (\bibinfo {year}
  {2005})%
  \bibAnnoteFile{NoStop}{Zhang2005Nat438}%
\bibitem{Berger2006Sci312}%
  \BibitemOpen
  \bibfield{author}{%
  \bibinfo {author} {\bibfnamefont{C.}~\bibnamefont{Berger}}, \bibinfo {author}
  {\bibfnamefont{Z.}~\bibnamefont{Song}}, \bibinfo {author}
  {\bibfnamefont{X.}~\bibnamefont{Li}, \bibfnamefont{X.~et al.~Wu}}, \bibinfo
  {author} {\bibfnamefont{N.}~\bibnamefont{Brown}}, \bibinfo {author}
  {\bibfnamefont{C.}~\bibnamefont{Naud}}, \bibinfo {author}
  {\bibfnamefont{D.}~\bibnamefont{Mayou}}, \bibinfo {author}
  {\bibfnamefont{T.}~\bibnamefont{Li}}, \bibinfo {author}
  {\bibfnamefont{J.}~\bibnamefont{Hass}}, \bibinfo {author}
  {\bibfnamefont{A.~N.}\ \bibnamefont{Marchenkov}}, \bibinfo {author}
  {\bibfnamefont{E.~H.}\ \bibnamefont{Conrad}}, \bibinfo {author}
  {\bibfnamefont{P.~N.}\ \bibnamefont{First}},\ and\ \bibinfo {author}
  {\bibfnamefont{W.~A.}\ \bibnamefont{de~Heer}},\ }%
  \bibfield{journal}{%
  \bibinfo {journal} {Science}\ }%
  \textbf{\bibinfo {volume} {312}},\ \bibinfo {pages} {1191} (\bibinfo {year}
  {2006})%
  \bibAnnoteFile{NoStop}{Berger2006Sci312}%
\bibitem{Wang2010NC2}%
  \BibitemOpen
  \bibfield{author}{%
  \bibinfo {author} {\bibfnamefont{X.}~\bibnamefont{Wang}}\ and\ \bibinfo
  {author} {\bibfnamefont{H.}~\bibnamefont{Dai}},\ }%
  \bibfield{journal}{%
  \bibinfo {journal} {Nature Chem.}\ }%
  \textbf{\bibinfo {volume} {2}},\ \bibinfo {pages} {661} (\bibinfo {year}
  {2010})%
  \bibAnnoteFile{NoStop}{Wang2010NC2}%
\bibitem{Xu2009APL95}%
  \BibitemOpen
  \bibfield{author}{%
  \bibinfo {author} {\bibfnamefont{Y.}~\bibnamefont{Xu}}, \bibinfo {author}
  {\bibfnamefont{X.}~\bibnamefont{Chen}}, \bibinfo {author}
  {\bibfnamefont{B.-L.}\ \bibnamefont{Gu}},\ and\ \bibinfo {author}
  {\bibfnamefont{W.}~\bibnamefont{Duan}},\ }%
  \bibfield{journal}{%
  \bibinfo {journal} {Appl. Phys. Lett.}\ }%
  \textbf{\bibinfo {volume} {95}},\ \bibinfo {pages} {233116} (\bibinfo {year}
  {2009})%
  \bibAnnoteFile{NoStop}{Xu2009APL95}%
\bibitem{Xu2010PRB81}%
  \BibitemOpen
  \bibfield{author}{%
  \bibinfo {author} {\bibfnamefont{Y.}~\bibnamefont{Xu}}, \bibinfo {author}
  {\bibfnamefont{X.}~\bibnamefont{Chen}}, \bibinfo {author}
  {\bibfnamefont{J.}~\bibnamefont{Wang}}, \bibinfo {author}
  {\bibfnamefont{B.-L.}\ \bibnamefont{Gu}},\ and\ \bibinfo {author}
  {\bibfnamefont{W.}~\bibnamefont{Duan}},\ }%
  \bibfield{journal}{%
  \bibinfo {journal} {Phys. Rev. B}\ }%
  \textbf{\bibinfo {volume} {81}},\ \bibinfo {pages} {195425} (\bibinfo {year}
  {2010})%
  \bibAnnoteFile{NoStop}{Xu2010PRB81}%
\bibitem{Evans2010APL96}%
  \BibitemOpen
  \bibfield{author}{%
  \bibinfo {author} {\bibfnamefont{W.~J.}\ \bibnamefont{Evans}}, \bibinfo
  {author} {\bibfnamefont{L.}~\bibnamefont{Hu}},\ and\ \bibinfo {author}
  {\bibfnamefont{P.}~\bibnamefont{Keblinski}},\ }%
  \bibfield{journal}{%
  \bibinfo {journal} {Appl. Phys. Lett.}\ }%
  \textbf{\bibinfo {volume} {96}},\ \bibinfo {pages} {203112} (\bibinfo {year}
  {2010})%
  \bibAnnoteFile{NoStop}{Evans2010APL96}%
\bibitem{Ni2011JPCM23}%
  \BibitemOpen
  \bibfield{author}{%
  \bibinfo {author} {\bibfnamefont{X.}~\bibnamefont{Ni}}, \bibinfo {author}
  {\bibfnamefont{G.}~\bibnamefont{Zhang}},\ and\ \bibinfo {author}
  {\bibfnamefont{B.}~\bibnamefont{Li}},\ }%
  \bibfield{journal}{%
  \bibinfo {journal} {J. Phys.: Condens. Matter}\ }%
  \textbf{\bibinfo {volume} {23}},\ \bibinfo {pages} {215301} (\bibinfo {year}
  {2011})%
  \bibAnnoteFile{NoStop}{Ni2011JPCM23}%
\bibitem{Chien2011APL98}%
  \BibitemOpen
  \bibfield{author}{%
  \bibinfo {author} {\bibfnamefont{S.-K.}\ \bibnamefont{Chien}}, \bibinfo
  {author} {\bibfnamefont{Y.-T.}\ \bibnamefont{Yang}},\ and\ \bibinfo {author}
  {\bibfnamefont{C.-K.}\ \bibnamefont{Chen}},\ }%
  \bibfield{journal}{%
  \bibinfo {journal} {Appl. Phys. Lett.}\ }%
  \textbf{\bibinfo {volume} {98}},\ \bibinfo {pages} {033107} (\bibinfo {year}
  {2011})%
  \bibAnnoteFile{NoStop}{Chien2011APL98}%
\bibitem{Pei2011carbon49}%
  \BibitemOpen
  \bibfield{author}{%
  \bibinfo {author} {\bibfnamefont{Q.-X.}\ \bibnamefont{Pei}}, \bibinfo
  {author} {\bibfnamefont{Z.-D.}\ \bibnamefont{Sha}},\ and\ \bibinfo {author}
  {\bibfnamefont{Y.-W.}\ \bibnamefont{Zhang}},\ }%
  \bibfield{journal}{%
  \bibinfo {journal} {Carbon}\ }%
  \textbf{\bibinfo {volume} {49}},\ \bibinfo {pages} {4752} (\bibinfo {year}
  {2011})%
  \bibAnnoteFile{NoStop}{Pei2011carbon49}%
\bibitem{Hu2010APL97}%
  \BibitemOpen
  \bibfield{author}{%
  \bibinfo {author} {\bibfnamefont{J.}~\bibnamefont{Hu}}, \bibinfo {author}
  {\bibfnamefont{S.}~\bibnamefont{Schiffli}}, \bibinfo {author}
  {\bibfnamefont{A.}~\bibnamefont{Vallabhaneni}}, \bibinfo {author}
  {\bibfnamefont{X.}~\bibnamefont{Ruan}},\ and\ \bibinfo {author}
  {\bibfnamefont{Y.~P.}\ \bibnamefont{Chen}},\ }%
  \bibfield{journal}{%
  \bibinfo {journal} {Appl. Phys. Lett.}\ }%
  \textbf{\bibinfo {volume} {97}},\ \bibinfo {pages} {133107} (\bibinfo {year}
  {2010})%
  \bibAnnoteFile{NoStop}{Hu2010APL97}%
\bibitem{Aksamija2011APL98}%
  \BibitemOpen
  \bibfield{author}{%
  \bibinfo {author} {\bibfnamefont{Z.}~\bibnamefont{Aksamija}}\ and\ \bibinfo
  {author} {\bibfnamefont{I.}~\bibnamefont{Knezevic}},\ }%
  \bibfield{journal}{%
  \bibinfo {journal} {Appl. Phys. Lett.}\ }%
  \textbf{\bibinfo {volume} {98}},\ \bibinfo {pages} {141919} (\bibinfo {year}
  {2011})%
  \bibAnnoteFile{NoStop}{Aksamija2011APL98}%
\bibitem{Ouyang2009EPL88}%
  \BibitemOpen
  \bibfield{author}{%
  \bibinfo {author} {\bibfnamefont{T.}~\bibnamefont{Ouyang}}, \bibinfo {author}
  {\bibfnamefont{Y.~P.}\ \bibnamefont{Chen}}, \bibinfo {author}
  {\bibfnamefont{K.~K.}\ \bibnamefont{Yang}},\ and\ \bibinfo {author}
  {\bibfnamefont{J.~X.}\ \bibnamefont{Zhong}},\ }%
  \bibfield{journal}{%
  \bibinfo {journal} {Europhys. Lett.}\ }%
  \textbf{\bibinfo {volume} {88}},\ \bibinfo {pages} {28002} (\bibinfo {year}
  {2009})%
  \bibAnnoteFile{NoStop}{Ouyang2009EPL88}%
\bibitem{Jiang2011APL98}%
  \BibitemOpen
  \bibfield{author}{%
  \bibinfo {author} {\bibfnamefont{J.-W.}\ \bibnamefont{Jiang}}, \bibinfo
  {author} {\bibfnamefont{B.-S.}\ \bibnamefont{Wang}},\ and\ \bibinfo {author}
  {\bibfnamefont{J.-S.}\ \bibnamefont{Wang}},\ }%
  \bibfield{journal}{%
  \bibinfo {journal} {Appl. Phys. Lett.}\ }%
  \textbf{\bibinfo {volume} {98}},\ \bibinfo {pages} {113114} (\bibinfo {year}
  {2011})%
  \bibAnnoteFile{NoStop}{Jiang2011APL98}%
\bibitem{Xie2011JPCM23}%
  \BibitemOpen
  \bibfield{author}{%
  \bibinfo {author} {\bibfnamefont{Z.-X.}\ \bibnamefont{Xie}}, \bibinfo
  {author} {\bibfnamefont{K.-Q.}\ \bibnamefont{Chen}},\ and\ \bibinfo {author}
  {\bibfnamefont{W.}~\bibnamefont{Duan}},\ }%
  \bibfield{journal}{%
  \bibinfo {journal} {J. Phys.: Condens. Matter}\ }%
  \textbf{\bibinfo {volume} {23}},\ \bibinfo {pages} {315302} (\bibinfo {year}
  {2011})%
  \bibAnnoteFile{NoStop}{Xie2011JPCM23}%
\bibitem{Ni2009APL95}%
  \BibitemOpen
  \bibfield{author}{%
  \bibinfo {author} {\bibfnamefont{X.}~\bibnamefont{Ni}}, \bibinfo {author}
  {\bibfnamefont{G.}~\bibnamefont{Liang}}, \bibinfo {author}
  {\bibfnamefont{J.-S.}\ \bibnamefont{Wang}},\ and\ \bibinfo {author}
  {\bibfnamefont{B.}~\bibnamefont{Li}},\ }%
  \bibfield{journal}{%
  \bibinfo {journal} {Appl. Phys. Lett.}\ }%
  \textbf{\bibinfo {volume} {95}},\ \bibinfo {pages} {192114} (\bibinfo {year}
  {2009})%
  \bibAnnoteFile{NoStop}{Ni2009APL95}%
\bibitem{Zhai2011EPL96}%
  \BibitemOpen
  \bibfield{author}{%
  \bibinfo {author} {\bibfnamefont{X.}~\bibnamefont{Zhai}}\ and\ \bibinfo
  {author} {\bibfnamefont{G.}~\bibnamefont{Jin}},\ }%
  \bibfield{journal}{%
  \bibinfo {journal} {Europhys. Lett.}\ }%
  \textbf{\bibinfo {volume} {96}},\ \bibinfo {pages} {16002} (\bibinfo {year}
  {2011})%
  \bibAnnoteFile{NoStop}{Zhai2011EPL96}%
\bibitem{Wei2011Nano22}%
  \BibitemOpen
  \bibfield{author}{%
  \bibinfo {author} {\bibfnamefont{N.}~\bibnamefont{Wei}}, \bibinfo {author}
  {\bibfnamefont{L.}~\bibnamefont{Xu}}, \bibinfo {author}
  {\bibfnamefont{H.-Q.}\ \bibnamefont{Wang}},\ and\ \bibinfo {author}
  {\bibfnamefont{J.-C.}\ \bibnamefont{Zheng}},\ }%
  \bibfield{journal}{%
  \bibinfo {journal} {Nanotechnology}\ }%
  \textbf{\bibinfo {volume} {22}},\ \bibinfo {pages} {105705} (\bibinfo {year}
  {2011})%
  \bibAnnoteFile{NoStop}{Wei2011Nano22}%
\bibitem{Guo2011PRB84}%
  \BibitemOpen
  \bibfield{author}{%
  \bibinfo {author} {\bibfnamefont{Z.-X.}\ \bibnamefont{Guo}}, \bibinfo
  {author} {\bibfnamefont{D.}~\bibnamefont{Zhang}},\ and\ \bibinfo {author}
  {\bibfnamefont{X.-G.}\ \bibnamefont{Gong}},\ }%
  \bibfield{journal}{%
  \bibinfo {journal} {Phys. Rev. B}\ }%
  \textbf{\bibinfo {volume} {84}},\ \bibinfo {pages} {075470} (\bibinfo {year}
  {2011})%
  \bibAnnoteFile{NoStop}{Guo2011PRB84}%
\bibitem{Ong2011PRB84}%
  \BibitemOpen
  \bibfield{author}{%
  \bibinfo {author} {\bibfnamefont{Z.-Y.}\ \bibnamefont{Ong}}\ and\ \bibinfo
  {author} {\bibfnamefont{E.}~\bibnamefont{Pop}},\ }%
  \bibfield{journal}{%
  \bibinfo {journal} {Phys. Rev. B}\ }%
  \textbf{\bibinfo {volume} {84}},\ \bibinfo {pages} {075471} (\bibinfo {year}
  {2011})%
  \bibAnnoteFile{NoStop}{Ong2011PRB84}%
\bibitem{zhang2011NS3}%
  \BibitemOpen
  \bibfield{author}{%
  \bibinfo {author} {\bibfnamefont{G.}~\bibnamefont{Zhang}}\ and\ \bibinfo
  {author} {\bibfnamefont{H.}~\bibnamefont{Zhang}},\ }%
  \bibfield{journal}{%
  \bibinfo {journal} {Nanoscale}\ }%
  \textbf{\bibinfo {volume} {3}},\ \bibinfo {pages} {4604} (\bibinfo {year}
  {2011})%
  \bibAnnoteFile{NoStop}{zhang2011NS3}%
\bibitem{Yang2012APL100}%
  \BibitemOpen
  \bibfield{author}{%
  \bibinfo {author} {\bibfnamefont{N.}~\bibnamefont{Yang}}, \bibinfo {author}
  {\bibfnamefont{X.}~\bibnamefont{Ni}}, \bibinfo {author}
  {\bibfnamefont{J.-W.}\ \bibnamefont{Jiang}},\ and\ \bibinfo {author}
  {\bibfnamefont{B.}~\bibnamefont{Li}},\ }%
  \bibfield{journal}{%
  \bibinfo {journal} {Appl. Phys. Lett.}\ }%
  \textbf{\bibinfo {volume} {100}},\ \bibinfo {pages} {093107} (\bibinfo {year}
  {2012})%
  \bibAnnoteFile{NoStop}{Yang2012APL100}%
\bibitem{Zhong2011APL98}%
  \BibitemOpen
  \bibfield{author}{%
  \bibinfo {author} {\bibfnamefont{W.-R.}\ \bibnamefont{Zhong}}, \bibinfo
  {author} {\bibfnamefont{M.-P.}\ \bibnamefont{Zhang}}, \bibinfo {author}
  {\bibfnamefont{B.-Q.}\ \bibnamefont{Ai}},\ and\ \bibinfo {author}
  {\bibfnamefont{D.-Q.}\ \bibnamefont{Zheng}},\ }%
  \bibfield{journal}{%
  \bibinfo {journal} {Appl. Phys. Lett.}\ }%
  \textbf{\bibinfo {volume} {98}},\ \bibinfo {pages} {113107} (\bibinfo {year}
  {2011})%
  \bibAnnoteFile{NoStop}{Zhong2011APL98}%
\bibitem{Cao2012PLA376}%
  \BibitemOpen
  \bibfield{author}{%
  \bibinfo {author} {\bibfnamefont{H.}~\bibnamefont{Cao}}, \bibinfo {author}
  {\bibfnamefont{Z.}~\bibnamefont{Guo}}, \bibinfo {author}
  {\bibfnamefont{H.}~\bibnamefont{Xiang}},\ and\ \bibinfo {author}
  {\bibfnamefont{X.}~\bibnamefont{Gong}},\ }%
  \bibfield{journal}{%
  \bibinfo {journal} {Phys. Lett. A}\ }%
  \textbf{\bibinfo {volume} {376}},\ \bibinfo {pages} {525} (\bibinfo {year}
  {2012})%
  \bibAnnoteFile{NoStop}{Cao2012PLA376}%
\bibitem{Yang2008APL93}%
  \BibitemOpen
  \bibfield{author}{%
  \bibinfo {author} {\bibfnamefont{N.}~\bibnamefont{Yang}}, \bibinfo {author}
  {\bibfnamefont{G.}~\bibnamefont{Zhang}},\ and\ \bibinfo {author}
  {\bibfnamefont{B.}~\bibnamefont{Li}},\ }%
  \bibfield{journal}{%
  \bibinfo {journal} {Appl. Phys. Lett.}\ }%
  \textbf{\bibinfo {volume} {93}},\ \bibinfo {pages} {243111} (\bibinfo {year}
  {2008})%
  \bibAnnoteFile{NoStop}{Yang2008APL93}%
\bibitem{Yang2009APL95}%
  \BibitemOpen
  \bibfield{author}{%
  \bibinfo {author} {\bibfnamefont{N.}~\bibnamefont{Yang}}, \bibinfo {author}
  {\bibfnamefont{G.}~\bibnamefont{Zhang}},\ and\ \bibinfo {author}
  {\bibfnamefont{B.}~\bibnamefont{Li}},\ }%
  \bibfield{journal}{%
  \bibinfo {journal} {Appl. Phys. Lett.}\ }%
  \textbf{\bibinfo {volume} {95}},\ \bibinfo {pages} {033107} (\bibinfo {year}
  {2009})%
  \bibAnnoteFile{NoStop}{Yang2009APL95}%
\bibitem{Hu2009NL9}%
  \BibitemOpen
  \bibfield{author}{%
  \bibinfo {author} {\bibfnamefont{J.}~\bibnamefont{Hu}}, \bibinfo {author}
  {\bibfnamefont{X.}~\bibnamefont{Ruan}},\ and\ \bibinfo {author}
  {\bibfnamefont{Y.~P.}\ \bibnamefont{Chen}},\ }%
  \bibfield{journal}{%
  \bibinfo {journal} {Nano Lett.}\ }%
  \textbf{\bibinfo {volume} {9}},\ \bibinfo {pages} {2730} (\bibinfo {year}
  {2009})%
  \bibAnnoteFile{NoStop}{Hu2009NL9}%
\bibitem{Guo2009APL95}%
  \BibitemOpen
  \bibfield{author}{%
  \bibinfo {author} {\bibfnamefont{Z.}~\bibnamefont{Guo}}, \bibinfo {author}
  {\bibfnamefont{D.}~\bibnamefont{Zhang}},\ and\ \bibinfo {author}
  {\bibfnamefont{X.-G.}\ \bibnamefont{Gong}},\ }%
  \bibfield{journal}{%
  \bibinfo {journal} {Appl. Phys. Lett.}\ }%
  \textbf{\bibinfo {volume} {95}},\ \bibinfo {pages} {163103} (\bibinfo {year}
  {2009})%
  \bibAnnoteFile{NoStop}{Guo2009APL95}%
\bibitem{poster}%
  \BibitemOpen
  \bibfield{author}{%
  \bibinfo {author} {\bibfnamefont{A.}~\bibnamefont{Eletskii}}, \bibinfo
  {author} {\bibfnamefont{I.}~\bibnamefont{Inskandarova}}, \bibinfo {author}
  {\bibfnamefont{A.}~\bibnamefont{Knizhnik}},\ and\ \bibinfo {author}
  {\bibfnamefont{D.}~\bibnamefont{Krasikov}},\ }%
  in\ \emph{\bibinfo {booktitle} {Graphene 2012 International Conference}}\
  (\bibinfo {year} {April 10-13, 2012, Brussels})%
  \bibAnnoteFile{NoStop}{}%
\bibitem{PR.68.Allen}%
  \BibitemOpen
  \bibfield{author}{%
  \bibinfo {author} {\bibfnamefont{K.~R.}\ \bibnamefont{Allen}}\ and\ \bibinfo
  {author} {\bibfnamefont{J.}~\bibnamefont{Ford}},\ }%
  \bibfield{journal}{%
  \Doi{10.1103/PhysRev.176.1046}{\bibinfo {journal} {Phys. Rev.}}\ }%
  \textbf{\bibinfo {volume} {176}},\ \bibinfo {pages} {1046} (\bibinfo {year}
  {1968})%
  \bibAnnoteFile{NoStop}{PR.68.Allen}%
\bibitem{PTPS.70.Matsuda}%
  \BibitemOpen
  \bibfield{author}{%
  \bibinfo {author} {\bibfnamefont{H.}~\bibnamefont{Matsuda}}\ and\ \bibinfo
  {author} {\bibfnamefont{K.}~\bibnamefont{Ishii}},\ }%
  \bibfield{journal}{%
  \Doi{10.1143/PTPS.45.56}{\bibinfo {journal} {Prog. Theor. Phys. Suppl.}}\ }%
  \textbf{\bibinfo {volume} {45}},\ \bibinfo {pages} {56} (\bibinfo {year}
  {1970})%
  \bibAnnoteFile{NoStop}{PTPS.70.Matsuda}%
\bibitem{JMP.71.Casher}%
  \BibitemOpen
  \bibfield{author}{%
  \bibinfo {author} {\bibfnamefont{A.}~\bibnamefont{Casher}}\ and\ \bibinfo
  {author} {\bibfnamefont{J.~L.}\ \bibnamefont{Lebowitz}},\ }%
  \bibfield{journal}{%
  \Doi{10.1063/1.1665794}{\bibinfo {journal} {J. Math. Phys}}\ }%
  \textbf{\bibinfo {volume} {12}},\ \bibinfo {pages} {1701} (\bibinfo {year}
  {1971})%
  \bibAnnoteFile{NoStop}{JMP.71.Casher}%
\bibitem{JMP.71.Rubin}%
  \BibitemOpen
  \bibfield{author}{%
  \bibinfo {author} {\bibfnamefont{R.~J.}\ \bibnamefont{Rubin}}\ and\ \bibinfo
  {author} {\bibfnamefont{W.~L.}\ \bibnamefont{Greer}},\ }%
  \bibfield{journal}{%
  \Doi{10.1063/1.1665793}{\bibinfo {journal} {J. Math. Phys}}\ }%
  \textbf{\bibinfo {volume} {12}},\ \bibinfo {pages} {1686} (\bibinfo {year}
  {1971})%
  \bibAnnoteFile{NoStop}{JMP.71.Rubin}%
\bibitem{CMP.79.Verheggen}%
  \BibitemOpen
  \bibfield{author}{%
  \bibinfo {author} {\bibfnamefont{T.}~\bibnamefont{Verheggen}},\ }%
  \bibfield{journal}{%
  \bibinfo {journal} {Commun. Math. Phys.}\ }%
  \textbf{\bibinfo {volume} {68}},\ \bibinfo {pages} {69} (\bibinfo {year}
  {1979})%
  \bibAnnoteFile{NoStop}{CMP.79.Verheggen}%
\bibitem{PRL.01.Dhara}%
  \BibitemOpen
  \bibfield{author}{%
  \bibinfo {author} {\bibfnamefont{A.}~\bibnamefont{Dhar}},\ }%
  \bibfield{journal}{%
  \Doi{10.1103/PhysRevLett.86.5882}{\bibinfo {journal} {Phys. Rev. Lett.}}\ }%
  \textbf{\bibinfo {volume} {86}},\ \bibinfo {pages} {5882} (\bibinfo {year}
  {2001})%
  \bibAnnoteFile{NoStop}{PRL.01.Dhara}%
\bibitem{PRE.08.Roy}%
  \BibitemOpen
  \bibfield{author}{%
  \bibinfo {author} {\bibfnamefont{D.}~\bibnamefont{Roy}}\ and\ \bibinfo
  {author} {\bibfnamefont{A.}~\bibnamefont{Dhar}},\ }%
  \bibfield{journal}{%
  \Doi{10.1103/PhysRevE.78.051112}{\bibinfo {journal} {Phys. Rev. E}}\ }%
  \textbf{\bibinfo {volume} {78}},\ \bibinfo {pages} {051112} (\bibinfo {year}
  {2008})%
  \bibAnnoteFile{NoStop}{PRE.08.Roy}%
\bibitem{PRL.08.Dhar}%
  \BibitemOpen
  \bibfield{author}{%
  \bibinfo {author} {\bibfnamefont{A.}~\bibnamefont{Dhar}}\ and\ \bibinfo
  {author} {\bibfnamefont{J.~L.}\ \bibnamefont{Lebowitz}},\ }%
  \bibfield{journal}{%
  \Doi{10.1103/PhysRevLett.100.134301}{\bibinfo {journal} {Phys. Rev. Lett.}}\
  }%
  \textbf{\bibinfo {volume} {100}},\ \bibinfo {pages} {134301} (\bibinfo {year}
  {2008})%
  \bibAnnoteFile{NoStop}{PRL.08.Dhar}%
\bibitem{PRB.83.John}%
  \BibitemOpen
  \bibfield{author}{%
  \bibinfo {author} {\bibfnamefont{S.}~\bibnamefont{John}}, \bibinfo {author}
  {\bibfnamefont{H.}~\bibnamefont{Sompolinsky}},\ and\ \bibinfo {author}
  {\bibfnamefont{M.~J.}\ \bibnamefont{Stephen}},\ }%
  \bibfield{journal}{%
  \Doi{10.1103/PhysRevB.27.5592}{\bibinfo {journal} {Phys. Rev. B}}\ }%
  \textbf{\bibinfo {volume} {27}},\ \bibinfo {pages} {5592} (\bibinfo {year}
  {1983})%
  \bibAnnoteFile{NoStop}{PRB.83.John}%
\bibitem{PRL.05.Lee}%
  \BibitemOpen
  \bibfield{author}{%
  \bibinfo {author} {\bibfnamefont{L.~W.}\ \bibnamefont{Lee}}\ and\ \bibinfo
  {author} {\bibfnamefont{A.}~\bibnamefont{Dhar}},\ }%
  \bibfield{journal}{%
  \Doi{10.1103/PhysRevLett.95.094302}{\bibinfo {journal} {Phys. Rev. Lett.}}\
  }%
  \textbf{\bibinfo {volume} {95}},\ \bibinfo {pages} {094302} (\bibinfo {year}
  {2005})%
  \bibAnnoteFile{NoStop}{PRL.05.Lee}%
\bibitem{PRL.02.Yang}%
  \BibitemOpen
  \bibfield{author}{%
  \bibinfo {author} {\bibfnamefont{L.}~\bibnamefont{Yang}},\ }%
  \bibfield{journal}{%
  \Doi{10.1103/PhysRevLett.88.094301}{\bibinfo {journal} {Phys. Rev. Lett.}}\
  }%
  \textbf{\bibinfo {volume} {88}},\ \bibinfo {pages} {094301} (\bibinfo {year}
  {2002})%
  \bibAnnoteFile{NoStop}{PRL.02.Yang}%
\bibitem{PRE.98.Lepri}%
  \BibitemOpen
  \bibfield{author}{%
  \bibinfo {author} {\bibfnamefont{S.}~\bibnamefont{Lepri}},\ }%
  \bibfield{journal}{%
  \Doi{10.1103/PhysRevE.58.7165}{\bibinfo {journal} {Phys. Rev. E}}\ }%
  \textbf{\bibinfo {volume} {58}},\ \bibinfo {pages} {7165} (\bibinfo {year}
  {1998})%
  \bibAnnoteFile{NoStop}{PRE.98.Lepri}%
\bibitem{PRE.03.Pereverzev}%
  \BibitemOpen
  \bibfield{author}{%
  \bibinfo {author} {\bibfnamefont{A.}~\bibnamefont{Pereverzev}},\ }%
  \bibfield{journal}{%
  \Doi{10.1103/PhysRevE.68.056124}{\bibinfo {journal} {Phys. Rev. E}}\ }%
  \textbf{\bibinfo {volume} {68}},\ \bibinfo {pages} {056124} (\bibinfo {year}
  {2003})%
  \bibAnnoteFile{NoStop}{PRE.03.Pereverzev}%
\bibitem{JCP.54.Green}%
  \BibitemOpen
  \bibfield{author}{%
  \bibinfo {author} {\bibfnamefont{M.~S.}\ \bibnamefont{Green}},\ }%
  \bibfield{journal}{%
  \Doi{10.1063/1.1740082}{\bibinfo {journal} {J. Chem. Phys.}}\ }%
  \textbf{\bibinfo {volume} {22}},\ \bibinfo {pages} {398} (\bibinfo {year}
  {1954})%
  \bibAnnoteFile{NoStop}{JCP.54.Green}%
\bibitem{JPSJ.57.Kubo}%
  \BibitemOpen
  \bibfield{author}{%
  \bibinfo {author} {\bibfnamefont{R.}~\bibnamefont{Kubo}}, \bibinfo {author}
  {\bibfnamefont{M.}~\bibnamefont{Yokota}},\ and\ \bibinfo {author}
  {\bibfnamefont{S.}~\bibnamefont{Nakajima}},\ }%
  \bibfield{journal}{%
  \Doi{10.1143/JPSJ.12.1203}{\bibinfo {journal} {J. Phys. Soc. Jpn.}}\ }%
  \textbf{\bibinfo {volume} {12}},\ \bibinfo {pages} {1203} (\bibinfo {year}
  {1957})%
  \bibAnnoteFile{NoStop}{JPSJ.57.Kubo}%
\bibitem{JSM.07.Delfini}%
  \BibitemOpen
  \bibfield{author}{%
  \bibinfo {author} {\bibfnamefont{L.}~\bibnamefont{Delfini}}, \bibinfo
  {author} {\bibfnamefont{S.}~\bibnamefont{Lepri}}, \bibinfo {author}
  {\bibfnamefont{R.}~\bibnamefont{Livi}},\ and\ \bibinfo {author}
  {\bibfnamefont{A.}~\bibnamefont{Politi}},\ }%
  \bibfield{journal}{%
  \Doi{10.1088/1742-5468/2007/02/P02007}{\bibinfo {journal} {J. Stat. Mech.}}\
  }%
  \textbf{\bibinfo {volume} {2007}},\ \bibinfo {pages} {P02007} (\bibinfo
  {year} {2007})%
  \bibAnnoteFile{NoStop}{JSM.07.Delfini}%
\bibitem{PRE.06.Mai}%
  \BibitemOpen
  \bibfield{author}{%
  \bibinfo {author} {\bibfnamefont{T.}~\bibnamefont{Mai}}\ and\ \bibinfo
  {author} {\bibfnamefont{O.}~\bibnamefont{Narayan}},\ }%
  \bibfield{journal}{%
  \Doi{10.1103/PhysRevE.73.061202}{\bibinfo {journal} {Phys. Rev. E}}\ }%
  \textbf{\bibinfo {volume} {73}},\ \bibinfo {pages} {061202} (\bibinfo {year}
  {2006})%
  \bibAnnoteFile{NoStop}{PRE.06.Mai}%
\bibitem{PRL.06.Hurtado}%
  \BibitemOpen
  \bibfield{author}{%
  \bibinfo {author} {\bibfnamefont{P.~I.}\ \bibnamefont{Hurtado}},\ }%
  \bibfield{journal}{%
  \Doi{10.1103/PhysRevLett.96.010601}{\bibinfo {journal} {Phys. Rev. Lett.}}\
  }%
  \textbf{\bibinfo {volume} {96}},\ \bibinfo {pages} {010601} (\bibinfo {year}
  {2006})%
  \bibAnnoteFile{NoStop}{PRL.06.Hurtado}%
\bibitem{CPAM.08.Lukkarinen}%
  \BibitemOpen
  \bibfield{author}{%
  \bibinfo {author} {\bibfnamefont{J.}~\bibnamefont{Lukkarinen}}\ and\ \bibinfo
  {author} {\bibfnamefont{H.}~\bibnamefont{Spohn}},\ }%
  \bibfield{journal}{%
  \Doi{10.1002/cpa.20243}{\bibinfo {journal} {Comm. Pure Appl. Math.}}\ }%
  \textbf{\bibinfo {volume} {61}},\ \bibinfo {pages} {1753} (\bibinfo {year}
  {2008})%
  \bibAnnoteFile{NoStop}{CPAM.08.Lukkarinen}%
\bibitem{JSP.06.Aoki}%
  \BibitemOpen
  \bibfield{author}{%
  \bibinfo {author} {\bibfnamefont{K.}~\bibnamefont{Aoki}}, \bibinfo {author}
  {\bibfnamefont{J.}~\bibnamefont{Lukkarinen}},\ and\ \bibinfo {author}
  {\bibfnamefont{H.}~\bibnamefont{Spohn}},\ }%
  \bibfield{journal}{%
  \bibinfo {journal} {J. Stat. Phys.}\ }%
  \textbf{\bibinfo {volume} {124}},\ \bibinfo {pages} {1105} (\bibinfo {year}
  {2006})%
  \bibAnnoteFile{NoStop}{JSP.06.Aoki}%
\bibitem{PLA.181.85}%
  \BibitemOpen
  \bibfield{author}{%
  \bibinfo {author} {\bibfnamefont{H.}~\bibnamefont{Kaburaki}}\ and\ \bibinfo
  {author} {\bibfnamefont{M.}~\bibnamefont{Machida}},\ }%
  \bibfield{journal}{%
  \Doi{10.1016/0375-9601(93)91129-S}{\bibinfo {journal} {Physics Letters A}}\
  }%
  \textbf{\bibinfo {volume} {181}},\ \bibinfo {pages} {85 } (\bibinfo {year}
  {1993})%
  \bibAnnoteFile{NoStop}{PLA.181.85}%
\bibitem{PRL.07.Mai}%
  \BibitemOpen
  \bibfield{author}{%
  \bibinfo {author} {\bibfnamefont{T.}~\bibnamefont{Mai}}, \bibinfo {author}
  {\bibfnamefont{A.}~\bibnamefont{Dhar}},\ and\ \bibinfo {author}
  {\bibfnamefont{O.}~\bibnamefont{Narayan}},\ }%
  \bibfield{journal}{%
  \Doi{10.1103/PhysRevLett.98.184301}{\bibinfo {journal} {Phys. Rev. Lett.}}\
  }%
  \textbf{\bibinfo {volume} {98}},\ \bibinfo {pages} {184301} (\bibinfo {year}
  {2007})%
  \bibAnnoteFile{NoStop}{PRL.07.Mai}%
\bibitem{EL.11.Wang}%
  \BibitemOpen
  \bibfield{author}{%
  \bibinfo {author} {\bibfnamefont{L.}~\bibnamefont{Wang}}\ and\ \bibinfo
  {author} {\bibfnamefont{T.}~\bibnamefont{Wang}},\ }%
  \bibfield{journal}{%
  \Doi{10.1209/0295-5075/93/54002}{\bibinfo {journal} {Europhys. Lett.}}\ }%
  \textbf{\bibinfo {volume} {93}},\ \bibinfo {pages} {54002} (\bibinfo {year}
  {2011})%
  \bibAnnoteFile{NoStop}{EL.11.Wang}%
\bibitem{PRE.12.Xiong}%
  \BibitemOpen
  \bibfield{author}{%
  \bibinfo {author} {\bibfnamefont{D.}~\bibnamefont{Xiong}}, \bibinfo {author}
  {\bibfnamefont{J.}~\bibnamefont{Wang}}, \bibinfo {author}
  {\bibfnamefont{Y.}~\bibnamefont{Zhang}},\ and\ \bibinfo {author}
  {\bibfnamefont{H.}~\bibnamefont{Zhao}},\ }%
  \bibfield{journal}{%
  \Doi{10.1103/PhysRevE.85.020102}{\bibinfo {journal} {Phys. Rev. E}}\ }%
  \textbf{\bibinfo {volume} {85}},\ \bibinfo {pages} {020102} (\bibinfo {year}
  {2012})%
  \bibAnnoteFile{NoStop}{PRE.12.Xiong}%
\bibitem{JCP.97.Muller-Plathe}%
  \BibitemOpen
  \bibfield{author}{%
  \bibinfo {author} {\bibfnamefont{F.}~\bibnamefont{Muller-Plathe}},\ }%
  \bibfield{journal}{%
  \Doi{10.1063/1.473271}{\bibinfo {journal} {J. Chem. Phys.}}\ }%
  \textbf{\bibinfo {volume} {106}},\ \bibinfo {pages} {6082} (\bibinfo {year}
  {1997})%
  \bibAnnoteFile{NoStop}{JCP.97.Muller-Plathe}%
\bibitem{PRE.98.Hu}%
  \BibitemOpen
  \bibfield{author}{%
  \bibinfo {author} {\bibfnamefont{B.}~\bibnamefont{Hu}}, \bibinfo {author}
  {\bibfnamefont{B.}~\bibnamefont{Li}},\ and\ \bibinfo {author}
  {\bibfnamefont{H.}~\bibnamefont{Zhao}},\ }%
  \bibfield{journal}{%
  \Doi{10.1103/PhysRevE.57.2992}{\bibinfo {journal} {Phys. Rev. E}}\ }%
  \textbf{\bibinfo {volume} {57}},\ \bibinfo {pages} {2992} (\bibinfo {year}
  {1998})%
  \bibAnnoteFile{NoStop}{PRE.98.Hu}%
\bibitem{PLA.00.Aoki}%
  \BibitemOpen
  \bibfield{author}{%
  \bibinfo {author} {\bibfnamefont{K.}~\bibnamefont{Aoki}}\ and\ \bibinfo
  {author} {\bibfnamefont{D.}~\bibnamefont{Kusnezov}},\ }%
  \bibfield{journal}{%
  \Doi{10.1016/S0375-9601(99)00899-3}{\bibinfo {journal} {Phys. Lett. A}}\ }%
  \textbf{\bibinfo {volume} {265}},\ \bibinfo {pages} {250} (\bibinfo {year}
  {2000})%
  \bibAnnoteFile{NoStop}{PLA.00.Aoki}%
\bibitem{PRE.00.Hu}%
  \BibitemOpen
  \bibfield{author}{%
  \bibinfo {author} {\bibfnamefont{B.}~\bibnamefont{Hu}}, \bibinfo {author}
  {\bibfnamefont{B.}~\bibnamefont{Li}},\ and\ \bibinfo {author}
  {\bibfnamefont{H.}~\bibnamefont{Zhao}},\ }%
  \bibfield{journal}{%
  \Doi{10.1103/PhysRevE.61.3828}{\bibinfo {journal} {Phys. Rev. E}}\ }%
  \textbf{\bibinfo {volume} {61}},\ \bibinfo {pages} {3828} (\bibinfo {year}
  {2000})%
  \bibAnnoteFile{NoStop}{PRE.00.Hu}%
\bibitem{PR.67.Payton}%
  \BibitemOpen
  \bibfield{author}{%
  \bibinfo {author} {\bibfnamefont{D.~N.}\ \bibnamefont{Payton}}\ and\ \bibinfo
  {author} {\bibfnamefont{W.~M.}\ \bibnamefont{Visscher}},\ }%
  \bibfield{journal}{%
  \Doi{10.1103/PhysRev.156.1032}{\bibinfo {journal} {Phys. Rev.}}\ }%
  \textbf{\bibinfo {volume} {156}},\ \bibinfo {pages} {1032} (\bibinfo {year}
  {1967})%
  \bibAnnoteFile{NoStop}{PR.67.Payton}%
\bibitem{PRL.01.Li}%
  \BibitemOpen
  \bibfield{author}{%
  \bibinfo {author} {\bibfnamefont{B.}~\bibnamefont{Li}}, \bibinfo {author}
  {\bibfnamefont{H.}~\bibnamefont{Zhao}},\ and\ \bibinfo {author}
  {\bibfnamefont{B.}~\bibnamefont{Hu}},\ }%
  \bibfield{journal}{%
  \Doi{10.1103/PhysRevLett.86.63}{\bibinfo {journal} {Phys. Rev. Lett.}}\ }%
  \textbf{\bibinfo {volume} {86}},\ \bibinfo {pages} {63} (\bibinfo {year}
  {2001})%
  \bibAnnoteFile{NoStop}{PRL.01.Li}%
\bibitem{PRE.08.Dhar}%
  \BibitemOpen
  \bibfield{author}{%
  \bibinfo {author} {\bibfnamefont{A.}~\bibnamefont{Dhar}}\ and\ \bibinfo
  {author} {\bibfnamefont{K.}~\bibnamefont{Saito}},\ }%
  \bibfield{journal}{%
  \Doi{10.1103/PhysRevE.78.061136}{\bibinfo {journal} {Phys. Rev. E}}\ }%
  \textbf{\bibinfo {volume} {78}},\ \bibinfo {pages} {061136} (\bibinfo {year}
  {2008})%
  \bibAnnoteFile{NoStop}{PRE.08.Dhar}%
\bibitem{JSP.00.Lippi}%
  \BibitemOpen
  \bibfield{author}{%
  \bibinfo {author} {\bibfnamefont{A.}~\bibnamefont{Lippi}}\ and\ \bibinfo
  {author} {\bibfnamefont{R.}~\bibnamefont{Livi}},\ }%
  \bibfield{journal}{%
  \bibinfo {journal} {J. Stat. Phys.}\ }%
  \textbf{\bibinfo {volume} {100}},\ \bibinfo {pages} {1147} (\bibinfo {year}
  {2000})%
  \bibAnnoteFile{NoStop}{JSP.00.Lippi}%
\bibitem{A.02.Grassberger}%
  \BibitemOpen
  \bibfield{author}{%
  \bibinfo {author} {\bibfnamefont{P.}~\bibnamefont{Grassberger}}\ and\
  \bibinfo {author} {\bibfnamefont{L.}~\bibnamefont{Yang}},\ }%
  \bibfield{journal}{%
  \bibinfo {journal} {arXiv:cond-mat/0204247v1}}%
   (\bibinfo {year} {2002})%
  \bibAnnoteFile{NoStop}{A.02.Grassberger}%
\bibitem{JPSJ.08.Shiba}%
  \BibitemOpen
  \bibfield{author}{%
  \bibinfo {author} {\bibfnamefont{H.}~\bibnamefont{Shiba}}\ and\ \bibinfo
  {author} {\bibfnamefont{N.}~\bibnamefont{Ito}},\ }%
  \bibfield{journal}{%
  \Doi{10.1143/JPSJ.77.054006}{\bibinfo {journal} {J. Phys. Soc. Jpn.}}\ }%
  \textbf{\bibinfo {volume} {77}},\ \bibinfo {pages} {054006} (\bibinfo {year}
  {2008})%
  \bibAnnoteFile{NoStop}{JPSJ.08.Shiba}%
\bibitem{PRL.99.Alonso}%
  \BibitemOpen
  \bibfield{author}{%
  \bibinfo {author} {\bibfnamefont{D.}~\bibnamefont{Alonso}}, \bibinfo {author}
  {\bibfnamefont{R.}~\bibnamefont{Artuso}}, \bibinfo {author}
  {\bibfnamefont{G.}~\bibnamefont{Casati}},\ and\ \bibinfo {author}
  {\bibfnamefont{I.}~\bibnamefont{Guarneri}},\ }%
  \bibfield{journal}{%
  \Doi{10.1103/PhysRevLett.82.1859}{\bibinfo {journal} {Phys. Rev. Lett.}}\ }%
  \textbf{\bibinfo {volume} {82}},\ \bibinfo {pages} {1859} (\bibinfo {year}
  {1999})%
  \bibAnnoteFile{NoStop}{PRL.99.Alonso}%
\bibitem{PRL.02.Li}%
  \BibitemOpen
  \bibfield{author}{%
  \bibinfo {author} {\bibfnamefont{B.}~\bibnamefont{Li}}, \bibinfo {author}
  {\bibfnamefont{L.}~\bibnamefont{Wang}},\ and\ \bibinfo {author}
  {\bibfnamefont{B.}~\bibnamefont{Hu}},\ }%
  \bibfield{journal}{%
  \Doi{10.1103/PhysRevLett.88.223901}{\bibinfo {journal} {Phys. Rev. Lett.}}\
  }%
  \textbf{\bibinfo {volume} {88}},\ \bibinfo {pages} {223901} (\bibinfo {year}
  {2002})%
  \bibAnnoteFile{NoStop}{PRL.02.Li}%
\bibitem{PRE.02.Alonso}%
  \BibitemOpen
  \bibfield{author}{%
  \bibinfo {author} {\bibfnamefont{D.}~\bibnamefont{Alonso}}, \bibinfo {author}
  {\bibfnamefont{A.}~\bibnamefont{Ruiz}},\ and\ \bibinfo {author}
  {\bibfnamefont{I.}~\bibnamefont{de~Vega}},\ }%
  \bibfield{journal}{%
  \Doi{10.1103/PhysRevE.66.066131}{\bibinfo {journal} {Phys. Rev. E}}\ }%
  \textbf{\bibinfo {volume} {66}},\ \bibinfo {pages} {066131} (\bibinfo {year}
  {2002})%
  \bibAnnoteFile{NoStop}{PRE.02.Alonso}%
\bibitem{PRE.03.Li}%
  \BibitemOpen
  \bibfield{author}{%
  \bibinfo {author} {\bibfnamefont{B.}~\bibnamefont{Li}}, \bibinfo {author}
  {\bibfnamefont{G.}~\bibnamefont{Casati}},\ and\ \bibinfo {author}
  {\bibfnamefont{J.}~\bibnamefont{Wang}},\ }%
  \bibfield{journal}{%
  \Doi{10.1103/PhysRevE.67.021204}{\bibinfo {journal} {Phys. Rev. E}}\ }%
  \textbf{\bibinfo {volume} {67}},\ \bibinfo {pages} {021204} (\bibinfo {year}
  {2003})%
  \bibAnnoteFile{NoStop}{PRE.03.Li}%
\bibitem{PRA.89.Blumen}%
  \BibitemOpen
  \bibfield{author}{%
  \bibinfo {author} {\bibfnamefont{A.}~\bibnamefont{Blumen}}, \bibinfo {author}
  {\bibfnamefont{G.}~\bibnamefont{Zumofen}},\ and\ \bibinfo {author}
  {\bibfnamefont{J.}~\bibnamefont{Klafter}},\ }%
  \bibfield{journal}{%
  \Doi{10.1103/PhysRevA.40.3964}{\bibinfo {journal} {Phys. Rev. A}}\ }%
  \textbf{\bibinfo {volume} {40}},\ \bibinfo {pages} {3964} (\bibinfo {year}
  {1989})%
  \bibAnnoteFile{NoStop}{PRA.89.Blumen}%
\bibitem{PRL.99.Dhar}%
  \BibitemOpen
  \bibfield{author}{%
  \bibinfo {author} {\bibfnamefont{A.}~\bibnamefont{Dhar}}\ and\ \bibinfo
  {author} {\bibfnamefont{D.}~\bibnamefont{Dhar}},\ }%
  \bibfield{journal}{%
  \Doi{10.1103/PhysRevLett.82.480}{\bibinfo {journal} {Phys. Rev. Lett.}}\ }%
  \textbf{\bibinfo {volume} {82}},\ \bibinfo {pages} {480} (\bibinfo {year}
  {1999})%
  \bibAnnoteFile{NoStop}{PRL.99.Dhar}%
\bibitem{PRL.05.Cipriani}%
  \BibitemOpen
  \bibfield{author}{%
  \bibinfo {author} {\bibfnamefont{P.}~\bibnamefont{Cipriani}}, \bibinfo
  {author} {\bibfnamefont{S.}~\bibnamefont{Denisov}},\ and\ \bibinfo {author}
  {\bibfnamefont{A.}~\bibnamefont{Politi}},\ }%
  \bibfield{journal}{%
  \Doi{10.1103/PhysRevLett.94.244301}{\bibinfo {journal} {Phys. Rev. Lett.}}\
  }%
  \textbf{\bibinfo {volume} {94}},\ \bibinfo {pages} {244301} (\bibinfo {year}
  {2005})%
  \bibAnnoteFile{NoStop}{PRL.05.Cipriani}%
\bibitem{PA.93.Klafter}%
  \BibitemOpen
  \bibfield{author}{%
  \bibinfo {author} {\bibfnamefont{J.}~\bibnamefont{Klafter}}\ and\ \bibinfo
  {author} {\bibfnamefont{G.}~\bibnamefont{Zumofen}},\ }%
  \bibfield{journal}{%
  \Doi{10.1016/0378-4371(93)90086-J}{\bibinfo {journal} {Physica A}}\ }%
  \textbf{\bibinfo {volume} {196}},\ \bibinfo {pages} {102} (\bibinfo {year}
  {1993})%
  \bibAnnoteFile{NoStop}{PA.93.Klafter}%
\bibitem{PRL.06.Zhao}%
  \BibitemOpen
  \bibfield{author}{%
  \bibinfo {author} {\bibfnamefont{H.}~\bibnamefont{Zhao}},\ }%
  \bibfield{journal}{%
  \Doi{10.1103/PhysRevLett.96.140602}{\bibinfo {journal} {Phys. Rev. Lett.}}\
  }%
  \textbf{\bibinfo {volume} {96}},\ \bibinfo {pages} {140602} (\bibinfo {year}
  {2006})%
  \bibAnnoteFile{NoStop}{PRL.06.Zhao}%
\bibitem{PRL.11.Zaburdaev}%
  \BibitemOpen
  \bibfield{author}{%
  \bibinfo {author} {\bibfnamefont{V.}~\bibnamefont{Zaburdaev}}, \bibinfo
  {author} {\bibfnamefont{S.}~\bibnamefont{Denisov}},\ and\ \bibinfo {author}
  {\bibfnamefont{P.}~\bibnamefont{H\"anggi}},\ }%
  \bibfield{journal}{%
  \Doi{10.1103/PhysRevLett.106.180601}{\bibinfo {journal} {Phys. Rev. Lett.}}\
  }%
  \textbf{\bibinfo {volume} {106}},\ \bibinfo {pages} {180601} (\bibinfo {year}
  {2011})%
  \bibAnnoteFile{NoStop}{PRL.11.Zaburdaev}%
\bibitem{s.12.Liu}%
  \BibitemOpen
  \bibfield{author}{%
  \bibinfo {author} {\bibfnamefont{S.}~\bibnamefont{Liu}}, \bibinfo {author}
  {\bibfnamefont{N.}~\bibnamefont{Li}}, \bibinfo {author}
  {\bibfnamefont{J.}~\bibnamefont{Ren}},\ and\ \bibinfo {author}
  {\bibfnamefont{B.}~\bibnamefont{Li}},\ }%
  \bibfield{journal}{%
  \bibinfo {journal} {arXiv:1103.2835}}%
   (\bibinfo {year} {2012})%
  \bibAnnoteFile{NoStop}{s.12.Liu}%
\bibitem{PRB.09.Henry}%
  \BibitemOpen
  \bibfield{author}{%
  \bibinfo {author} {\bibfnamefont{A.}~\bibnamefont{Henry}}\ and\ \bibinfo
  {author} {\bibfnamefont{G.}~\bibnamefont{Chen}},\ }%
  \bibfield{journal}{%
  \Doi{10.1103/PhysRevB.79.144305}{\bibinfo {journal} {Phys. Rev. B}}\ }%
  \textbf{\bibinfo {volume} {79}},\ \bibinfo {pages} {144305} (\bibinfo {year}
  {2009})%
  \bibAnnoteFile{NoStop}{PRB.09.Henry}%
\bibitem{PRB.08.Xu}%
  \BibitemOpen
  \bibfield{author}{%
  \bibinfo {author} {\bibfnamefont{Y.}~\bibnamefont{Xu}}, \bibinfo {author}
  {\bibfnamefont{J.-S.}\ \bibnamefont{Wang}}, \bibinfo {author}
  {\bibfnamefont{W.}~\bibnamefont{Duan}}, \bibinfo {author}
  {\bibfnamefont{B.-L.}\ \bibnamefont{Gu}},\ and\ \bibinfo {author}
  {\bibfnamefont{B.}~\bibnamefont{Li}},\ }%
  \bibfield{journal}{%
  \Doi{10.1103/PhysRevB.78.224303}{\bibinfo {journal} {Phys. Rev. B}}\ }%
  \textbf{\bibinfo {volume} {78}},\ \bibinfo {pages} {224303} (\bibinfo {month}
  {Dec}\ \bibinfo {year} {2008})%
  \bibAnnoteFile{NoStop}{PRB.08.Xu}%
\end{thebibliography}%

\end{document}